\documentclass[twocolumn,aps,pra,amsmath,amssymb,nofootinbib,superscriptaddress]{revtex4-1}

\usepackage{amsmath,amsfonts,amssymb}
\usepackage[usenames]{color}
\usepackage{hyperref}
\usepackage{graphicx}

\newcommand{\be}{\begin{equation}}
\newcommand{\ee}{\end{equation}}
\newcommand{\bea}{\begin{eqnarray}}
\newcommand{\eea}{\end{eqnarray}}
\newcommand{\rr}{\mathbf{r}}
\newcommand{\kk}{\mathbf{k}}
\newcommand{\qq}{\mathbf{q}}

\newcommand{\zero}{\mathbf{0}}

\newcommand{\ii}{\mathrm{i}}
\newcommand{\dd}{\mathrm{d}}
\newcommand{\eee}{\mathrm{e}}

\newcommand{\meanv}[1]{\langle #1 \rangle}

\newcommand{\meanvlr}[1]{\left\langle #1 \right\rangle}

\newcommand{\bb}[1]{\left( #1 \right)}

\newcommand{\bbcro}[1]{\left[ #1 \right]}
\newcommand{\bbcror}[1]{\left. #1 \right]}
\newcommand{\bbcrol}[1]{\left[ #1 \right.}
\newcommand{\bbaco}[1]{\left\{ #1 \right\}}

\renewcommand{\Re}{\textrm{Re}\, }
\renewcommand{\Im}{\textrm{Im}\, }

\begin{document}
\title{Cumulant theory of the unitary Bose gas:  Prethermal and Efimovian dynamics}

\author{V. E. Colussi}
\email[Corresponding author:  ]{colussiv@gmail.com}
\affiliation{Eindhoven University of Technology, PO Box 513, 5600 MB Eindhoven, The Netherlands}
\author{H. Kurkjian}
\affiliation{TQC, Universiteit Antwerpen, Universiteitsplein 1, B-2610 Antwerp, Belgium}
\author{M. Van Regemortel}
\affiliation{TQC, Universiteit Antwerpen, Universiteitsplein 1, B-2610 Antwerp, Belgium}
\affiliation{Joint Quantum Institute, National Institute of Standards and Technology and the University of Maryland, Gaithersburg, MD 20899, USA}
\author{S. Musolino}
\affiliation{Eindhoven University of Technology, PO Box 513, 5600 MB Eindhoven, The Netherlands}
\author{J. van de Kraats}
\affiliation{Eindhoven University of Technology, PO Box 513, 5600 MB Eindhoven, The Netherlands}
\author{M. Wouters}
\affiliation{TQC, Universiteit Antwerpen, Universiteitsplein 1, B-2610 Antwerp, Belgium}
\author{S. J. J. M. F. Kokkelmans}
\affiliation{Eindhoven University of Technology, PO Box 513, 5600 MB Eindhoven, The Netherlands}

\begin{abstract}  
We study the quench of a degenerate ultracold Bose gas to the unitary regime, where interactions are as strong as allowed by quantum mechanics. {We lay the foundations of a cumulant theory able
to capture {simultaneously the three-body Efimov effect} and {ergodic evolution}}.  After an initial period of rapid quantum depletion, a universal prethermal stage is established characterized by a kinetic temperature and an emergent Bogoliubov dispersion law while the {microscopic degrees of freedom} remain far-from-equilibrium.   Integrability is then broken by higher-order interaction terms in the many-body Hamiltonian, leading to a momentum-dependent departure from power law to decaying exponential behavior of the occupation numbers at large momentum.  We find also signatures of the Efimov effect in the many-body dynamics and make a precise identification between the observed beating phenomenon and the binding energy of an Efimov trimer.   Throughout the work, our predictions for a uniform gas are quantitatively compared with experimental results for quenched unitary Bose gases in uniform potentials. 
\end{abstract}

\maketitle

\section{Introduction}
Precision control of interatomic interactions in dilute ultracold quantum gases has made possible remarkable progress in our understanding of strongly-correlated many-body systems. Here, strongly-interacting quantum fluids can be studied in the laboratory, with a great flexibility in the way in which the system is manipulated and probed. Ultracold quantum gases are typically dilute with respect to the range of the specific interatomic interaction and sensitive only to the two-body $s$-wave scattering length $a$, which sets the effective interaction strength \cite{pitaevskii2016bose}. Experiments have typically focused on measuring equilibrium or near-equilibrium properties, such as the equation-of-state or elementary excitations. This picture is realized in two-component Fermi gases 
\cite{Jin2003,Ketterle2003,Grimm2003} even in the unitary regime $n|a|^3\gg1$, where $n$ is the atomic density, \cite{Navon729,Horikoshi442,Ku563,zwerger2011bcs,PhysRevLett.98.020604,Cao58}.  Here, system properties behave universally, scaling continously as powers of the remaining density (Fermi) scales $k_\mathrm{n}=(6\pi^2 n)^{1/3}$, 
$E_\mathrm{n}=\hbar^2k_\mathrm{n}^2/2m$, and $t_\mathrm{n}=\hbar/E_\mathrm{n}$ and can be related to other strongly-interacting Fermi systems such as the inner crust of neutron stars \cite{PhysRevLett.92.160404,PhysRevLett.88.090402,PhysRevLett.87.120406,houcke2012,Levinsen_2017}.  

In ultracold quantum gases where multi-body effects are not suppressed by the Pauli exclusion principle, an infinite number of three-body bound Efimov states form whose finite size and discrete scaling leads to a spectacular departure from this universal paradigm \cite{efimov1971weakly,efimov1979low,BRAATEN2006259,D_Incao_2018,Naidon_2017,RevModPhys.89.035006}.  This includes three-component Fermi gases, whose rich phase diagram is predicted to contain a trimer phase at low densities reminiscent of Quantum Chromodynamics \cite{Nishida2012,Naidon2019}.  It includes also (single-component) Bose gases, where quasi-equilibrated states have been recently achieved through fast ramps onto the unitary regime before loss-induced heating dominates \cite{makotyn2014universal,klauss2017observation,eigen2017universal,eigen2018prethermal}.  Here, the conversion of correlation dynamics into a mixture of free atoms, Feshbach dimers, 
and Efimov trimers was observed in an ultracold Bose gas of $^{85}$Rb 
by sweeping the unitary gas back onto the weakly-interacting regime \cite{klauss2017observation}.  Within a three-body model, this conversion was shown to be dominated by the Efimov trimer 
with size comparable to the interparticle spacing \cite{PhysRevLett.121.023401}, which also leads to an enhanced 
growth of triplet correlations at early-times after the quench \cite{PhysRevLett.120.100401,PhysRevA.99.043604}.  Extending these early-time, few-body studies to Fermi timescales requires that the Efimov effect be woven into a many-body framework, which remains an outstanding theoretical challenge.  

At the same time, performing a deep quench leaves these strongly-interacting systems in a highly-excited state.  Here, different quantities can effectively prethermalize, equilibrating before the system has relaxed to the true thermal equilibrium \cite{PhysRevLett.93.142002}.  Experimentally, signs of a universal prethermal state 
characterized by Bogoliubov scalings (phonons
and free particles at low and high momenta, respectively) 
were observed in a quenched ultracold Bose gas of $^{39}$K \cite{eigen2018prethermal}.  
Whether this prethermal steady-state is due to integrable dephasing dynamics, as in the weakly-interacting regime \cite{PhysRevA.98.053612},
or to ergodic mechanisms is unclear. State-of-the-art integrable theories of the post-quench evolution \cite{PhysRevA.89.021601,PhysRevLett.124.040403,PhysRevA.99.023623,Foster2015} are
by definition unable to capture the relaxation dynamics which must occur in this ergodic system.
Additionally, the usual perturbative inclusion of such processes using Boltzmannian approaches \cite{huang1987statistical} is not justified in this regime where the distinctness of collisions is blurred and all rates are on the order of the Fermi scale.  The challenge of constructing a many-body framework, both ergodic and strongly-interacting, therefore remains central.

In this work, we establish the foundations of a general approach able to capture both the Efimov effect and ergodicity
in far-from-equilibrium, strongly-interacting ultracold Bose gases.  Using the method 
of cumulants, we track the sequential growth of genuine few-body correlations systematically encoded in the cumulants of the quenched many-body system \cite{fricke1996transport,burnett2002,kira2011semiconductor,kira2014excitation,KIRA2015185,kirancomm,Kokkelmans2018,PhysRevA.99.043604,PhysRevLett.120.100401}.  Containing only two-body correlations, the cumulant theory at the doublet level is equivalent to the time-dependent Hartree-Fock 
Bogoliubov and Nozi\`eres-Saint James variational approaches 
studied in Refs.~\cite{PhysRevA.100.013612,Kokkelmans2018,PhysRevA.89.021601,PhysRevA.91.013616,PhysRevLett.124.040403,PhysRevA.99.023623}.  
Here, we show how a universal prethermal stage emerges from the integrable dynamics, providing a framework for the conceptual and 
quantitative understanding of the universal Bogoliubov scalings observed experimentally.  We find that the next truncation level that includes higher-order correlations while respecting the underlying conservation laws is the cumulant theory at the quadruplet level.  Although we provide explicit expressions for the cumulant equations of motion in the quadruplet model, its full simulation remains numerically intractable.  Therefore, we simulate the cumulant theory truncated at the triplet level, which contains already the Efimov effect, as demonstrated in a study of the embedded few-body problem in Ref.~\cite{Kokkelmans2018}.  Within the triplet model, we explore the various manifestations of Efimov physics in the many-body observables, including the instantaneous chemical potential, quantum depletion, pairing field, and two and three-body contacts.  This analysis is performed at times before the violation of energy conservation muddies the long-time dynamics and any physical connections with thermalization.

The organization of this work is as follows.  In Sec.~\ref{sec:mbmodel}, we outline the many-body model, calibrated to reproduce finite-range corrections near resonance and reformulated in the symmetry-breaking picture to describe Bose-condensation in the system.  In Sec.~\ref{sec:eom}, the method of cumulants is introduced and explicit expressions for the cumulant equations of motion are derived, connected with the underlying few-body physics, and the interplay between their truncation and conservation laws is detailed.  In Sec.~\ref{sec:hfb}, the prethermal stage that emerges in the doublet model is analyzed and compared with experiment.  In Sec.~\ref{sec:triplet}, the departure from the prethermal stage and Efimovian dynamics are analyzed in the triplet model, and we conclude in Sec.~\ref{sec:conclusion}.  The more formal and technical discussions in this work can be found in the Appendices.  In Appendix~\ref{app:few}, details of the calibrated, finite-range potential are given along with the resulting Efimov spectrum.  In Appendix~\ref{app:SimEqns}, the cumulant equations are given in a form more suitable for simulation and their numerical implementation is discussed.  In Appendix~\ref{app:quad}, we provide the formal, explicit expressions for the quadruplet equations of motion and discuss their solution.  In Appendix~\ref{app:connect}, we connect the cumulant theory outlined in this work with alternative approaches found in the literature.

\section{Many-Body Model}\label{sec:mbmodel}
In this work, we study a quenched uniform gas of degenerate bosons in a cubic volume $V$.  We consider short-range single-channel interactions that capture the broad, entrance-channel dominated Feshbach resonances used experimentally \cite{RevModPhys.82.1225,makotyn2014universal,klauss2017observation,eigen2018prethermal,eigen2017universal}.  First, we introduce the many-body Hamiltonian in Sec.~\ref{sec:hamiltonian} and discuss the potential parameters calibrated to match finite-range corrections near resonance, referring the interested reader to Appendix~\ref{app:few} for more details.  In Sec.~\ref{sec:breaking} we move to the symmetry-breaking picture to describe Bose condensation in the system.  In Sec.~\ref{sec:expansion}, the many-body Hamiltonian is reformulated in preparation for the cumulant expansion in the following section (Sec.~\ref{sec:eom}).

\subsection{Hamiltonian}\label{sec:hamiltonian}
In an ultracold Bose gas, atoms interact through a local $s$-wave pairwise potential $\langle {\bf r}_{\rm in} |\hat{V}|{\bf r}_{\rm out}\rangle=V(|{\bf r}_{\rm in}|)\delta^{(3)}({\bf r}_{\rm out}-{\bf r}_{\rm in})$ with relative coordinates ${\bf r}_{\rm in}$ and ${\bf r}_{\rm out}$ \cite{taylor2006scattering} of the two incoming and outgoing atoms.  The corresponding many-body Hamiltonian is given by
\begin{align}
\hat{H} =& \int d^3 r \hat{\psi}^\dagger(\rr) \left( - \frac{\hbar^2}{2m}  \Delta_{\rr}\right)  \hat{\psi}(\rr) \nonumber\\
&+ \frac{1}{2} \int \dd^3 r \dd^3 r'   \hat{\psi}^\dagger(\rr) \hat{\psi}^\dagger(\rr') V(|\rr-\rr'|) \hat{\psi}(\rr'){\hat{\psi}}(\rr) ,
\label{hamiltonianR}
\end{align}
where ${\bf r}_{\rm in}=\rr-\rr'$ is the relative position for incoming particles located at $\rr$ and $\rr'$.
To diagonalize the kinetic energy part of this hamiltonian, we introduce the Fourier
operators $\hat\psi(\rr)=(1/\sqrt{V})\sum_\kk \hat{a}_\kk \eee^{\ii \kk\cdot\rr}$ for a uniform gas occupying a cubic volume $V$, which can be taken to infinity in the thermodynamic limit.
In Fourier space, this Hamiltonian reads
\bea
\hat{H} = \sum_{\kk} \epsilon_\kk \hat{a}_\kk^\dagger \hat{a}_\kk + \frac{1}{2V}\sum_{\kk,\kk',\qq} V_\qq \hat{a}_{\kk'+\qq}^\dagger \hat{a}_{\kk-\qq}^\dagger \hat{a}_\kk \hat{a}_{\kk'} ,
\label{hamiltonianF}
\eea
where $\epsilon_{\bf k}=\hbar^2 k^2/2m$ is the one-body kinetic energy, and the Fourier components of the local potential are given by 
\begin{equation}
\langle {\bf k}|\hat{V}|{\bf k'}\rangle=\quad V_{{\bf k'}-{\bf k}}=\int d^3 r\ e^{\ii\rr\cdot({\bf k-k'})}V(|\rr|) \label{Vk},
\end{equation}
which depends only on the magnitude of the difference in relative momenta ${\bf k}$ and ${\bf k'}$.

The physical properties of ultracold Bose gases are typically characterized by a single parameter, the two-body $s$-wave scattering length $a$, which sets the effective strength of two-body interactions and can be adjusted precisely by tuning the binding energy of a Feshbach molecule via external magnetic fields \cite{pitaevskii2016bose,RevModPhys.82.1225}.  On resonance, the cross section becomes independent of the scattering length in the unitarity limit $\sigma=8\pi/k^2$ \cite{taylor2006scattering}.  The gas is both dilute with respect to the range of the interatomic interaction parametrized by the van der Waals length $r_\mathrm{vdW}=(mC_6/\hbar^2)^{1/4}/2$, where $m$ is the atomic mass and $C_6$ is the dispersion coefficient associated with the van der Waals interaction between neutral ground-state atoms \cite{RevModPhys.82.1225}, while being simultaneously strongly-interacting $|a|/r_\mathrm{vdW}\gg1$.  The short-range details of the potential are therefore relatively unimportant, and there is freedom in choosing the potential.  All formulas in the main text are therefore given in terms of a local potential for concision, but the numerical calculations are actually performed with a separable pairwise potential (see Appendix~\ref{app:few}) with renormalized effective interaction strength $g=U_0\Gamma$ where $U_0=4\pi\hbar^2a/m$ and $\Gamma=1/(1-2a\Lambda/\pi)$.  To match finite-range effects in the vicinity of the Feshbach resonance, the relative momentum cutoff is calibrated as $\Lambda=2/\pi\bar{a}$, where $\bar{a}=4\pi r_\mathrm{vdW}/\Gamma(1/4)^2\approx 0.956r_\mathrm{vdW}$ is the mean scattering length and $\Gamma(x)$ is the Gamma function \cite{Kokkelmans2018,burnett2003,PhysRevA.59.1998}.  This gives $g=-\pi^3 \hbar^2\bar{a}/m$ for the effective interaction strength at unitarity.  Importantly, this calibration has consequences on the three-body level for the spectrum of Efimov states, and we refer the interested reader to Appendix~\ref{app:few} for more details on the few-body physics contained in this model.

\subsection{Symmetry-breaking picture}\label{sec:breaking}
The gas is initially condensed in the $k=0$ mode, which means that the population
\be
N_0\equiv\meanv{\hat a_0^\dagger \hat a_0} \lesssim N
\label{condensate}
\ee
is macroscopic. We describe only evolution that preserves this property, which, for very energetic quenches
where all particles are eventually ejected out of the condensate, restricts us to short times.
We use the symmetry-breaking picture \footnote{In Appendix \ref{app:NC}, we show 
that our symmetry-breaking approach is equivalent to a number-conserving one in the thermodynamic limit.} to
describe the dynamics the condensate: the condensate operator $\hat a_0$ is replaced by a wavefunction $\psi_0=\langle \hat{a}_0\rangle/\sqrt{V}$
acting as an order parameter. The Gross-Pitaevskii equation (GPE) describing the dynamics of this wavefunction $\psi_0$ is
obtained by treating $\hat H$ as a classical hamiltonian, and $\psi_0$ and $\psi_0^*$ as 
canonically conjugated variables:
\begin{align}
\ii\hbar\partial_t \psi_0=\meanvlr{\frac{\partial\hat H}{\partial\psi_0^*}}=&V_{\bf 0}n\psi_0+\psi_0\frac{1}{V}\sum_{\kk\neq0} V_\kk \meanv{\hat a_\kk^\dagger\hat a_\kk}\nonumber\\
&+ \psi_0^*\frac{1}{V}\sum_{\kk\neq0} V_\kk \meanv{\hat a_{-\kk}\hat a_\kk} \nonumber\\
&+ \frac{1}{V^{3/2}}\sum_{\kk,\qq\neq0} V_\qq \meanv{\hat a_{\kk+\qq}^\dagger\hat a_{\kk}\hat a_{\qq}},
\label{GPE}
\end{align}
where $n=N/V$ is the total atomic density ($n_0=N_0/V=|\psi_0|^2$ being the density of condensed particles).

To eliminate the condensate variables and focus on the dynamics of
the excited fraction, we decompose the condensate wavefunction into its modulus and phase, 
\be
\psi_0 = \sqrt{n_0}\eee^{\ii\theta_0},
\ee
and introduce the operators unrotated by the condensate phase:
\be
\hat b_\kk = \eee^{-\ii\theta_0}\hat a_\kk.
\ee
The dynamics of the $\hat b$ operators now incorporates the evolution of $\theta_0$ \footnote{To avoid restrictions on the summations, we use the convention $\hat b_\zero=0$.}
\begin{align}
&\ii\hbar\partial_t \hat b_\kk=[\hat b_\kk,\hat H_b],\nonumber\\
&\quad\mbox{with} \quad \hat{H}_b=\hat{H} + \hbar\partial_t \theta_0\sum_{\kk} \bb{\hat b_\kk^\dagger\hat b_\kk - \meanv{\hat b_\kk^\dagger\hat b_\kk}}.
\label{dtb}
\end{align}
We note that the summation involving $\meanv{\hat b_\kk^\dagger\hat b_\kk}$ has been trivially added to $\hat H_b$ to ensure that $\meanv{\hat H_b}=\meanv{\hat H}$.
In $\hat H_b$, the number of particles in the condensate is no longer treated as an independent variable and is related to the $\hat b$ field
and total number of particles by the conservation equation:
\be
 N_0=N-\sum_{\kk} \meanv{\hat b_\kk^\dagger\hat b_\kk}.
 \label{N0}
 \ee
The phase derivative can also be expressed in terms of the $\hat b$, which
finally eliminates the condensate variables from the dynamics:
\begin{align}
\hbar\frac{\dd\theta_0}{\dd t}=&-\frac{1}{2n_0} \bb{\psi_0^*\ii\hbar \frac{\dd\psi_0}{\dd t}-\ii\hbar \frac{\dd\psi_0^*}{\dd t}\psi_0},\\
=&-\Bigg[V_\zero n+\frac{1}{V}\sum_\kk \bbcro{V_\kk \meanv{\hat b_\kk^\dagger \hat b_\kk} +{V_\kk}\Re \meanv{\hat b_{-\kk} \hat b_\kk}}\nonumber\\
&+\frac{1}{\sqrt{n_0V^3}} \sum_{\kk\qq} V_\qq {\Re\meanv{\hat b_{\kk+\qq}^\dagger \hat b_\kk \hat b_\qq}}\Bigg].\label{thetapoint}
\end{align} 
In section \ref{sec:hfb}, we use the interpretation of this equation as a second Josephson relation $\hbar{\partial_t \theta_0}=-\mu(t)$
to generalize the notion of an instantaneous chemical potential to our out-of-equilibrium system \cite{Witkowska2009,Sinatra2016}.

\subsection{Expansion of the Hamiltonian}\label{sec:expansion}
We start by expanding the many-body Hamiltonian $\hat H_b$ in powers of the non-condensed field $\hat b$
\begin{align}
\hat H_b =& E_0(t)+\hat H_2 +\hat H_3 +\hat H_4\label{Hb}, \\
E_0 =& \frac{V_\zero N_0^2}{2V}-\hbar\partial_t \theta_0(N-N_0), \\
\hat H_2 =& \sum_{\kk} \Big(\bbcro{\epsilon_\kk+(V_\kk+V_\zero) n_0 +\hbar\partial_t \theta_0 }\hat b_\kk^\dagger\hat b_\kk \nonumber\\
&+ \frac{V_\kk n_0}{2}[\hat b_{-\kk}\hat b_\kk + \hat b_\kk^\dagger\hat b_{-\kk}^\dagger]\Big), \\
\hat H_3 =&   \sqrt{\frac{n_0}{V}}\sum_{\kk,\qq} V_\qq \bb{\hat b_{\kk+\qq}^\dagger\hat b_{\kk}\hat b_{\qq}+\mbox{h.c.}}, \\
\hat H_4 =& \frac{1}{2V}\sum_{\kk,\kk',\qq} V_\qq b_{\kk'+\qq}^\dagger \hat{b}_{\kk-\qq}^\dagger \hat{b}_\kk \hat{b}_{\kk'}. 
\end{align}
The usual Bogoliubov approach ($\hat H_3=\hat H_4=0$) reduces the many-body Hamiltonian to quadratic form, and is justified by an expansion in powers of $na^3$ \cite{pitaevskii2016bose,PhysRevA.98.053612}.  This approach describes two-body processes at the level of the Born approximation, which will not give the correct unitarity limit $\sigma(k)=8\pi/k^2$ of the $s$-wave partial cross-section \cite{taylor2006scattering}.  To overcome this and produce a theory that reproduces the Hartree-Fock Bogoliubov equations at lowest order \cite{Ripka1985}, we rewrite the many-body Hamiltonian by adding and subtracting the partial contraction \footnote{A partial contraction is the replacement 
$\hat a \hat b \hat c \hat d \to \meanv{\hat a \hat b} \hat c \hat d + \hat a \hat b \meanv{\hat c \hat d} + \meanv{\hat a \hat c} \hat b \hat d+\ldots$} 
of $\hat H_4$ defined as
\begin{equation}
\delta \hat H_2 = \frac{1}{2} \sum_{\kk} \bbcro{ \delta\Delta_\kk^* \hat  b_{-\kk} \hat  b_{\kk} + \mbox{cc.}} +\sum_{\kk} \delta\epsilon_\kk \hat  b_{\kk}^\dagger \hat  b_{\kk},
\end{equation}
with
\begin{align}
&\Delta_\kk = \frac{1}{V}\sum_\qq V_\qq \meanv{\hat  b_{-\kk-\qq} \hat  b_{\kk+\qq} } ,\\
&\delta\epsilon_\kk  = \frac{1}{V}\sum_\qq (V_{\bf 0}+V_\qq) \meanv{\hat  b_{\kk+\qq}^\dagger \hat  b_{\kk+\qq}}.
\label{defHHFB}
\end{align}
This yields an effective quartic $\hat H_4^{\rm eff}=\hat H_4-\delta\hat H_2$ and quadratic hamiltonian:
\be
\hat{H}_2^{\rm eff}=\hat{H}_2+ \delta\hat H_2\equiv \sum_{\kk} \bb{{E_\kk}\hat b_\kk^\dagger\hat b_\kk + \bbcro{\frac{\Delta_\kk^*}{2}\hat b_{-\kk}\hat b_\kk +\textrm{h.c.}}},
\ee
whose diagonal and anomalous matrix elements are respectively
\bea
E_\kk&=& \epsilon_\kk+(V_{\bf 0}+V_\kk) n_0+\delta\epsilon_\kk +\hbar\partial_t\theta_0, \label{Ek} \\
\Delta_\kk &=& V_\kk n_0 +\delta\Delta_\kk. \label{Deltak}
\eea
In the following section, we use the cumulant expansion method to construct equations of motion from this reformulated many-body Hamiltonian.

\section{Equations of Motion}\label{sec:eom}
Prior to the quench, all $N$ bosons in the gas are prepared in a non-interacting uniform Bose condensate at zero temperature such that $n_0=n$.  A sudden projection of the pure condensate into the unitary regime approximates the effect of the rapid interaction quench.  The fully-condensed initial state is actually a highly-excited state in the strong-coupling regime (for comparison, the ground state of superfluid $^{4}$He has a condensed fraction of the order of 0.07 \cite{Stirling2000}), and the gas begins to rapidly quantum deplete such that $N-N_0$ becomes comparable to $N$. As the gas evolves, correlations begin to develop amongst excitations, and the system becomes strongly-correlated.  Correlations that intrinsically relate larger numbers of excitations however develop {\it sequentially}, beginning from the generation of correlated pairs out of the condensate \cite{kira2014excitation,KIRA2015185,kirancomm,PhysRevA.99.043604,PhysRevLett.120.100401}.  We can use this picture to construct a many-body description of this far-from-equilibrium, strongly-interacting system by systematically including intrinsically higher-order effects into our theory, using the method of cumulants.  In this section, we outline the cumulant theory, beginning in Sec.~\ref{sec:hierarchy} with an introduction to the cumulant hierarchy.  In Sec.~\ref{sec:conslaws} we detail how truncating this hierarchy impacts the underlying conservation laws.  In Sec.~\ref{sec:explicit}, the cumulant equations of motion are given explicitly, and in Sec.~\ref{sec:formal} we discuss how they may be solved in a way that reveals the underlying few-body physics at each level of the hierarchy.  

\subsection{Hierarchy of cumulants}\label{sec:hierarchy}

\begin{table}[ht!]
\caption{Relations between the cumulant $\meanv{\hat O}_c$ and the quantum average value (the moment) $\meanv{\hat O}$ for operators up to the quadruplet level. The one-body operators $\hat{a}$, $\hat{b}$, $\hat{c}$, and $\hat{d}$ $\in\{\hat{b}_\kk,\hat{b}_\kk^\dagger,\kk\neq 0 \}$ are normally ordered. The cancelation of the singlets $\meanv{\hat{a}}_c=0$ (used implicitly in the third and fourth line of the table) is a consequence of the spatial homogeneity of the gas.}
\label{table:cumulants}
\centering
\begin{ruledtabular}
\begin{tabular}{c  c }

Cumulant Order & Moment Expansion\\
\noalign{\smallskip}
\hline
\noalign{\smallskip}
Singlet & $\langle \hat{a}\rangle_c=\langle \hat{a}\rangle=0 $\\
\noalign{\smallskip}
Doublet & $\langle \hat{a}\hat{b}\rangle_c=\langle \hat{a}\hat{b}\rangle-\langle \hat{a}\rangle_c\langle \hat{b}\rangle_c=\langle \hat{a}\hat{b}\rangle$ \\
\noalign{\smallskip}
Triplet & $\langle \hat{a}\hat{b}\hat{c}\rangle_c=\langle \hat{a}\hat{b}\hat{c}\rangle $\\
\noalign{\smallskip}
Quadruplet & $\langle \hat{a}\hat{b}\hat{c}\hat{d}\rangle_c=\langle \hat{a}\hat{b}\hat{c}\hat{d}\rangle -\langle\hat{a}\hat{b}\rangle_c\langle\hat{c}\hat{d}\rangle_c-\langle\hat{a}\hat{c}\rangle_c\langle\hat{b}\hat{d}\rangle_c$\\
&$-\langle\hat{a}\hat{d}\rangle_c\langle\hat{b}\hat{c}\rangle_c$\\
\noalign{\smallskip}
\dots & \dots\\
\end{tabular}
\end{ruledtabular}
\end{table}
To describe the coupled-correlation dynamics, we introduce the cumulant of a $p$-body operator as 
\begin{align}
\meanvlr{\prod_{i=0}^l\hat{b}_{\kk_i}^\dagger {\prod_{j=0}^m \hat{b}_{\kk_j'}}}_c&=(-1)^m \prod_{i=0}^l \frac{\partial}{\partial x_i} \prod_{j=0}^m \frac{\partial}{\partial y_j^*}\nonumber\\
&\times \ln\left. \meanvlr{\eee^{\sum_{i=0}^l x_i \hat{b}_{\kk_i}^\dagger} \eee^{\sum_{j=0}^m y_j^*\hat{b}_{\kk_j'}}}\right\vert_{{\bf x},{\bf y}=0}. \label{formalcumulant}
\end{align}
We call the cumulant of an $p$-body operator (here $p=l+m$), a ``$p$-uplet''.
In practice, a $p$-uplet is obtained by subtracting from the quantum average value 
(the ``moment'' of the $p$-body operator) all the possible contractions
into products of $n$-body operator average, with $n<p$ \cite{fricke1996transport,burnett2002}.  
This recursive definition of the cumulants is shown in Table~\ref{table:cumulants} up to the quadruplet level.
In the homogeneous system considered here, only the cumulants that conserve the total momentum
(that is, verify $\sum_i \kk_i=\sum_j \kk_j'$, in the notations of Eq.~\eqref{formalcumulant}) can become nonzero
during the time evolution. This implies in particular that the singlets $\meanv{\hat b_{\kk}}_{\kk\neq 0}$ remain zero at all times.

Due to the cubic and quartic parts of the many-body Hamiltonian ($\hat H_3$ and $\hat H_4^{\rm eff}$, respectively) the doublet dynamics couple to triplets and quadruplets.  Therefore, the depletion of the condensate into opposite momentum pairs in turn will sequentially generate higher-order few-body correlations, beginning at the three and four-body levels.  At the next level of the hierarchy, the triplets couple to doublets, quadruplets, and quintuplets, and this trend is repeated to all orders.  In practice, this hierarchy must be truncated, which limits the range of validity of the model to times before higher-order few-body correlations become non-negligible \cite{KIRA2015185}. We address truncation of the cumulant hierarchy in the following section.

\subsection{Truncation scheme and conservation laws}\label{sec:conslaws}
When the time-evolution of the many-body system is described only approximately, namely, in a truncated cumulant expansion, it is unclear whether the same constants of motion associated with the many-body Hamiltonian arise \cite{Ripka1985}.  Therefore, it is not guaranteed {\it a priori} that truncation at a given level of cumulants results in a theory which respects all of the underlying conservation laws. With that caveat, we note that all of the truncation schemes studied in this
work conserve the average number of atoms by construction (see Eq.~\eqref{N0}). We discuss now in detail the interplay between truncation order and the conservation of energy.  

\paragraph{Doublet truncation}  
The simplest model within the cumulant theory (the ``doublet model''), which corresponds to the Hartree-Fock-Bogoliubov (HFB) theory \cite{Ripka1985}, can be constructed by keeping only the doublets while
setting all higher-order cumulants
to zero\footnote{In the formal equations of motion \eqref{doublets0}, \eqref{doublets2}, \eqref{triplets2} and \eqref{doublets1}--\eqref{quadruplets1},
the operators $\hat a, \hat b, \hat c, \hat d \in \{\hat b_\kk^\dagger,\hat b_\kk\}_{\kk\neq 0}$ are normally ordered}. This yields the equations of motion:
\bea
\ii\hbar\partial_t \meanv{\hat a \hat b} &\underset{\rm Doub.}{\simeq}& \meanv{[\hat a \hat b,\hat H_2^{\rm eff}]}, \label{doublets0}
\eea 
where we have used the abbreviation ``Doub.'' to indicate this particular truncation scheme.
The total energy $E\equiv\meanv{\hat H}=\meanv{\hat H_b}$ is here approximated by $E\underset{\rm Doub.}{\simeq}\meanv{\hat H_2^{\rm eff}}$,
and its time-derivative vanishes as can by checked by summing over the doublets 
$\hat a \hat b$ in Eq.~\eqref{doublets0} to form the derivative of $\hat H_2^{\rm eff}$. 
Alternatively, these conclusions are anticipated by the variational derivation \cite{Ripka1985}. Simulation results for the doublet model are the subject of Sec.~\ref{sec:hfb}.

\paragraph{Triplet truncation}  
To go beyond the doublet model, one can first choose to retain also the triplets (the ``triplet model'') in the truncation scheme.  This yields the equations of motion:
\bea
\ii\hbar\partial_t \meanv{\hat a \hat b} &\underset{\rm Tri.}{\simeq}& \meanv{[\hat a \hat b,\hat H_2^{\rm eff}+\hat H_3]}, \label{doublets2} \\ 
\ii\hbar\partial_t\meanv{\hat a \hat b \hat c} &\underset{\rm Tri.}{\simeq}& \meanv{[\hat a \hat b \hat c,\hat H_b]}-\meanv{[\hat a \hat b \hat c,\hat H_3+\hat H_4^{\rm eff}]}_c, \label{triplets2} 
\eea
where we have used the abbreviation ``Tri.'' to indicate truncation at the triplet level.  
From the exact time derivative $\ii\hbar\partial_t \meanv{\hat a \hat b} =\meanv{[\hat a \hat b,\hat H_b]}$, the triplet truncation subtracts $\meanv{[\hat a \hat b,\hat H_4^{\rm eff}]}$, which is by construction composed only of quadruplets, resulting in Eq.~\eqref{doublets2}.  From the exact time derivative $\ii\hbar\partial_t \meanv{\hat a \hat b \hat c} =\meanv{[\hat a \hat b \hat c,\hat H_b]}$, it subtracts the quadruplets and quintuplets contained in $\meanv{[\hat a \hat b \hat c ,\hat H_3]}$ and $\meanv{[\hat a \hat b \hat c ,\hat H_4^{\rm eff}]}$, respectively, while the corresponding doublet-doublet and triplet-doublet contributions remain in Eq.~\eqref{triplets2}.  Additionally, the triplet truncation of the total energy is $E\underset{\rm Tri.}{\simeq}\meanv{\hat H_2^{\rm eff}+\hat H_3}$, and its time derivative does {\it not} vanish:
\begin{align}
\ii\hbar\partial_t E  &\underset{\rm Tri.}{\simeq} \meanv{[\hat H_3,\hat H_4^{\rm eff}]} - \meanv{[\hat H_3,\hat H_4^{\rm eff}]}_c\neq0.\label{energy2}
\end{align}
This can be obtained by summing over the doublets and triplets in Eqs.~(\ref{doublets2} and (\ref{triplets2}) to form the time derivatives of $\hat H_2^{\rm eff}$ and $\hat H_3$ respectively.
From the above remarks, the origin of this violation is therefore clear:  whereas the cumulant equations of motion (Eqs.~\eqref{doublets2} and \eqref{triplets2}) follow from the full Hamiltonian $\hat{H}_b$, the energy is computed from the truncated Hamiltonian $\hat{ H}_2^{\rm eff}+\hat{H}_3$.
Simulation results for the triplet model are the subject of Sec.~\ref{sec:triplet}, and energy violation results can be found in Appendix~\ref{app:SimEqns}.  

\paragraph{Quadruplet truncation}  
Going beyond the doublet model in a way that does not violate
energy-conservation therefore requires the addition of quadruplets (the ``quadruplet model'') so that the energy is computed from the full Hamiltonian $\hat{H}_b$.  This yields the equations of motion: 
\bea
\ii\hbar\partial_t \meanv{\hat a \hat b} &=& \meanv{[\hat a \hat b,\hat H_b]} ,\label{doublets1} \\ 
\ii\hbar\partial_t\meanv{\hat a \hat b \hat c} &\underset{\rm Quad.}{\simeq}& \meanv{[\hat a \hat b \hat c,\hat H_b]}-\meanv{[\hat a \hat b \hat c,\hat H_4^{\rm eff}]}_c, \label{triplets1} \\
\ii\hbar\partial_t\meanv{\hat a \hat b \hat c \hat d} &\underset{\rm Quad.}{\simeq}& \meanv{[\hat a \hat b \hat c \hat d,\hat H_b]} -\meanv{[\hat a \hat b \hat c \hat d,\hat H_3]}_{c}\nonumber\\
&&-\meanv{[ \hat a \hat b \hat c \hat d,\hat H_4^{\rm eff}]}_{\rm c},  \label{quadruplets1}
\eea
where we have used the abbreviation ``Quad.'' to indicate truncation at the quadruplet level\footnote{The equation of motion of the cumulant $\meanv{\hat a \hat b \hat c \hat d}_c$
is deduced from that of the moment Eq.~\eqref{quadruplets1} using
$
\ii\hbar\partial_t\meanv{\hat a \hat b \hat c \hat d}-\ii\hbar\partial_t\meanv{\hat a \hat b \hat c \hat d}_c=\meanvlr{\bbcro{ \meanv{\hat a \hat b}\hat c \hat d+\meanv{\hat c \hat d}\hat a \hat b ,\hat H_b}}
+\meanvlr{\bbcro{\meanv{\hat a \hat c}\hat b \hat d+\meanv{\hat b \hat d}\hat a \hat c+ \meanv{\hat a \hat d}  \hat b \hat c+\meanv{\hat b \hat c} \hat a \hat d ,\hat H_b}}.
$
In practice, this removes from \eqref{quadruplets1} $(i)$ the non quadruplet part of $\meanv{[\hat a \hat b \hat c \hat d,\hat H_2^{\rm eff}]}$ and $(ii)$ the ``reducible'' contractions in $\meanv{[\hat a \hat b \hat c \hat d,\hat H_3+\hat H_4^{\rm eff}]}$, \textit{i.e} those where a doublet is formed from two elements in $\{\hat a, \hat b, \hat c, \hat d\}$.}.  Although the doublet equations of motion are now exact, the quadruplet truncation scheme subtracts $\meanv{[\hat a \hat b \hat c,\hat H_4^{\rm eff}]}_c$ from the exact time derivative $\ii\hbar\partial_t\meanv{\hat a \hat b \hat c} = \meanv{[\hat a \hat b \hat c,\hat H_b]}$ to produce Eq.~\eqref{triplets1} and subtracts the quintuplets $\meanv{[\hat a \hat b \hat c \hat d,\hat H_3]}_c$ and sextuplets $\meanv{[\hat a \hat b \hat c \hat d,\hat H_4^{\rm eff}]}_c$ from the exact time derivative $\ii\hbar\partial_t\meanv{\hat a \hat b \hat c \hat d} = \meanv{[\hat a \hat b \hat c \hat d,\hat H_b]}$ to produce Eq.~\eqref{quadruplets1}.  The quadruplet model trivially conserves
 the total energy because the full Hamiltonian $\hat H_b$ is used to evolve both the energy and cumulants.  Although the quadruplet model is not simulated in this work due to the large resource requirements, with the size of a $p$-dimensional cumulant array scaling roughly as $\Lambda^{p-1}$ (see Appendix~\ref{app:SimEqns}), we give the general cumulant equations in the following section.  

\subsection{Cumulant equations of motion}\label{sec:explicit}
We now give the equations of motion for the doublet, triplet
and quadruplet cumulants (Eqs.~(\ref{doublets1}--\ref{quadruplets1}) within the quadruplet model. We use Greek letters $\alpha$, $\beta$,
$\gamma$\ldots to denote the wavevector indices of the considered cumulants,
and we keep the bold letters $\kk$, $\qq$ for the wavevectors which are summed over.
The cumulants that compose the closed system of equations of motion are denoted:
\begin{align}
n_\alpha &= \meanv{\hat b_\alpha^\dagger \hat b_\alpha},  &           c_\alpha &=\meanv{\hat b_{-\alpha} \hat b_\alpha},  \nonumber\\
M_{\alpha,\beta} &= \meanv{\hat b_{\alpha-\beta}^\dagger \hat b_\beta^\dagger\hat b_{\alpha}},  &    R_{\alpha,\beta} &=\meanv{\hat b_{\beta-\alpha} \hat b_\alpha \hat b_{-\beta}}, \nonumber\\
Q_{\alpha,\beta;\gamma} &= \meanv{\hat b_{\alpha+\beta-\gamma}^\dagger \hat b_\gamma^\dagger \hat b_\alpha \hat b_\beta}_c,
& P_{\alpha,\beta,\gamma} &=\meanv{\hat b_{\alpha+\beta+\gamma}^\dagger \hat b_\alpha \hat b_\beta \hat b_\gamma}_c,\nonumber\\
 T_{\alpha,\beta,\gamma} &=\meanv{\hat b_{-\alpha-\beta-\gamma} \hat b_\alpha \hat b_\beta \hat b_\gamma}_c.&&
\end{align}
To obtain compact and readable expressions,
one should exploit the invariance of the cumulants under permutation
of their indices (for example $M_{\alpha+\beta,\beta}$ is invariant under the exchange
of $\alpha$ and $\beta$, $P_{\alpha,\beta,\gamma}$ is invariant under the exchange
of $\alpha$, $\beta$ and $\gamma$). For this purpose, we introduce the symmetrizer
$\mathcal{S}_{\alpha_1,\ldots,\alpha_n}$ which sums all the values of a function
$f(\alpha_1,\ldots,\alpha_n)$ obtained after permutation of its arguments:
\be
\mathcal{S}_{\{\alpha_1,\ldots,\alpha_n\}}\bbcro{f(\alpha_1,\ldots,\alpha_n)}=\sum_{\sigma\in \mathfrak{S}(n)} f(\alpha_{\sigma(1)},\ldots,\alpha_{\sigma(n)}),
\label{permu}
\ee
where $\mathfrak{S}(n)$ is the set of permutations of $\{1,\ldots n\}$.
For the cumulant $Q$, which obeys the symmetry relation 
$Q_{\alpha,\beta;\gamma}^*=Q_{\gamma,\alpha+\beta-\gamma;\alpha}$,
we will also need the antisymmetrizer:
\be
\mathcal{A}_{\{(\alpha,\beta),(\gamma,\delta)\}} [f(\alpha,\beta;\gamma,\delta)]=  f(\alpha,\beta;\gamma,\delta) -  [f(\gamma,\delta;\alpha,\beta)]^*.
\ee
All the equations of motion we give here can be checked using the computer algebra program available online at~\cite{repository}.

Let us first reexpress the coefficients of $\hat H_2^{\rm eff}$ (Eqs.~\eqref{Ek}--\eqref{Deltak}) and the phase derivative (Eq.~\eqref{thetapoint})
in terms of the doublets and triplets
\begin{align}
E_\alpha=& \epsilon_\alpha+V_{\bf 0}n+V_\alpha n_0+\frac{1}{V}\sum_{\qq}V_\qq n_{\alpha+\qq} +\hbar\partial_t\theta_0,\label{eq:kerenorm} \\
\Delta_\alpha =& V_\alpha n_0 + \frac{1}{V} \sum_{\qq} V_\qq c_{\alpha+\qq}, \label{eq:delta}\\
\hbar\frac{\dd\theta_0}{\dd t}=&-\Bigg[V_{\bf 0}n+\frac{1}{V}\sum_{\qq} V_\qq (n_\qq +\Re c_\qq)\nonumber\\
&+\frac{1}{\sqrt{n_0V^3}}\sum_{\kk,\qq} V_\qq \Re M_{\kk+\qq,\kk}^*\Bigg], \label{eq:thetadot}
\end{align}
where $E_\mathrm{\alpha}$ and $\Delta_\alpha$ are the expressions for the Hartree-Fock hamiltonian and pairing field, respectively, in the rotating frame \cite{proukakis1996generalized,Ripka1985}. 

For the doublet equations of motion (assuming the invariance 
of the triplets under parity, $M_{-\alpha,-\beta}=M_{\alpha,\beta}$ and 
$R_{-\alpha,-\beta}=R_{\alpha,\beta}$), we have:
\begin{widetext}
\bea
\ii\hbar\partial_t n_\alpha &=& 2\ii\Im\bb{\Delta_\alpha c_\alpha^*+\sqrt{\frac{n_0}{V}}\sum_{\qq} \bbcro{V_\qq M_{\alpha,\qq}^*-(V_\qq+V_\alpha)M_{\alpha+\qq,\alpha}^*}+\frac{1}{V}\sum_{\kk,\qq}V_\qq Q_{\alpha+\qq,\kk;\alpha}},  \label{n}\\
\ii\hbar\partial_t c_\alpha &=& 2E_\alpha c_\alpha+\Delta_\alpha (2n_\alpha+1)+2\sqrt{\frac{n_0}{V}}\sum_\qq\bbcro{V_\qq R_{\alpha,-\qq}+(V_\alpha+V_\qq)M_{\qq,\alpha}^*}+\frac{2}{V}\sum_{\kk,\qq}V_\qq P_{\alpha,\qq-\alpha,\kk}\label{c}.\label{eq:kineticeq}
\eea
\end{widetext}
We note that these doublet equations of motion are equivalent to the Hyperbolic Bloch equations discussed in Ref.~\cite{KIRA2015185}. 

For the triplet equations of motion, we have
\begin{widetext}
\begin{align}
\ii\hbar\partial_t M_{\alpha+\beta,\beta} =&\left(E_{\alpha+\beta}-E_\alpha-E_{\beta}\right)M_{\alpha+\beta,\beta}-\Delta_{\alpha}^* M^*_{\beta,\alpha+\beta}-\Delta_\beta^* M^*_{\alpha,\alpha+\beta}+\Delta_{\alpha+\beta} R_{\alpha,\alpha+\beta}^*\nonumber\\
& +\mathcal{M}^{H_3}_{\alpha,\beta}+\mathcal{M}^{H_4}_{\alpha,\beta}, \label{eq:Mdot}\\
\ii\hbar\partial_t R_{\alpha,\alpha+\beta} =& \left(E_\alpha+E_\beta+E_{\alpha+\beta}\right)R_{\alpha,\alpha+\beta}+\Delta_\alpha M_{-\alpha,\beta}^*+\Delta_{\beta} M_{-\beta,\alpha} +\Delta_{\alpha+\beta} M_{\alpha+\beta,\alpha}^* \nonumber\\
&+\mathcal{R}^{H_3}_{\alpha,\beta}+\mathcal{R}^{H_4}_{\alpha,\beta},\label{eq:Rdot}
\end{align}
\end{widetext}
where we have written separately the contribution of the cubic and
quartic Hamiltonians. The former contains both doublet products and quadruplets
\begin{widetext}
\bea
\frac{\mathcal{M}^{H_3}_{\alpha,\beta}}{\sqrt{n_0/V}}&=&\mathcal{S}_{\{\alpha,\beta\}} \Bigg{[}V_\alpha\bbcro{n_\alpha n_\beta-n_{\gamma}(1+n_\alpha+n_\beta)}
- c_{\gamma} {c_\alpha^*\bb{V_{\gamma}+V_\beta}}  - n_{\gamma}{c_\alpha^*\bb{V_{\gamma}+V_\alpha}}+n_\alpha c_\beta^* (V_\beta+V_{\gamma}) \notag
\\ && \bbcror{+\sum_{\qq}\bbaco{\frac{V_{\gamma}+V_\qq}{2}P_{\alpha,\beta,\qq}^* - (V_\beta+V_\qq)Q_{\gamma,\qq;\alpha}-V_\qq\bbcro{P_{\alpha,\qq,\beta-\qq}^*-\frac{1}{2}Q_{\gamma-\qq,\qq;\alpha}}}},\label{eq:MH3}\\
\frac{\mathcal{R}^{H_3}_{\alpha,\beta}}{\sqrt{n_0/V}}&=&\mathcal{S}_{\{\alpha,\beta,\gamma'\}}\bbcro{V_\beta\bbaco{c_\beta c_{\gamma'}+c_\alpha(1+n_\beta+n_{\gamma'})}
+\sum_{\qq}\bbaco{\frac{V_\alpha+V_\qq}{2}P_{\alpha+\qq,\beta,\gamma'}+\frac{V_\qq}{2}{T_{\beta,\gamma',\qq}}}},\label{eq:RH3} 
\eea
\end{widetext}
while the latter contains products of doublets and triplets
\begin{widetext}
\bea
\mathcal{M}^{H_4}_{\alpha,\beta}&=&-\frac{\mathcal{S}_{\{\alpha,\beta\}}}{V}\bbcrol{\sum_{\qq} V_\qq \frac{1+n_\alpha+n_\beta}{2} M_{\gamma,\alpha-\qq} + (V_\alpha+V_\qq)(n_\gamma-n_\beta)  M_{\gamma-\qq,\alpha}} \notag \\
&&  \bbcror{\vphantom{\sum_\qq}+(V_\gamma+V_{\beta+\qq})c_\beta^* M_{\gamma+\qq,\gamma}^*+ V_\qq\bbaco{ c_\gamma R_{\beta,\qq-\alpha}^* - c_\beta^* M_{\alpha,\qq-\beta}^*)}}, \label{eq:MH4}\\
\mathcal{R}^{H_4}_{\alpha,\beta}&=&\frac{{\mathcal{S}_{\{\alpha,\beta,\gamma'\}}}}{V}\bbcro{\sum_{\qq} \frac{V_\qq}{2} (1+n_\alpha+n_\beta) R_{\alpha-\qq,-\gamma'} + (V_\alpha+V_{\qq-\gamma'}) c_\beta M_{\alpha+\qq,\alpha}^*}.  \label{eq:RH4}
\eea
\end{widetext}
In these expressions $\gamma$ (in $\mathcal{M}^{H_3}$ and $\mathcal{M}^{H_4}$) and $\gamma'$
(in $\mathcal{R}^{H_3}$ and $\mathcal{R}^{H_4}$), which denote
the third wavevector deduced from $\alpha$ and $\beta$ by
momentum conservation, should be replaced respectively by $\gamma=\alpha+\beta$ and 
$\gamma'=-\alpha-\beta$ after the action of the symmetrizer $\mathcal{S}$.

Finally, for the quadruplets, using the notations $\delta=\alpha+\beta-\gamma$,
$\delta'=\alpha+\beta+\gamma$ and $\delta''=-\alpha-\beta-\gamma$ for the fourth
wavevector of respectively $Q_{\alpha\beta;\gamma}$, $P_{\alpha,\beta,\gamma}$ and
$T_{\alpha,\beta,\gamma}$,  we have
\begin{widetext}
\bea
\ii\hbar\partial_t Q_{\alpha,\beta;\gamma} &=& \bb{E_\alpha+E_\beta-E_{\gamma}-E_{\delta}}Q_{\alpha\beta;\gamma}+\mathcal{S}_{\{\alpha,\beta\}}[\Delta_\alpha P^*_{-\alpha,\gamma,\delta}] - \mathcal{S}_{\{\gamma,\delta\}}[\Delta_\gamma^* P^*_{\alpha,\beta,-\gamma}]+\mathcal{Q}^{H_3}_{\alpha,\beta;\gamma}+\mathcal{Q}^{H_4}_{\alpha,\beta;\gamma} ,\label{Q} \\
\ii\hbar\partial_t P_{\alpha,\beta,\gamma} &=& \bb{E_\alpha+E_\beta+E_{\gamma}-E_{\delta'}}P_{\alpha,\beta,\gamma}+\mathcal{S}_{\{\alpha,\beta,\gamma\}}[\Delta_\alpha Q_{\beta,\gamma;\delta'}] - \Delta_{\delta'}^* T_{\alpha,\beta,\gamma}+\mathcal{P}^{H_3}_{\alpha,\beta,\gamma}+\mathcal{P}^{H_4}_{\alpha,\beta,\gamma}, \label{P} \\
\ii\hbar\partial_t T_{\alpha,\beta,\gamma} &=& \bb{E_\alpha+E_\beta+E_{\gamma}+E_{\delta''}}T_{\alpha,\beta,\gamma}+\mathcal{S}_{\{\alpha,\beta,\gamma,\delta''\}}[\Delta_\alpha P_{\beta,\gamma,\delta''}] +\mathcal{T}^{H_3}_{\alpha,\beta,\gamma}+\mathcal{T}^{H_4}_{\alpha,\beta,\gamma} \label{T}.
\eea
\end{widetext}
The lengthy expressions for $\mathcal{Q}^{H_3}_{\alpha,\beta;\gamma}$, $\mathcal{Q}^{H_4}_{\alpha,\beta;\gamma}$, $\mathcal{P}^{H_3}_{\alpha,\beta,\gamma}$, $\mathcal{P}^{H_4}_{\alpha,\beta,\gamma} $, $\mathcal{T}^{H_3}_{\alpha,\beta,\gamma}$, and $\mathcal{T}^{H_4}_{\alpha,\beta,\gamma}$ can be found in Appendix~\ref{app:quad}.  

For completeness, the cumulant equations of motion up to the level of triplets can be found given explicitly in Appendix~\ref{app:SimEqns} for a separable potential, which allows for modest simplifications important for numerical implementation.  Following this formal discussion of the cumulant equations in the quadruplet model, we now analyze their structure and solution at early-times following the quench in the following section.

\subsection{Few-body physics and the early-time structure of the cumulant hierarchy}\label{sec:formal}
In this section, we discuss in greater detail the sequential correlation buildup picture using the cumulant equations of motion outlined in Sec.~\ref{sec:explicit}.  This discussion also highlights the few-body physics contained at each level of the hierarchy and is therefore crucial to understanding how the Efimov effect is introduced into the many-body model.  The sequential buildup of correlations can be understood formally from the structure of the homogeneous and inhomogeneous (drive) terms in the cumulant equations of motion given in Sec.~\ref{sec:explicit}.  At the lowest level, the correlation buildup begins with the generation of $(\alpha,-\alpha)$ pairs from the drive term $V_\alpha n_0$ in Eq.~\eqref{c}.  Consequently, the occupation of momentum modes is reflected in the dynamics of $n_\alpha$, which remains small compared to unity at early-times such that the Bose-enhancement factors $(1+n_\alpha+n_\beta)\approx 1$ can be ignored and the exponentiation $(n_\alpha)^m$ in the drive terms of the higher-order cumulants vanishes as $m$ tends to infinity.  The three-excitation Beliaev-Landau type processes described by $M$ and $R$ cumulants, are the next level to be driven by terms of the form $V_\alpha n_\gamma\sqrt{n_0/V}$ and $V_\beta c_\alpha \sqrt{n_0/V}$ in Eqs.~\eqref{eq:Mdot} and \eqref{eq:Rdot}, respectively.  At the next level,  the quadruplet processes described by $Q$, $P$ and $T$ are driven by terms of the form $ M_{\gamma+\delta,\gamma}\sqrt{n_0/V}$ and $V_{\gamma-\alpha}n_\gamma n_\delta$, $V_\beta M^*_{\delta',\alpha}\sqrt{n_0/V}$ and $V_{\alpha+\gamma}c_\gamma n_{\delta'}$, and $V_\alpha R_{\gamma,-\delta''}\sqrt{n_0/V}$ and $V_{\alpha+\delta''}c_\gamma c_{\delta''}$ in Eqs.~\eqref{Q}, \eqref{P}, and \eqref{T} (see Appendix~\ref{app:quad}), respectively.  From these examples, it is clear that the sequential buildup behavior is a general property of the post-quench early-time dynamics of cumulants.  Indeed, this property serves as the motivation for using cumulants in the present study to describe the buildup of correlations even in the strongly-interacting regime where a natural truncation parameter is lacking.  

At early-times, these properties of the cumulant hierarchy can be used to generate solutions highlighting the underlying few-body physics in the many-body system.  First, the hierarchy is recast into a reduced `early-time' form by ignoring the $p+1$ and $p+2$ higher-order correlation functions in the equation of motion for cumulants of order $p$, identical to the truncation scheme in Ref.~\cite{burnett2002}. At the level of the doublets, the $c$ cumulant equation (Eq.~\eqref{c}) reduces to
\begin{equation}\label{eq:cindependent}
i\hbar\partial_t|c_t,c_t\rangle=\hat{H}_{\mathrm{12}}(t)|c_t,c_t\rangle+\hat{V}|\psi_{0,t},\psi_{0,t}\rangle,
\end{equation}
where $\hat{H}_{12}(t)=\hat{\epsilon}_1+\hat{\epsilon}_2-2\mu(t)+\hat{V}$ is the two-body Hamiltonian in the rotating frame of the condensate, written in terms of the one-body kinetic energy-operator $\hat{\epsilon}|\alpha\rangle=\epsilon_\alpha|\alpha\rangle$ and the pairwise potential $\hat{V}$.  The second Josephson relation gives the instantaneous chemical potential $\mu(t)\equiv-\hbar\dot{\theta}_0(t)$.  Additionally, the pair matrix has been cast into basis-independent symmetric state $\langle \alpha,\beta|c_t,c_t\rangle=c_\alpha(t)\delta_{\alpha,-\beta}$, which reflects its behavior under unitary transformations \cite{Ripka1985}.  We have also defined the generalized rank $(0,2)$ tensor $|\psi_{0,t},\psi_{0,t}\rangle=n_0(t)|{\bf 0},{\bf 0}\rangle$, where tensor subscripts in the ket indicate the time.  Equation~\eqref{eq:cindependent} can be solved formally using the two-body evolution operator $\hat{\mathcal{U}}_{12}(t-t_0)=\exp\left[-i\int_{t_0}^td\tau\hat{H}_{12}(\tau)/\hbar\right]$ as
\begin{align}
|c_t,c_t\rangle=&\hat{\mathcal{U}}_{12}(t-t_0)|c_{t_0},c_{t_0}\rangle\nonumber\\
&+\frac{1}{i\hbar}\int_{t_0}^td\tau\hat{\mathcal{U}}_{12}(t-\tau)\hat{V}|\psi_{0,\tau},\psi_{0,\tau}\rangle,\label{eq:csol}
\end{align}
where the initial conditions  at $t=t_0$ are encoded in the first term on the right-hand side of the above equality.  Analogously, the $M$-cumulant equation of motion (Eq.~\ref{eq:Mdot}) becomes 
\begin{align}\label{eq:mformal}
i\hbar\partial_t |M_t\rangle\langle M_t,M_t| =& \hat{H}_1(t)|M_t\rangle\langle M_t,M_t|\nonumber\\
&-|M_t\rangle\langle M_t,M_t|\hat{H}_{12}(t)\nonumber\\
&-|n_t\rangle\langle n_t,\psi_{0,t}|(1+\hat{P}_{12})\hat{V},
\end{align}
where $\hat{H}_1(t)=\hat{\epsilon}-\mu(t)$ is the one-body Hamiltonian in the rotating frame of the condensate, and $\hat{P}_{12}$ is the cyclic permutation operator.  We have defined rank (1,2) tensors $\langle \alpha |M_t\rangle\langle M_t,M_t|\beta,\gamma\rangle=M_{\alpha,\beta}(t)\delta_{\bf \alpha,\beta+\gamma}\sqrt{V}$ and $\langle \alpha,\beta|\psi_{0,t},n_t\rangle\langle n_t|\gamma\rangle=\delta_{\bf \gamma,\alpha}\delta_{\bf \beta}n_{\alpha}(t)\sqrt{n_0(t)}$.  Equation~\eqref{eq:mformal} can be solved formally as
\begin{widetext}
\begin{align}
&|M_t\rangle\langle M_t,M_t|=\hat{{\mathcal{U}}}_1(t-t_0)|M_{t_0} \rangle\langle M_{t_0},M_{t_0}|\hat{\mathcal{U}}_{12}(t_0-t)-\frac{1}{i\hbar}\int_{t_0}^t d\tau \hat{\mathcal{U}}_1(t-\tau)|n_\tau\rangle\langle n_\tau,\psi_{0,\tau}|(1+\hat{P}_{12})\hat{V}\hat{\mathcal{U}}_{12}(\tau-t),\label{eq:msol}
\end{align}
\end{widetext}
where $\hat{\mathcal{U}}_1(t-t_0)=\exp\left[-i\int_{t_0}^td\tau\hat{H}_1(\tau)/\hbar\right]$ is the one-body evolution operator.  We have chosen to write the cumulant equations of motion in basis-independent form in order to facilitate and emphasize the generality of the discussion that follows.  

From the formal integral relations (Eqs.~\eqref{eq:csol} and \eqref{eq:msol}), it is possible to solve for the dynamics of the energies $\mu(t)$, $E_{\alpha}(t)$, and $\Delta_{\alpha}(t)$.  Approximating the quench as a sudden projection of a pure, non-interacting condensate onto unitarity, the initial conditions in Eqs.~\eqref{eq:csol} and \eqref{eq:msol} are neglected.  Inserting the formal solutions into Eq.~\eqref{eq:thetadot}, we find the time-dependent expression for the chemical potential
\begin{widetext}
\begin{align}\label{eq:muformal}
N_0(t)\mu(t)=&\text{Tr}\left[\Re\left[\int_{t_0}^td\tau \hat{\mathcal{T}}_+(t-\tau)|\psi_{0,\tau},\psi_{0,\tau}\rangle\langle\psi_{0,t},\psi_{0,t}|+\hat{\mathcal{T}}_+(t-\tau)(1+\hat{P}_{12})|n_\tau,\psi_{0,\tau}\rangle\langle \psi_{0,t},n_\tau|\hat{\mathcal{U}}_1(\tau-t)\right]\right],\\
\mu(t)=&\Re\int_{t_0}^td\tau \mathcal{T}_+({\bf 0},{\bf 0},t-\tau)n_0(\tau)\nonumber\\
&+\Re\int_{t_0}^td\tau\int d^3 q\left[\mathcal{T}_+\left(\frac{{\bf q}}{2},\frac{{\bf q}}{2},t-\tau\right)+\mathcal{T}_+\left(\frac{{\bf q}}{2},-\frac{{\bf q}}{2},t-\tau\right)\right]\sqrt{\frac{n_0(\tau)}{n_0(t)}}n_{\bf q}(\tau)\mathcal{U}_1({\bf q},\tau-t),
\end{align}
\end{widetext}
where $\hat{\mathcal{U}}_1(t)|{\bf k}\rangle=\mathcal{U}_1({\bf k},t)|{\bf k}\rangle$, and  we have defined the retarded two-body $T$-operator in the rotating frame of the condensate \cite{newton2013scattering}
\begin{equation}
\hat{\mathcal{T}}_+(t)=\delta(t)\hat{V}+\frac{1}{i\hbar}\theta(t)\hat{V}\hat{\mathcal{U}}_{12}(t)\hat{V},
\end{equation}
with $\theta$ the function of Heaviside and $\langle {\bf k+ q},{\bf k-q}|\hat{\mathcal{T}}_+(t)|{\bf k'+p},{\bf k'-p}\rangle=\delta_{\bf k,k'}\mathcal{T}_+({\bf q,p},t)$.  
From Eq.~\eqref{eq:muformal}, we see that as $t-t_0\to 0^+$, $\hat{\mathcal{T}}_+(0)=\delta(0)\hat{V}$ and therefore $\mu(t)=V_{\bf 0} n_0$, which is the first Born approximation for the chemical potential of a pure condensate.  Analogously, inserting the formal solutions into Eq.~\eqref{eq:delta}, we find an expression for the time-dependent pairing field 
\begin{align}
|\Delta,\Delta\rangle&=\int_{t_0}^t d\tau \hat{\mathcal{T}}_+(t-\tau)|\psi_{0,\tau},\psi_{0,\tau}\rangle,\\
\Delta_{\bf k}(t)&=\int_{t_0}^td\tau \mathcal{T}_+({\bf k},{\bf 0},t-\tau)n_0(\tau),\label{eq:Deltaformal}
\end{align}
which has been written in basis-independent form $\langle \alpha,\beta|\Delta,\Delta\rangle=\Delta_{\alpha}\delta_{\alpha,-\beta}$.  As $t-t_0\to 0^+$, the first Born appoximation $\Delta_{\alpha}(t)=V_\alpha n_0$ is recovered.   As time evolves, the memory kernels in Eqs.~\eqref{eq:muformal} and \eqref{eq:Deltaformal} are integrated over larger intervals of time.  Here, the unitarity limit of the $s$-wave cross section $\sigma\propto1/k^2$ translates into the universal behavior $\hat{\mathcal{T}}_+(t)\propto \exp[-2i\theta_0(t)]/\sqrt{t}$ reflecting the gradual decay of resonant collisions.  The energy $E_\alpha(t)$ however contains the Hartree-Fock mean-field energy $E^{(\mathrm{HF})}_{\alpha}=(V_{\bf 0}+V_\alpha) n_0+\delta\epsilon_\alpha$ that remains at the level of the first Born approximation regardless of the system dynamics.  To estimate the relevance of the Hartree-Fock mean-field energies in the unitary regime, we rescale to the Fermi energy $E_\mathrm{n}$, finding in general $E^{(\mathrm{HF})}_{\alpha}/E_\mathrm{n}\propto n^{1/3}r_\mathrm{vdW}$ due to the calibration of the effective interaction strength $g\propto r_\mathrm{vdW}$ for the resonance coupling strength (see Appendix~\ref{app:few}).  Therefore, the Hartree-Fock mean-field energies can be neglected in the unitary regime for realistic systems where the criterion $nr_\mathrm{vdW}^3\ll1$ is well-satisfied. 

The presence of few-body operators in the solutions of the cumulant equations of motion reveals how few-body effects are woven into the early-time structure of the hierarchy.  To demonstrate this explicitly, we spectrally decompose the evolution operator $\hat{\mathcal{U}}_{12}(t)$ as
\begin{equation}\label{eq:2bspectral}
\hat{\mathcal{U}}_{12}(t)=\sum_i e^{-i\epsilon_i t/\hbar}|\phi_i\rangle\langle \phi_i|+\int d\epsilon\ e^{-i\epsilon t/\hbar}|\phi(\epsilon)\rangle\langle \phi(\epsilon)|,
\end{equation} 
into the vacuum bound states $|\phi_i\rangle$ with binding energy $\epsilon_i$ and two-body continuum states $|\phi(\epsilon)\rangle$.  Qualitatively, the response of the system at a dimer binding energy depends on the overlap between the $|\phi_i\rangle$'s and the driving terms of the memory kernels in Eqs.~\eqref{eq:csol} and \eqref{eq:msol}.  At unitarity, the $s$-wave dimer state is at threshold, however the system may still respond at any of the infinite number of bound three-body Efimov trimers that exist in vacuum.  To understand how the Efimov frequencies enter the cumulant hierarchy, we reduce the equation of motion for the $R$ cumulant (Eq.~\eqref{eq:Rdot}) to the early-time form  
\begin{align}
i\hbar\partial_t |R_t,R_t,R_t\rangle=&\hat{H}_\mathrm{123}(t)|R_t,R_t,R_t\rangle+(1+\hat{P}_++\hat{P}_-)\nonumber\\
&\times\left[(\hat{V}_{12}+\hat{V}_{13})|\psi_{0,t},c_t,c_t\rangle\right]\label{eq:Rdot2},
\end{align}
where $\langle \alpha,\beta,\gamma|\hat{V}_{12}|\gamma',\beta',\alpha'\rangle=\delta_{\alpha,\alpha'}\langle \beta,\gamma|\hat{V}|\gamma',\beta'\rangle$, and $\hat{H}_\mathrm{123}(t)=(1+\hat{P}_++\hat{P}_-)\hat{H}_{12}$ is the vacuum three-body Hamiltonian in the rotating frame of the condensate, written in terms of the cyclic and anticyclic permutation operators $\hat{P}_+\equiv\hat{P}_{123}$ and $\hat{P}_-\equiv\hat{P}_{132}$ with $\hat{P}_{123}|\alpha,\beta,\gamma\rangle=|\gamma,\alpha,\beta\rangle$, respectively.   In Refs.~\cite{kohler2002,Kokkelmans2018}, Eq.~\eqref{eq:Rdot2} was shown to yield generalized three-body $T$-matrices satisfying the Faddeev equations.  These $T$-matrices appear in the Gross-Pitaevskii equation as higher-order corrections due to effective three-body scattering, encapsulated in the scattering hypervolume \cite{PhysRevA.100.050702,kohler2002,mestrom2019van,PhysRevA.78.013636}.  We have also defined the rank (0,3) tensor $\langle \alpha,-\beta,\gamma|R_t,R_t,R_t\rangle=R_{\alpha,\beta}(t)\delta_{\alpha-\beta,-\gamma}\sqrt{V}$, whose formal solution is 
\begin{widetext}
\begin{equation}
|R_t,R_t,R_t\rangle=\hat{\mathcal{U}}_{123}(t-t_0)|R_{t_0},R_{t_0},R_{t_0}\rangle+\frac{1}{i\hbar}\int_{t_0}^t d\tau\hat{\mathcal{U}}_{123}(t-\tau)(1+\hat{P}_++\hat{P}_-)(\hat{V}_{12}+\hat{V}_{13})|\psi_{0,\tau},c_\tau,c_\tau\rangle,\label{eq:rsol}
\end{equation}
\end{widetext}
where $\hat{\mathcal{U}}_{123}(t-t_0)=\exp\left[-i\int_{t_0}^td\tau\hat{H}_{123}(\tau)/\hbar\right]$ is the three-body evolution operator in the rotating frame of the condensate.  The eigen-decomposition of the three-body evolution operator is \footnote{We have ignored the coupling between channels which is generally weak at unitarity \cite{PhysRevLett.97.150401}.}
\begin{align}\label{eq:3bspectral}
\hat{\mathcal{U}}_{123}(t)=&\sum_{s}\Bigg[\sum_n e^{-iE_{s,n}t/\hbar}|\Phi_{s,n}\rangle\langle\Phi_{s,n}|\nonumber\\
&+\int dE\ e^{-iE t/\hbar}|\Phi_s(E)\rangle\langle \Phi_s(E)|\Bigg],
\end{align} 
expressed in terms of the vacuum three-body continuum states $|\Phi_s(E)\rangle$ and vacuum three-body bound-states $|\Phi_{n,s}(E)\rangle$ with binding energy $E_{s,n}$.  The three-body spectrum can be decomposed into universal channels $s^2>0$ that do not support bound states and the Efimovian channel $s=is_0$ with $s_0\approx1.006$ that supports an infinite number of trimers.  The introduction of additional length scales in the Efimov channel due to the finite size of Efimov trimers can break the universal scaling of system properties with the density \cite{PhysRevLett.120.100401,PhysRevLett.121.023401,PhysRevA.99.043604}.  In principle the system can respond at any one of the infinity of Efimov trimer frequencies, determined by the overlap between the Efimov trimer wave functions $|\phi_{is_0,n}\rangle$ and the driving terms in the memory kernel of Eq.~\eqref{eq:rsol}, which will be studied in Sec.~\ref{sec:triplet}. 

What is the range of validity of the early-time form of the cumulant equations?  In Ref.~\cite{burnett2002}, this scheme was designed to include multiple scatterings in the cumulant equations of motion in order to extend their range of validity.  By design, such multiple scatterings are described by the vacuum $T$-operators--the so-called ``free dynamics".  In Sec.~\ref{sec:hfb}, we will see however that this picture is spoiled in the presence of strong quantum depletion.  In particular, the energy $\Delta_\alpha$, which is negligible and can be ignored at early-times, rapidly grows towards the Fermi scale at later times as correlations develop.  Consequently, the triplet cumulant dynamics become strongly coupled and therefore can no longer be treated separately as in Eqs.~\eqref{eq:msol} and \eqref{eq:rsol}, spoiling the appearance of vacuum operators and energies.

We note that the equations of motion for each of the quadruplets can also be reduced to their early-time forms and solved as integral equations.  As in Eqs.~\eqref{eq:csol} and \eqref{eq:rsol}, the vacuum one, two, three, and four-body evolution operators also appear in the memory kernels for the $Q$, $P$, and $T$ cumulants.  However, because the numerical simulation of the full quadruplet cumulant theory outlined in this section remains an outstanding numerical challenge, this mostly formal discussion can be found in Appendix~\ref{app:quad}.  Having established the cumulant equations, justified their truncation for quenched systems, and highlighted the underlying few-body physics, we now simulate the doublet model in Sec.~\ref{sec:hfb} and the triplet model in Sec.~\ref{sec:triplet}.

\section{Doublet Model of the Quenched Unitary Bose Gas}
\label{sec:hfb}

In this section, we study the quenched unitary Bose gas within the doublet model by neglecting all third and fourth-order cumulants in Sec.~\ref{sec:eom} such that only $n_{\bf k}$ and $c_{\bf k}$ remain.  To mimic the experimental sequence of Refs.~\cite{makotyn2014universal,klauss2017observation,eigen2017universal,eigen2018prethermal}, we make the sudden approximation and model the quench as infinitely fast.  An initially pure, non-interacting condensate is then evolved in the unitary regime for a variable amount of time up to $t\sim 2.5t_\mathrm{n}$ where $n_{\bf k}$ begins to exceed unity and the exclusion of strongly-driven higher-order cumulants cannot be justified \cite{PhysRevA.100.013612,Kokkelmans2018}.  The condensate is depleted by pairwise excitations $({\bf k},-{\bf k})$ described by the $c$ cumulant.  In this section, we compare the doublet model results to the experimental data from Refs.~\cite{eigen2017universal,eigen2018prethermal} for quenched unitary Bose gases in a uniform system.  The early-time agreement with experiment found in this section motivates an investigation of higher-order effects that will be addressed in Sec.~\ref{sec:triplet}
\begin{figure}[ht!]
\centering
\includegraphics[width=8.6cm]{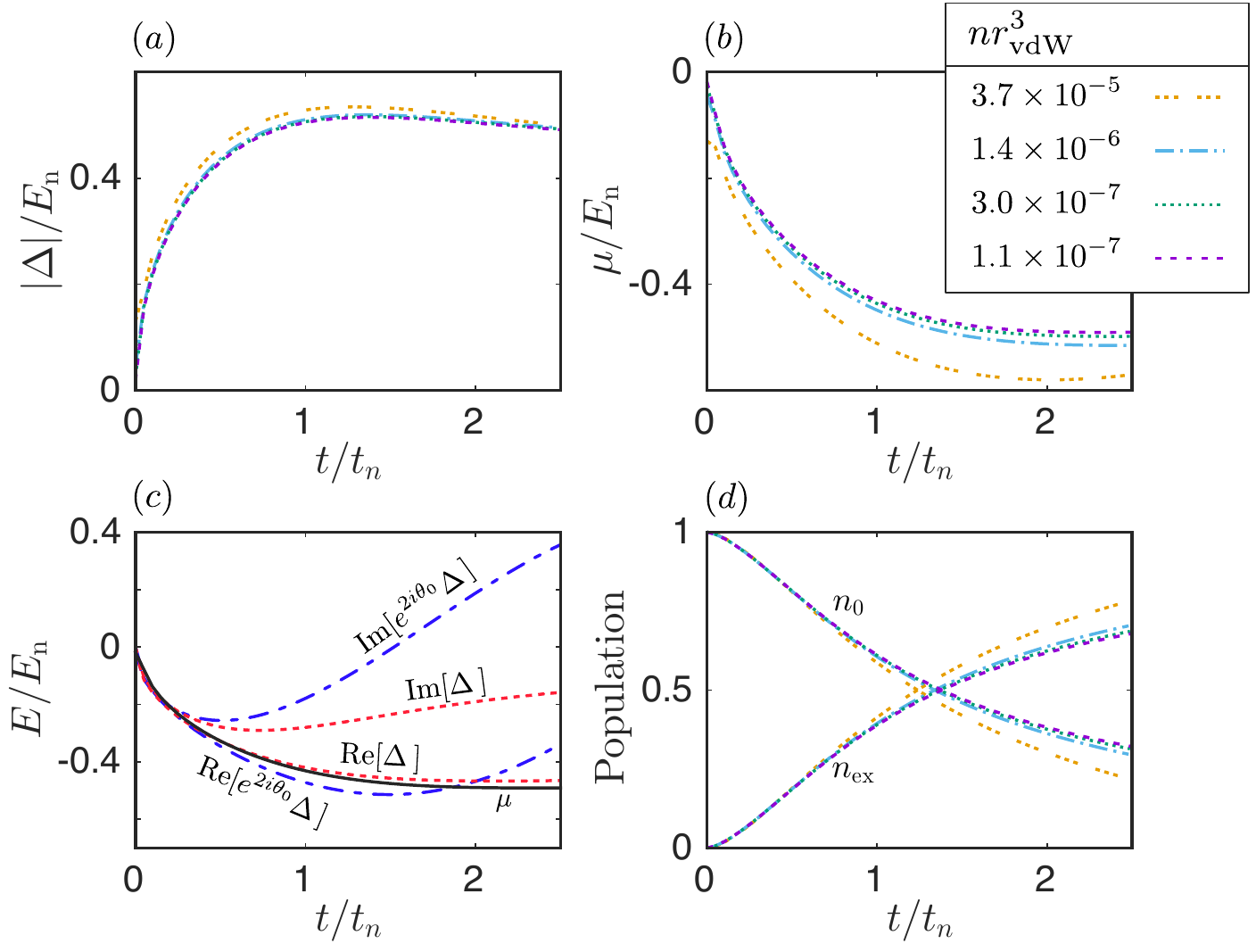}
\caption{Dynamics of (a-c) energy scales and (d) populations for different values of the van der Waals diluteness parameter within the doublet model.  The asymptotic values of the pairing field and phase derivative are roughly equal $\Delta\approx\mu\approx-0.5E_\mathrm{n}$, where the development of the real part and decay of the imaginary part of the pairing field in the lab ($\Delta\exp{2i\theta_0}$) and condensate ($\Delta$) frames are shown in (c) for density $nr_\mathrm{vdW}^3=1.1\times10^{-7}$.}\label{fig:energy_number_hfb}
\end{figure} 
\subsection{Energy and number dynamics}
Before comparing against experiment, we study the time-dependence of the characteristic energies  $\Delta$ and $\mu(t)$ (Eqs.~\eqref{eq:delta} and \eqref{eq:thetadot}, respectively) in the doublet model simulation as a function of the van der Waals diluteness parameter $nr_\mathrm{vdW}^3$.  The dynamics of these energies are shown in Fig.~\ref{fig:energy_number_hfb}(a-c), where we have used the fact that $\Delta_{\bf k}\equiv \Delta$ is independent of ${\bf k}$ within the regime of interest ($|{\bf k}|\leq\Lambda$).  Although not shown, the Hartree-Fock mean-field energies (Eq.~\eqref{eq:kerenorm}) are negligible behaving as finite-range effects which decrease relative to $E_\mathrm{n}$ as powers of $nr_\mathrm{vdW}^3$.  Such finite-range effects are responsible for the long-time differences between the population dynamics seen in Fig.~\ref{fig:energy_number_hfb}(d).  By $nr_\mathrm{vdW}^3\approx10^{-7}$, finite-range contributions to the population dynamics are negligible as the time-dependence is set purely by the Fermi scales characteristic of the universal regime.  We compare this with range of densities $10^{-7}\lesssim nr_\mathrm{vdW}^3\lesssim 10^{-9}$ studied experimentally in Refs.~\cite{eigen2017universal,eigen2018prethermal} for quenched unitary Bose gases in a uniform system.

The pairing field and instantaneous chemical potential are also initially non-universal, depending on finite-range physics as $\Delta(t=0^+)=\mu(t=0^+)=gn$.  However, these energies quickly evolve toward the Fermi scale and approach the universal steady-state $\mu(t)\approx\Delta\approx-0.5E_\mathrm{n}$.  We understand the universality of the $\mu$ and $\Delta$ steady-states from their evolution with the two-body $T$-matrix in Eqs.~\eqref{eq:muformal} and \eqref{eq:Deltaformal}, which is dominated by the unitarity limit of the $s$-wave partial cross section on resonance \cite{taylor2006scattering}.  Importantly, the reality of $\Delta$ at long-times \footnote{For evidence that this oscillation frequency is time-independent in the lab frame at very long times $(t\sim 100t_\mathrm{n})$, we direct the reader to Fig. 1 in the supplementary material of Ref.~\cite{PhysRevLett.124.040403}.} is due to working in the frame of the condensate as shown in Fig.~\ref{fig:energy_number_hfb}(c).  In the lab-frame description of this steady state, the pairing field rotates as $\sim \exp(-2i\mu t/\hbar)$ characteristic of the behavior at true equilibrium.  We understand the approximate equality of $\mu$ and $\Delta$ from the rapid growth of pairing correlations, which leads to dominance of the $c$-cumulant contributions in Eqs.~\eqref{eq:delta} and \eqref{eq:thetadot} (see also Appendix~\ref{app:SimEqns}) such that $\Re\Delta\approx (g/V)\sum_{\bf q} \Re c_{\bf q}\approx \mu$.

\begin{figure}[t!]
\centering
\includegraphics[width=8.6cm]{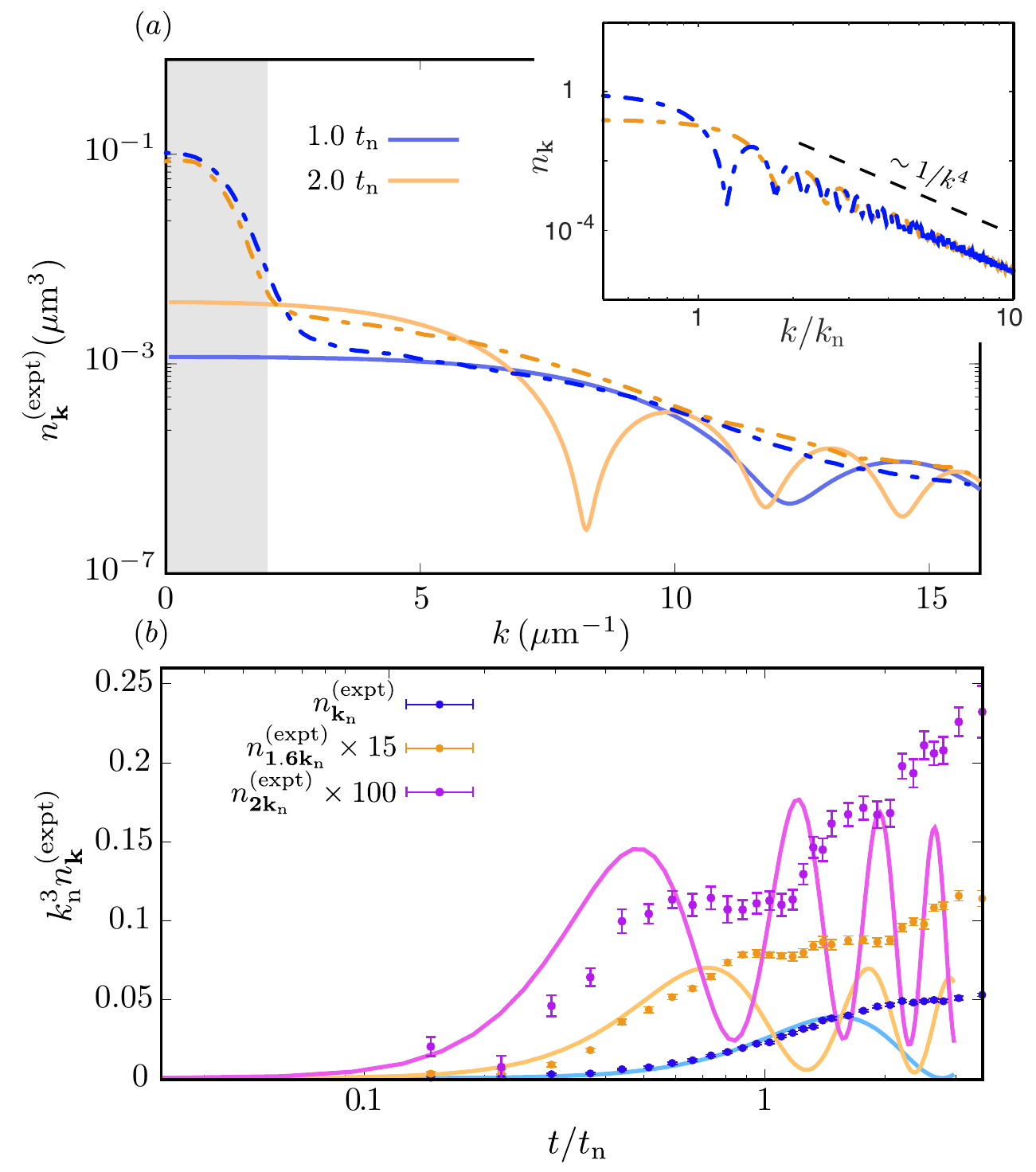}
\caption{Dynamics of the momentum distribution for $k_\mathrm{n}=6.7\mu$m$^{-1}$ with $t_\mathrm{n}=27\mu$s.  We note that due to different normalizations, the experimental momentum distribution is related as $k_\mathrm{n}^3 n_{\bf k}^{(\mathrm{expt})}=3n_{\bf k}/4\pi$.  The results of the doublet model simulation (solid lines) are compared against the experimental findings of Refs.~\cite{eigen2018prethermal,eigen_2018_data} (dashed-dotted lines connecting $\sim 200$ raw data points each following the presentation in that work) at times $t=t_\mathrm{n}$ (blue) and $t=2t_\mathrm{n}$ (orange).  Experimental results in the shaded region are not quantitatively reliable (see the Methods section of Ref.~\cite{eigen2018prethermal}).  The inset shows the $1/k^4$ power law behavior of $n_{\bf k}$ at large-$k$.  (b) Time dependence of the momentum distribution for at fixed-$k$, comparing the results of the doublet model (solid lines) with the experimental data points of Refs.~\cite{eigen2018prethermal,eigen_2018_data}.  Each line has been multiplied by a numerical factor to increase visibility.}\label{fig:nk_hfb}
\end{figure}
\begin{figure}[t!]
\centering
\includegraphics[width=8.6cm]{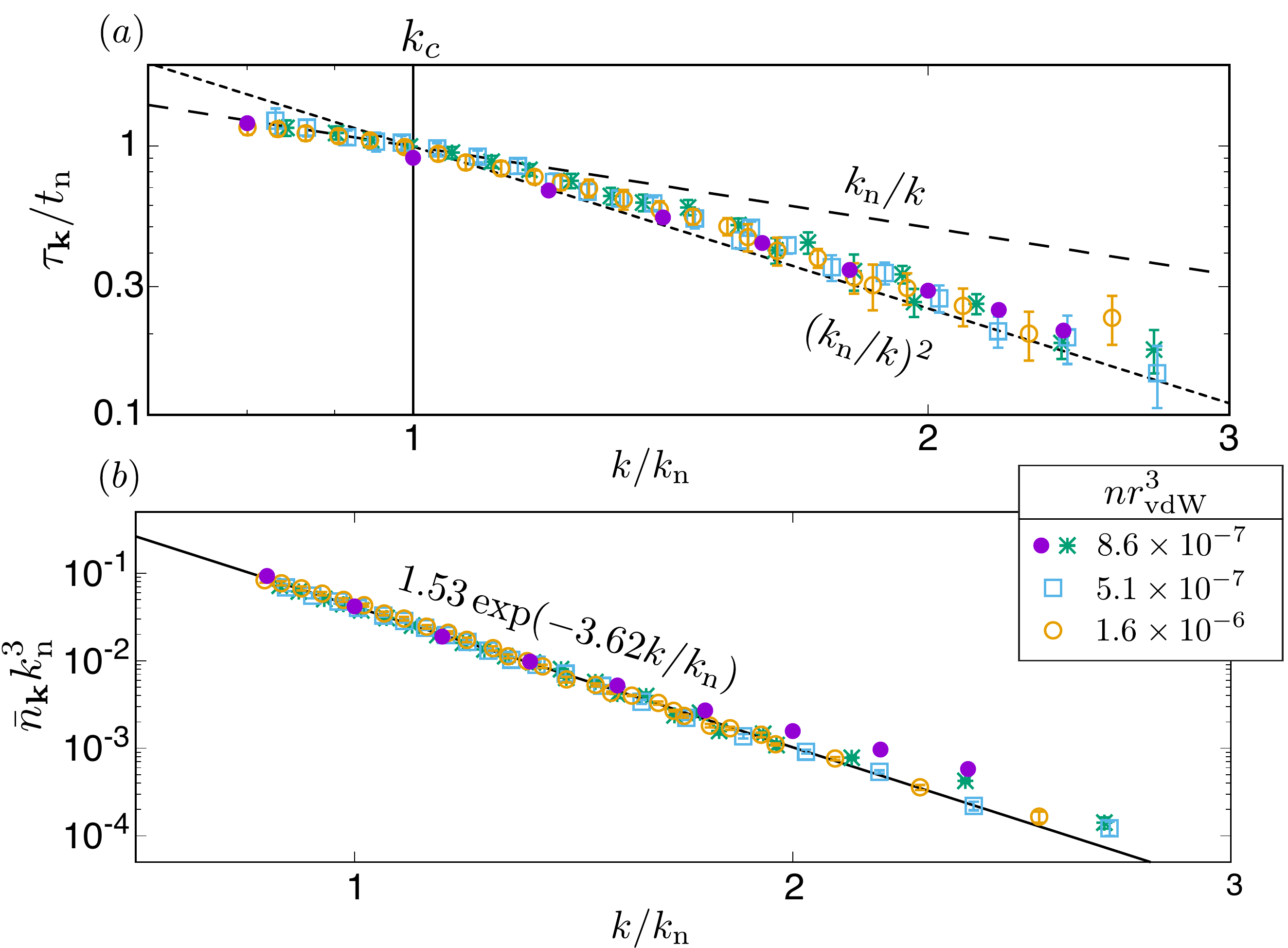}
\caption{(a)  The momentum dependent half-way time $\tau_{\bf k}$ (a) and the plateau value $\bar{n}_{\bf k}$ (b) for three different densities considered in Refs.~\cite{eigen2018prethermal,eigen_2018_data} compared against the results of the doublet model simulation (purple filled circles).  In (a) the asymptotic behaviors $t_{\bf k}=k_\mathrm{n}/k$ and $t_{\bf k}=(k_\mathrm{n}/k)^2$ for the characteristic prethermal timescale $t_{\bf k}$ found in Sec.~\ref{sec:prethermal} are indicated by the dashed and dotted lines, respectively. The two asymptotes cross at $k_c=2mc_\mathrm{pth}/\hbar=\sqrt{2}/\xi_{\rm pth}\approx k_{\rm n}$ (vertical solid).  In (b) we compare against the decaying exponential $1.53\exp(-3.62k/k_\mathrm{n})$ found experimentally indicated by the solid line.}\label{fig:tau_hfb}
\end{figure}

 
Even though these energies approach a steady-state, other observables in the system remain far from equilibrium as we now discuss.  In Fig.~\ref{fig:nk_hfb}(a), the doublet model results for $n_{\bf k}$ are compared to the relevant experimental results \footnote{We note that the experimental momentum distribution is normalized as $1=\int d^3k n_{\bf k}^{(\mathrm{expt})}$, which gives the conversion $k_\mathrm{n}^3 n_{\bf k}^{(\mathrm{expt})}=3n_{\bf k}/4$.} of Ref.~\cite{eigen2018prethermal}.  This comparison is not made in the grey shaded region $k<2\mu$m$^{-1}\approx 0.3k_\mathrm{n}$ where the experimental results are not quantitatively reliable due to initial cloud size and non-infinite time of flight \cite{eigen2018prethermal}.  Qualitatively, it is clear that the nodal pattern of the doublet model results is absent from the experimental data.  To quantify these results, we follow \cite{eigen2018prethermal} and fit the initial growth of $k_\mathrm{n}^3n_{\bf k}^{(\mathrm{expt})}=3n_{\bf k}/4\pi$ for fixed $k$ shown in Fig.~\ref{fig:nk_hfb}(b) to a sigmoid $f(t)=a+b/(\exp(-c(t+d))+1)$, obtaining plateau value $\bar{n}_{\bf k}=a+b$ and half-way time $\tau_{\bf k}$ defined as $\bar{n}_{\bf k}=2n_{\bf k}(\tau_{\bf k})$, finding generally good agreement \footnote{Although we have followed the fitting routine used on the experimental data in Ref.~\cite{eigen2018prethermal}, the dynamics of $n_{\bf k}$ in the doublet model oscillate in the prethermal state and do not plateau.  Therefore, we expect the experimental comparison of the plateau value $\bar{n}_{\bf k}$ in the doublet model to be qualitative.  We note recent quantitative predictions for the half-way time in the thermal regime in Ref.~\cite{sun2020high}.  Further complicating matters is the absence of a clear steady-state plateau for $k/k_\mathrm{n}<0.8$ in the experimental data of Ref.~\cite{eigen2018prethermal}, which adds ambiguity to the fitting routine without making additional assumptions about the timescale separation between heating and prethermalization. },
consistent with Refs.~\cite{PhysRevA.99.023623,PhysRevLett.124.040403}, as shown in Fig.~\ref{fig:tau_hfb}.  

\subsubsection{Prethermal state}\label{sec:prethermal}
The equilibration of many-body observables in a quenched system while the microscopic degrees of freedom remain strongly out-of-equilibrium is characteristic of prethermalization \cite{PhysRevLett.93.142002}. To describe this stage of the doublet model in the universal limit ($nr_\mathrm{vdW}^3\to 0$), we solve the doublet model using the asymptotic values for $\Delta$ and $\mu$ shown in Fig.~\ref{fig:energy_number_hfb}(a-b) while the doublets $n_{\bf k}$ and $c_{\bf k}$ remain periodic in time \footnote{In the limit $nr_\mathrm{vdW}^3\to 0$ the doublet equations of motion simplify to their universal form
$$
\ii\hbar\partial_t c_\kk = 2\bb{\epsilon_\kk-\mu(t)} c_\kk + \Delta(t) (1+2n_\kk),
$$
$$
\ii\hbar\partial_t n_\kk = \Delta(t) c_\kk^*-\textrm{c.c.}   
$$
Assuming that $\mu$ and $\Delta$ are real, negative, and time-independent (as observed numerically in Fig.~\ref{fig:energy_number_hfb}), this linear system 
solves into Eq.~\eqref{eq:nkHFB}.}
\begin{align}
n_{\bf k}(t)=&n_{\kk}(t_0)+\frac{\Delta\textrm{Re}\,\alpha_\kk(t_0)}{2\omega_\kk^2}\bbcro{1-\cos(2\omega_\kk (t-t_0))}\nonumber\\
&-\frac{\Delta\textrm{Im}\,\alpha_\kk(t_0)}{2\xi_\kk\omega_\kk}\sin(2\omega_\kk (t-t_0)),\label{eq:nkHFB}\\
|c_{\bf k}|^2=&n_{\bf k}(1+n_{\bf k}),
\end{align}
in agreement with Ref.~\cite{PhysRevLett.124.040403} (see in particular Eq.~(S32) therein).  The eigenfrequency of these oscillations matches the HFB spectrum \cite{Griffin1996}
\be
\omega_\kk=\sqrt{\xi_\kk^2-\Delta^2}, \label{omegak}
\ee
(with $\xi_\kk=\epsilon_\kk-\mu$ in the limit $nr_\mathrm{vdW}^3\to 0$ when $gn\ll\mu$)
and the energy $\alpha_\kk(t_0)=\Delta\bbcro{1+2n_\kk(t_0)}+2\xi_\kk c_\kk(t_0)$ 
encodes the initial condition at $t_0$ (with $t>t_0\gg t_n$). We note that $\alpha_\kk=0$
gives the HFB ground state \cite{James1982}.  

In the dilute limit where the van der Waals diluteness parameter $nr_\mathrm{vdW}^3$ tends towards 0, the mean-field energy $gn/E_\mathrm{n}$ also vanishes relative to the Fermi energy (see Sec.~\ref{sec:formal}) while $\Delta/E_\mathrm{n}$ remains finite.  According to Ref.~\cite{Griffin1996}, $\mu=gn+\Delta$, which implies $\mu=\Delta$ in the dilute strong-interacting limit.  This matches the long-time, steady-state dynamics shown in Fig.~\ref{fig:energy_number_hfb}. Remarkably, this condition $\mu=\Delta$ produces a {\it gapless} excitation spectrum ($\omega_{\bf 0}=0$), and the elementary excitations follow a Bogoliubov dispersion law $\omega_{\bf k}=\sqrt{\epsilon_{\bf k}(\epsilon_{\bf k}+2|\mu|)}$.  We therefore find long-wavelength phonons with energy  $\hbar ck$ and sound velocity $c_\mathrm{pth}=\sqrt{|\mu|/m}\simeq 0.5 \hbar k_\mathrm{n}/m$ in the unitary regime.  The smooth crossover to the particle-regime occurs for $\epsilon_{\bf k}\sim mc_\mathrm{pth}^2\simeq 0.5 E_\mathrm{n}$, which allows us to define a characteristic healing length \cite{pitaevskii2016bose} in the prethermal state ($t\gg t_n$), $k=1/\xi_{\rm pth}$, such that $k_\mathrm{n}\xi_{\rm pth}\simeq\sqrt{2}$.  This is to be contrasted against the usual Bogoliubov dispersion law at weak interactions $\omega_{\bf k}^{0}=\sqrt{\epsilon_{\bf k}(\epsilon_{\bf k}+2 U_0n)}$ with $U_0=4\pi\hbar^2a/m$, discussed in Sec.~\ref{sec:hamiltonian}, and $a\gtrsim0$ \cite{pitaevskii2016bose}.  The dispersion laws $\omega_{\bf k}$ and $\omega_{\bf k}^{0}$ are connected by replacing the usual mean-field energy $U_0n$ by $E_\mathrm{n}/2$, i.e. through a mapping of the form $a\to 1/k_\mathrm{n}$.  In Ref.~\cite{PhysRevA.93.033653}, $U_0n$ was replaced ad hoc by $4E_\mathrm{n}/3\pi$, by assuming a universal Bogoliubov excitation spectrum.  In the present work, this replacement is not assumed {\it a priori}, rather a universal Bogoliubov spectrum {\it emerges} within the prethermal steady-state at strong interactions. In a quasistationary picture, the mapping of quantities between vacuum and Fermi scales occurs smoothly as a result of the interplay between quantum depletion and few-body processes in the system \cite{Kokkelmans2018}.  

From the inverse of the excitation energy, we obtain the characteristic timescale $t_{\bf k}=\hbar/\omega_{\bf k}$, behaving asymptotically as $t_{\bf k}/t_\mathrm{n}= k_\mathrm{n}/k$ for $\xi_{\rm pth} k\ll1$ and $t_{\bf k}/t_\mathrm{n}= (k_\mathrm{n}/k)^2$ for $\xi_{\rm pth} k\gg 1$.  In Fig.~\ref{fig:tau_hfb}(a), these scalings (dashed and dotted lines) are compared directly against the numerical and experimental results (symbols) for the half-way times $\tau_{\bf k}$ and are in excellent quantitative agreement without adjustment.  This comparison assumes that the system has entered the prethermal stage on a timescale comparable to the range of $\tau_{\bf k}$ considered in Fig.~\ref{fig:tau_hfb}(a).  We address this assumption later in this section by defining a ``prethermalization time'' $t_\mathrm{pth}$ from the dynamics of the kinetic temperature following Ref.~\cite{PhysRevLett.93.142002}.  Qualitatively, the smooth crossover between sound and free-particle regimes takes place when $k\sim \hbar/\xi_{\rm pth}$, which is of order $\mathcal{O}(k_\mathrm{n})$ consistent with the experimental findings of Ref.~\cite{eigen2018prethermal}.

\begin{figure}[t!]
\centering
\includegraphics[width=8.6cm]{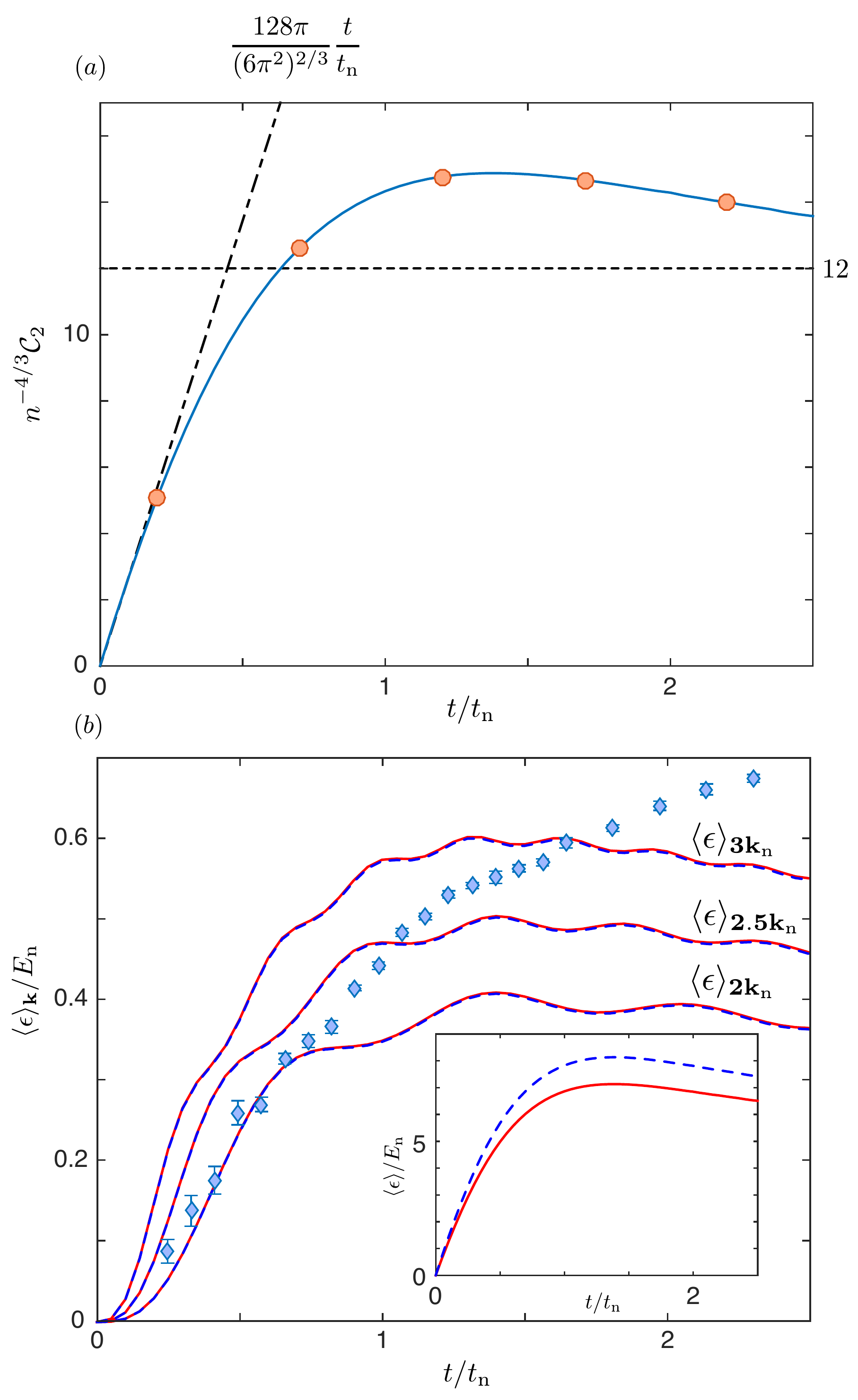}
\caption{(a) Universal two-body contact dynamics obtained via the $k^{-4}$ power law tail of $n_{\bf k}$ (solid blue) and from the interaction energy (circles) for $nr_\mathrm{vdW}^3=7.2\times10^{-8}$.  We compare also with the universal early-time growth rate (dash-dotted) obtained in Ref.~\cite{PhysRevA.91.013616} and the asymptotic result (dotted) obtained in Ref.~\cite{PhysRevA.89.021601}.  (b) Universal dynamics of the restricted kinetic energy per particle $\langle\epsilon\rangle_{{\bf k}=\{{\bf 2k_\mathrm{n}},{\bf 2.5k_\mathrm{n}},{\bf 3k_\mathrm{n}}\}}$ for $nr_\mathrm{vdW}^3=7.2\times10^{-8}$ (dashed blue) and $nr_\mathrm{vdW}^3=1.1\times10^{-7}$ (solid red) compared against the experimental data of Ref.~\cite{eigen2017universal,eigen_2017_data} (blue diamonds).  (Inset) Nonuniversal dynamics of the full kinetic energy per particle $\langle\epsilon\rangle$ as predicted in the doublet model.}\label{fig:c2_hfb}
\end{figure} 
\subsection{Dynamics of the two-body contact}\label{sec:c2doublet}
Whereas the decaying exponential in Fig.~\ref{fig:tau_hfb}(a) describes the full range of experimental data, the profile of $n_{\bf k}$ in the doublet model simulation transitions to a $1/k^4$ power law tail as can be seen in the inset of Fig.~\ref{fig:nk_hfb}(a).  We discuss this power-law behavior in the doublet model presently.  In an ultracold quantum gases, typical momentum scales ($k_\mathrm{n}$, $\lambda_\mathrm{dB}^{-1}$, etc...) are such that $k/\Lambda\ll1$, where $\Lambda$ corresponds to the inverse range of the potential.  In this regime, when two bosons separated by a distance $r=|{\bf r}_1-{\bf r}_2|$ come together such that $\Lambda^{-1}\ll r\ll\{ n^{-1/3},|a|,\lambda_\mathrm{dB}$\}, their relative wave function is proportional to $\phi(r)=(1-a/r)$, and the many-body wave function $|\Psi\rangle$ (normalized as $\langle\Psi|\Psi\rangle=N$) takes the form
 \begin{equation}\label{eq:bp}
\Psi({\bf r}_1,{\bf r}_2,\dots,{\bf r_\mathrm{N}})\approx\phi(r)\mathcal{A}({\bf c}_{12},{\bf r}_3\dots,{\bf r_\mathrm{N}}),
\end{equation}
with center of mass coordinate ${\bf c}_{12}=({\bf r_1}+{\bf r}_2)/2$.  This microscopic behavior of the many-body wave function can be used to derive a set of important relationships between system properties, revolving around the extensive quantity $C_2\underset{k\to\infty}{\equiv}Vk^4n_{\bf k}$ known as the two-body contact that measures the probability for pairs of atoms to be close together \cite{TAN20082952,TAN20082971,TAN20082987,PhysRevA.86.053633,PhysRevLett.106.153005}.  The intensive counterpart $\mathcal{C}_2$ is the two-body contact density related to the (extensive) two-body contact as $V \mathcal{C}_2=C_2$.  The two-body contact is also related to the total interaction energy $U=\langle\hat{H}_\mathrm{int}\rangle$ as $C_2=2m^2gU/\hbar^4$, where $\hat{H}_\mathrm{int}$ is the interaction part of the many-body Hamiltonian (Eq.~\eqref{hamiltonianF}).  Although these relations were derived for equilibrium states, they give consistent results for the dynamical two-body contact $\mathcal{C}_2(t)$ within the doublet model shown in Fig.~\ref{fig:c2_hfb}(a).  Additionally, these findings are consistent with previous studies, namely the universal early-time growth $n^{-4/3}\mathcal{C}_2(t)=128\pi t/(6\pi^2)^{2/3}t_\mathrm{n}$ and asymptotic value $n^{-4/3}\mathcal{C}_2(t)\approx12$ found in Refs.~\cite{PhysRevA.91.013616,PhysRevA.89.021601}.  

Although we have found consistent results for the dynamical two-body contact by blindly applying equilibrium relations within the doublet model, counterexamples from quenches in one-dimension \cite{PhysRevA.94.023604} highlight that care should be taken when generalizing these relations to non-equilibrium scenarios.  Therefore, we revisit the assumptions needed to derive the equilibrium contact relations.  The simple form of the microscopic two-body wave function $\phi(r)$ (Eq.~\eqref{eq:bp}) holds locally, regardless of whether the many-body system is in equilibrium or not, and one can define then the dynamical two-body contact $\mathcal{C}_2(t)$ density by integrating over the coordinates of the two-body regular part, $\mathcal{A}$, of the many-body wave function in Eq.~\eqref{eq:bp} to obtain 
\begin{eqnarray}
g^{(2)}({\bf r},t)&\equiv&\frac{\langle\hat{\psi}^\dagger({\bf r})\hat{\psi}^\dagger({\bf 0})\hat{\psi}({\bf 0})\hat{\psi}({\bf r})\rangle(t)}{n^2},\\
&\underset{{\bf r}\to0}{=}&\frac{\mathcal{C}_2(t)}{16\pi^2 n^2 r^2},\label{eq:c2g2}
\end{eqnarray}
for the functional form of the pair correlation function in a uniform system.  The interaction energy relation results then from balancing the divergence of $g^{(2)}({\bf 0},t)$ by powers of the potential $V({\bf 0})$ and neglecting sub-leading finite-range corrections decaying as powers of $1/\Lambda$~\cite{PhysRevA.86.013626,PhysRevA.86.053633,PhysRevLett.106.153005}.  To generalize the $k^{-4}$ power law tail equilibrium definition one must consider in addition the Fourier transform 
\begin{align}
n_{\bf k}(t)=&\frac{1}{V}\sum_i\int\left(\prod_{l\neq i}d^3 r_l\right)\nonumber\\
&\times\left|\int d^3 r_i e^{-\ii{\bf k}\cdot{\bf r}_i}\Psi({\bf r}_1,{\bf r}_2,\dots,r_\mathrm{N},t)\right|^2,\label{eq:fourier_nk}
\end{align}
where the sums are taken over all particles.  When the short-distance divergent behavior in Eq.~\eqref{eq:bp} dominates the large-$k$ limit of Eq.~\eqref{eq:fourier_nk}, one obtains the power-law behavior $n_{\bf k}\propto1/k^4$ and the equilibrium definition follows \cite{PhysRevLett.91.090401,PhysRevA.86.013626,PhysRevA.86.053633}.  Although this argument holds in equilibrium, it is not guaranteed in a dynamical system due to the possibility of energetic nonlocal physics as shown in one-dimension \cite{PhysRevA.94.023604}.  We note that this caveat also anticipates the difficulties encountered within the triplet model in the next section (Sec.~\ref{sec:triplet}.)

\subsection{Kinetic temperature}\label{sec:kintemp}
The two-body contact provides valuable insight into the dynamics of the interaction energy per particle $\langle u\rangle = U/N$ and the kinetic energy per particle $\langle\epsilon\rangle=\langle\hat{H}_\mathrm{kin}\rangle/N$, where $\hat{H}_\mathrm{kin}$ is the kinetic part of the many-body Hamiltonian (Eq.~\eqref{hamiltonianF}).  Within the sudden approximation, the quench generates correlation waves out to arbitrarily large energies \cite{PhysRevA.94.023604}.  Finite-range effects cure this ultraviolet divergence by providing a natural short-range cutoff at the scale of $\Lambda$, calibrated to the scale of the van der Waals energy $E_\mathrm{vdW}=\hbar^2/mr_\mathrm{vdW}^2$ in our model (see Appendix~\ref{app:few}).  Although the total energy per particle in the doublet model simulation $\langle e_\mathrm{tot}\rangle=\langle u\rangle+\langle \epsilon\rangle$ is negligible ($\langle e_\mathrm{tot}\rangle/E_\mathrm{n}\propto n^{1/3}r_\mathrm{vdW}$), both the interaction energy per particle $\langle u\rangle$ and kinetic energy per particle $\langle\epsilon\rangle$ diverge.  This divergence can be understood by collecting powers of $\Lambda$ in the contact relation $\langle u(t)\rangle=(\hbar^4/2gnm^2)\mathcal{C}_2(t)$.  Whereas $\mathcal{C}_2$ scales universally with the density, the bare interaction scales as $g\propto 1/\Lambda$, and therefore $\langle u(t)\rangle$ scales linearly with $\Lambda$.  In our model, this translates into a finite-range effect such that $\langle u(t)\rangle\propto r_\mathrm{vdW}^{-1}$.  This behaviors applies analogously to $\langle\epsilon\rangle$ due to energy conservation $\langle\epsilon\rangle/E_\mathrm{n}=-\langle u\rangle/E_\mathrm{n}+O(n^{1/3}r_\mathrm{vdW})$.  This explains the early-time linear growth, late-time asymptotics, and divergence with $r_\mathrm{vdW}^{-1}$ of $\langle \epsilon\rangle$ shown in the inset of Fig.~\ref{fig:c2_hfb}(b).

The rapid equipartion of kinetic and potential energies with $\langle\epsilon\rangle/\langle u\rangle\approx -1$ provides the basis for discussing, in the far-from-equilibrium many-body system, a ``kinetic temperature'' proportional to $\langle\epsilon(t)\rangle$ 
\cite{PhysRevLett.93.142002}.  In contrast to mode-specific quantities, $\langle\epsilon(t)\rangle$ provides a mode-averaged measure of the rate at which the system prethermalizes.  Therefore, following in the spirit of the original treatment in Ref.~\cite{PhysRevLett.93.142002}, we define a prethermal time $t_\mathrm{tph}$ from the criterion $|\langle\epsilon(t_\mathrm{pth})\rangle-\langle \epsilon\rangle_\mathrm{as.}|/\langle \epsilon\rangle_\mathrm{as.}\lesssim0.2$ for $t>t_\mathrm{pth}$ using the asymptotic estimate of $\mathcal{C}_2(t)$ \cite{PhysRevA.89.021601} to obtain $\langle \epsilon\rangle_\mathrm{as.}\approx(-\hbar^4/gm^2)6n^{1/3}$.  We find $t_\mathrm{pth}\simeq 0.4-0.5$, which is consistent with the saturation timescale estimate in Ref.~\cite{PhysRevA.89.021601} and the equilibration time of the largest momenta modes measured in Ref.~\cite{makotyn2014universal}.  Additionally, $\tau_{\bf k}>t_\mathrm{pth}$, for momenta in the crossover between sound and free-particle regimes shown in Fig.~\ref{fig:tau_hfb}(a), clarifying the assumptions made in Sec.~\ref{sec:prethermal}.

How can the dependence of $\langle\epsilon\rangle$ on the non-universal short-range scales be reconciled with the universal dynamics of the kinetic energy per particle observed in Ref.~\cite{eigen2017universal}?  We understand this discrepancy then from the comparatively limited range of experimentally accessible momenta (c.f. Figs.~\ref{fig:nk_hfb} and \ref{fig:tau_hfb}). To compare with experiment, we therefore define the {\it restricted} kinetic energy per particle $\langle\epsilon\rangle_{\bf k}=\int_0^{k}d^3k' n_{\bf k'}\epsilon_{\bf k'}/n$ and compare with the experiment as shown in Fig.~\ref{fig:c2_hfb}(b).  Here the doublet model simulation results are roughly consistent with the universal evolution of $\langle\epsilon\rangle_{\bf 2k_\mathrm{n}}$ for early times $t\lesssim t_\mathrm{n}$.  The oscillations of $\langle \epsilon\rangle_{\bf k}$ are due to the periodicities of the underlying $n_{\bf k}$ as discussed in Sec.~\ref{sec:prethermal}.  As the integration includes a larger range of modes, the oscillations dephase and are absent in $\langle \epsilon\rangle$.  We note that the time range studied is however still less than the time $t\sim 4t_\mathrm{n}$ where the kinetic temperature of the experimental data begins to follow the power law $\langle \epsilon\rangle\propto t^{2/13}$ for recombinative heating in the thermal regime \cite{PhysRevLett.110.163202}.  In the intermediate time $1\lesssim t/t_\mathrm{n}\lesssim 4$, however, the effects of heating and lossless correlation dynamics are difficult to differentiate, requiring a theoretical investigation of each contribution individually. 

\subsection{Summary}
In this section, the quenched unitary Bose gas was studied within the doublet model.  This theory describes the universal prethermal state that rapidly forms as the condensate is depleted by pairing excitations.  The signature of this prethermal state is the establishment of steady-state values for $\mu$ and $\Delta$ even while the momentum distribution dynamics remain far from equilibrium.  Within this steady-state, one finds the emergence of a universal Bogoliubov dispersion law, which quantitively matches the prethermal timescales observed experimentally.  This behavior at strong interactions is in stark contrast to quenches at weak interactions where the Bogoliubov dispersion law can be assumed \cite{PhysRevA.98.053612}.  Finding disagreement with the exponential tail of $n_{\bf k}$ found experimentally, we analyze the origin of the $1/k^4$ power law tail observed in the doublet model by studying the dynamical two-body contact.  In turn, the universal dynamics of the two-body contact were used to shed light on the non-universal growth of the kinetic temperature of the gas, which diverges for quenches treated within the sudden approximation.  To connect with experiment, we consider the kinetic temperature one would obtain with access to only a restricted range of momentum modes, finding agreement at early times.  In the next section (Sec.~\ref{sec:triplet}), we go beyond the doublet model and retain also the triplet cumulants in order to understand the impact of three-body correlations on the prethermal state and to search for non-universal signatures of the Efimov effect.  

\section{Triplet Model of the Quenched Unitary Bose Gas}
\label{sec:triplet}

In this section, we study the quenched unitary Bose gas within the triplet model by neglecting all fourth-order cumulants in Sec.~\ref{sec:eom} such that only $n_{\bf k}$, $c_{\bf k}$, $M_{\bf k,q}$, and $R_{\bf k,q}$ remain.  For consistency, we follow the same quench sequence as in Sec.~\ref{sec:hfb}, starting from an initially pure condensate.  As the gas evolves in the unitary regime, the condensate is depleted by both pairwise and three-body effects.  However, the triplet model suffers from a violation of energy conservation as discussed in Sec.~\ref{sec:conslaws}.  This violation leads to unphysical behavior of the triplet model at long times (see Appendix~\ref{app:SimEqns}).  We therefore limit our analysis to times $t\lesssim t_\mathrm{n}$ before these effects become significant.  In this section, we focus on $(i)$ departures from the prethermal state found in Sec.~\ref{sec:prethermal} due to the ergodic dynamics introduced by $\hat{H}_3$ and $\hat{H}_4^\mathrm{eff}$ and $(ii)$ signatures of the Efimov effect in the system, motivated by the few-body studies \cite{PhysRevLett.120.100401,PhysRevLett.121.023401,PhysRevA.99.043604}, the discussion in Sec.~\ref{sec:formal}, and the experimental observation of a macroscopic population of Efimov trimers in Ref.~\cite{klauss2017observation}.  

Due to the limitations of the triplet model to times $t\lesssim t_\mathrm{n}$, we simulate only the dominant parts of the drive terms $\mathcal{M}^{H_4}_{\bf k,q}$ and $\mathcal{R}^{H_4}_{\bf k,q}$ (Eqs.~\eqref{eq:MH4} and \eqref{eq:RH4}) so that
\begin{align}
\mathcal{M}^{H_4}_{\alpha,\beta}&\approx-\frac{1}{V}\mathcal{S}_{\{\alpha,\beta\}}\left[\sum_{\qq} V_\qq \frac{1+n_\alpha+n_\beta}{2} M_{\gamma,\alpha-\qq} \right],\label{eq:MH4_approx} \\
\mathcal{R}^{H_4}_{\alpha,\beta}&\approx{\frac{1}{V}\mathcal{S}_{\{\alpha,\beta,\gamma'\}}}\left[\sum_{\qq} \frac{V_\qq}{2} (1+n_\alpha+n_\beta) R_{\alpha-\qq,-\gamma'} \right].\label{eq:RH4_approx}
\end{align}
This contains the vacuum contribution (the ``1'' in ``$1+n+n$''), which dominates at short times and ensures that few-body interactions at unitarity (see Sec.~\ref{sec:formal}) are correctly described.  Additionally, due to the increased computation resources required to simulating the triplet model, the results in this section are limited to densities $nr_\mathrm{vdW}^3\leq6.9\times10^{-6}$, which includes a portion of the density range studied in Ref.~\cite{klauss2017observation} but is more dense than the range considered in Refs.~\cite{makotyn2014universal,eigen2017universal,eigen2018prethermal}.  We refer the interested reader to App.~\ref{app:convergence} where technical details related to convergence of the triplet model simulations and the computational hardware used are discussed.

%
\begin{figure}[t!]
\centering
\includegraphics[width=8.6cm]{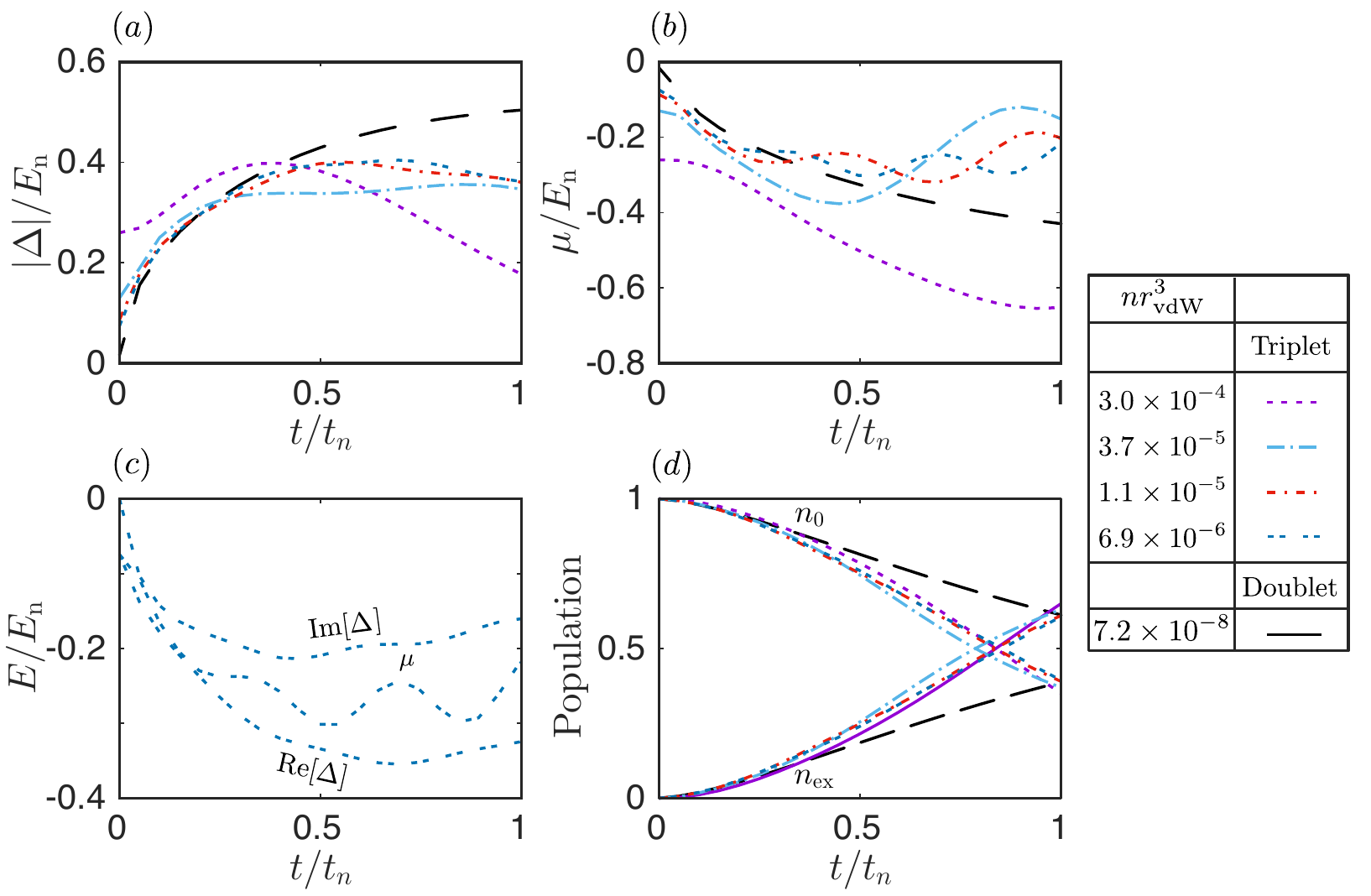}
\caption{Dynamics of (a-c) energy scales and (d) populations for different values of the van der Waals diluteness parameter within the triplet model.  We compare with the doublet model results for density $nr_\mathrm{vdW}^3=7.2\times10^{-8}$ (black dashed) in this section, which is sufficiently dilute to be universal.  (c) The dynamics of the instantaneous chemical potential $\mu(t)$ and the real and imaginary parts of the pairing field shown for density $nr_\mathrm{vdW}^3=6.9\times10^{-6}$.}\label{fig:energy_number_triplet}
\end{figure} 
\subsection{Energy and number dynamics}
We begin by revisiting the time dependence of the characteristic energies $\Delta$ and $\mu(t)$ in the triplet model as a function of the van der Waals diluteness parameter $nr_\mathrm{vdW}^3$ shown in Fig.~\ref{fig:energy_number_triplet}(a-c).  Compared to the doublet model, $\mu(t)$ now has an additional contribution from the triplet $M$-cumulant (see Eq.~\eqref{eq:thetadot}), whereas the expression for $\Delta$ (Eq.~\eqref{eq:delta}) remains unchanged.  This has the effect of introducing oscillations into the dynamics of $\mu(t)$, which can be see in Fig.~\ref{fig:energy_number_triplet}(b-c).  These oscillations, which are absent in the doublet model results, black dashed lines in Fig.~\ref{fig:energy_number_triplet}, are signatures of Efimov states.  The oscillation frequency is set by the three-body parameter $\kappa_*r_\mathrm{vdW}=0.211$ (see Appendix~\ref{app:few}) and is therefore non-universal (density-independent).  This behavior is in contrast with the dynamics of $\Delta$ shown in Fig~\ref{fig:energy_number_triplet}(a), with oscillations that are comparatively less visible and therefore weakly-dependent on $\kappa_*$.  By $t\sim t_\mathrm{n}$, we see that $\Delta$ seems to converge to $\Delta\sim -0.4E_\mathrm{n}$ with decreasing imaginary component visible in Fig~\ref{fig:energy_number_triplet}(c).  The population dynamics shown in Fig.~\ref{fig:energy_number_triplet}(d) also depend weakly on $\kappa_*$, and we see that the addition of three-body effects lead to more rapid depletion of the condensate than in the doublet model (compare with Fig.~\ref{fig:energy_number_hfb}(d)).   Although not shown, the Hartree-Fock mean-field energies remain negligible as $nr_\mathrm{vdW}^3$ is descreased, which follows from the general conclusions in Sec.~\ref{sec:formal}.  

\begin{figure}[t!]
\centering
\includegraphics[width=8.6cm]{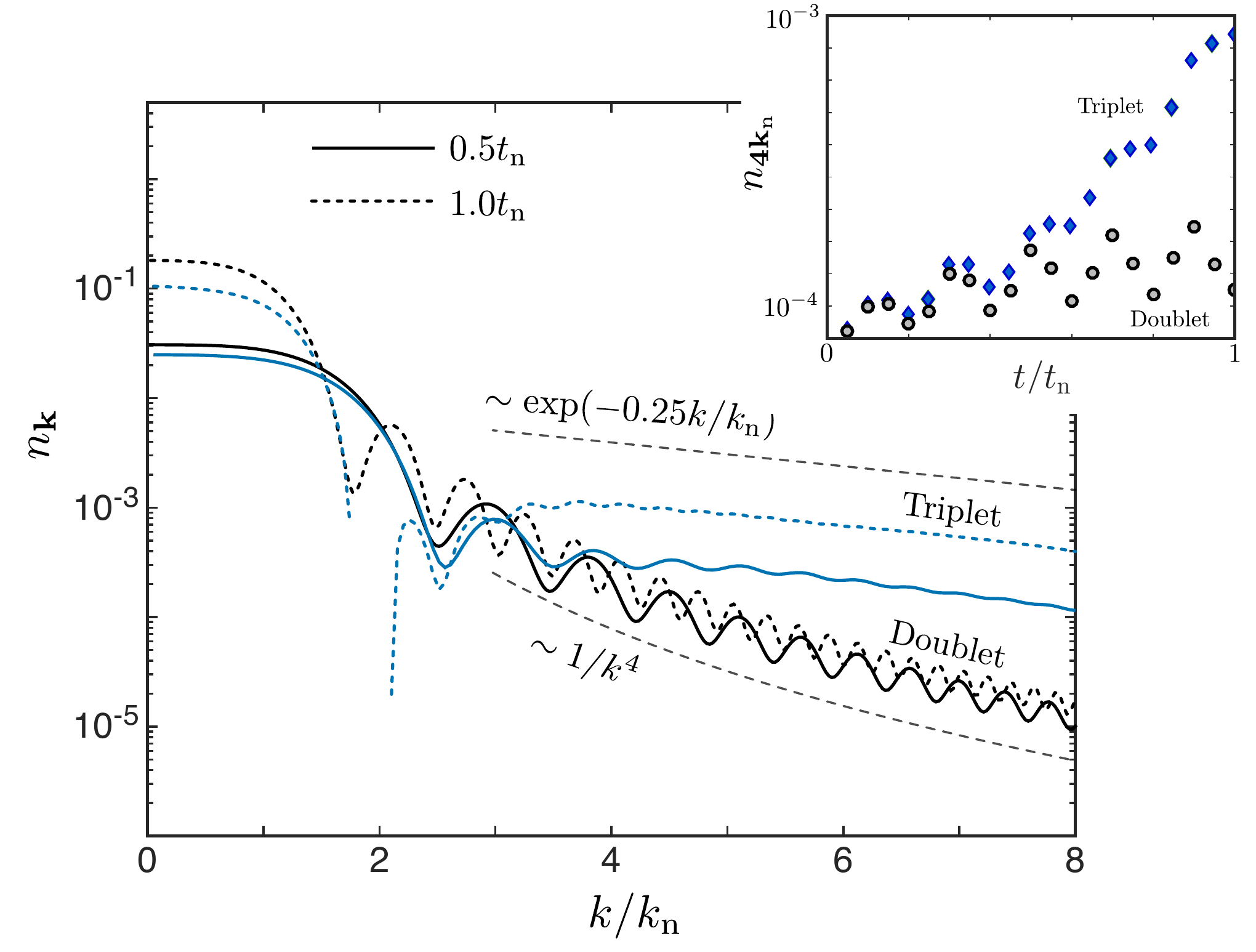}
\caption{Evolution of the single-particle momentum distribution within the triplet model at density $nr_\mathrm{vdW}^3=6.9\times 10^{-6}$ (blue) and the universal doublet model (black).  (inset) Dynamics of $n_{\bf k}$ at fixed $k=4k_\mathrm{n}$ from triplet and doublet theories illustrating the transition out of the prethermal state. }\label{fig:nk_dynamics_triplet}
\end{figure} 
As the energies $\Delta$ and $\mu$ begin to display steady-state and periodic behaviors, the dynamics of $n_{\bf k}$ remain far-from-equilibrium.  The triplet and doublet model dynamics of $n_{\bf k}$ are shown in Fig.~\ref{fig:nk_dynamics_triplet}.  By $t=0.5t_\mathrm{n}$, we see already a departure in the large momentum behavior of $n_{\bf k}$ from the $1/k^4$ power law tail towards a decaying exponential.  The formation of a decaying exponential tail in $n_{\bf k}$ is a robust feature of the triplet model, and can be found even at much later times (even though positivity of $n_{\bf k}$ becomes violated at low momenta due to violation of energy conservation.)  Although the amplitude of the exponential tail grows in time, the decay rate remains roughly constant as $n_{\bf k}\propto \exp(-0.25 k/k_\mathrm{n})$ for $nr_\mathrm{vdW}^3=7.0\times10^{-6}$, which is more gradual than experimentally observed decay $n_{\bf k}\propto \exp(-3.62 k/k_\mathrm{n})$ \cite{eigen2018prethermal}.  Although not shown, similar exponential decay of $n_{\bf k}$ at large momentum can be found over the full range of densities considered in this section.  As noted in Ref.~\cite{eigen2018prethermal}, the development of a decaying exponential tail at large momenta is not consistent with the power-law tail predictions from local short-range physics.  Ultimately, due to its absence in the doublet model, the decaying exponential is necessarily due to three-body processes.

\subsubsection{Departure from the prethermal state}
Such deviations from the integrable doublet model dynamics also signal the departure from the prethermal state.  Physically, this is expected due to the ergodic dynamics introduced by $\hat{H}_3$ and $\hat{H}_4^\mathrm{eff}$, which take the system towards true thermalization.  For quenches in the weakly-interacting regime, the timescales between the prethermal and thermal stages are separated by orders of magnitude \cite{PhysRevA.98.053612} as nonintegrable Beliaev-Landau scatterings drive the system
toward full thermalization.  On resonance, this picture of distinct on-shell quasiparticle scatterings begins to breakdown, and one expects generically that all rates scale with the Fermi time so that distinct stages in the evolution of the gas may not be well-separated.  Indeed, we see from Fig.~\ref{fig:nk_dynamics_triplet} that the departure from the integrable doublet model dynamics occurs at a momentum-dependent rate evidenced by the widening gap between power law and exponential tails at large momentum.  This departure from the prethermal dynamics is shown explicitly in the inset of Fig.~\ref{fig:nk_dynamics_triplet} in the dynamics of $n_{ 4k_\mathrm{n}}$ where the enchanced growth and damped oscillations in the triplet model are clearly visible at later times.  We note that because the triplet model dynamics inherently violate energy conservation, the breaking of integrability removes the system from the initial phase-space manifold, which muddies the physical connection between the long-time dynamics and thermalization.  

\begin{figure}[t!]
\centering
\includegraphics[width=8.6cm]{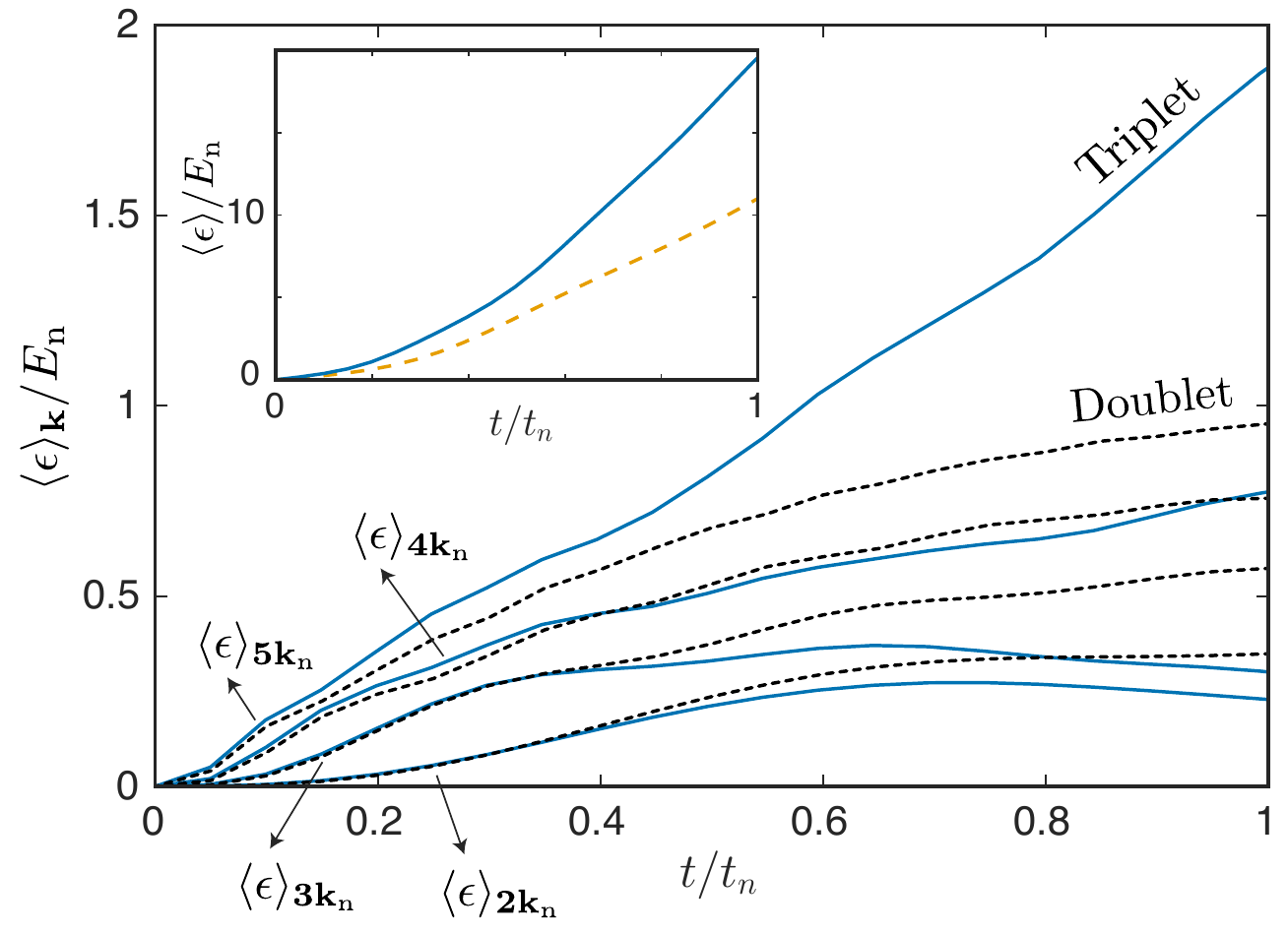}
\caption{Dynamics of the restricted kinetic energy per particle $\langle\epsilon\rangle_{{\bf k}=\{{\bf 2k_\mathrm{n}},{\bf 3k_\mathrm{n}},{\bf 4k_\mathrm{n}},{\bf 5k_\mathrm{n}}\}}$ for the triplet model with $nr_\mathrm{vdW}^3=6.9\times10^{-6}$ (solid blue) and the universal doublet model (dotted black).  (Inset) Nonuniversal dynamics of the full kinetic energy per particle $\langle \epsilon\rangle$ within the triplet model for $nr_\mathrm{vdW}^3=6.9\times10^{-6}$ (solid blue) and $nr_\mathrm{vdW}^3=1.9\times 10^{-5}$ (dashed orange).}\label{fig:ek_dynamics_triplet}
\end{figure} 

As the system shifts away from the prethermal state, the gradual decay in the occupation of large momentum modes leads to an increase in the average kinetic energy per particle $\langle \epsilon\rangle$ relative to the doublet model as shown in Fig.~\ref{fig:ek_dynamics_triplet}.  To understand which modes are responsible for this growth, we examine in Fig.~\ref{fig:ek_dynamics_triplet} how the dynamics of the restricted kinetic-energy per particle $\langle \epsilon\rangle_{\bf k}$ change as the large momentum modes transition into an exponentially decaying tail.  First, we observe that the momentum dependent departure of $n_{\bf k}$ from the prethermal doublet dynamics is mirrored also in $\langle \epsilon\rangle_{\bf k}$.  Second, the kinetic energy per particle for modes $k\lesssim0.3k_\mathrm{n}$ decreases relative to the doublet model dynamics, which illustrates the large pileup of kinetic energy in the decaying exponential tail and draining of kinetic energy from the low momentum modes.  This accumulation of kinetic energy in the exponential tail is a signature of imbalanced three-body kinetics within the triplet model.  Specifically the three-body processes in the kinetic equation (Eq.~\eqref{n}) require contributions from the quadruplet in order to satisfy the condition of detailed balance, which will be demonstrated in a forthcoming publication \cite{kineticsTBP}.  Because the total energy is not conserved, the rapid equipartition of energy observed in the doublet model (see Sec.~\ref{sec:kintemp}) is not observed distinctly in the triplet model, such that a kinetic temperature cannot be so clearly defined as before.  Here, the growth of the total kinetic energy results instead in an effective heating of the system.  Due to this violation, the dynamics of $\langle \epsilon\rangle$ cannot be connected to the dynamical two-body contact $\mathcal{C}_2(t)$ as was done in Sec.~\ref{sec:hfb}, and we must turn to other methods as we discuss now.  

\subsection{Dynamics of the contacts}
In addition to having pairs of correlated bosons close together, in the triplet model it is also possible to have triplets of bosons clustered together, experiencing the attractive $1/R^2$ effective three-body potential in the Efimov channel.  When three bosons in a configuration parameterized by hyperradius $R=\sqrt{(r^2+\rho^2)/2}$ and hyperangles $\Omega=\{ \hat{\boldsymbol{\rho}},{\bf \hat{r}},\alpha=\arctan(r/\rho)\}$, with Jacobi coordinates ${\bf r}={\bf r_1}-{\bf r_2}$ and $\boldsymbol{\rho}=(2\boldsymbol{r_3}-\boldsymbol{r_1}-\boldsymbol{r_2})/\sqrt{3}$ and spherical angles $\hat{\boldsymbol{\rho}}$ and $\hat{\boldsymbol{r}}$, come together $\Lambda^{-1}\ll R\ll \{n^{-1/3},|a|,\lambda_\mathrm{dB}\}$) their relative wave function is proportional to \cite{PhysRevA.86.053633}
\begin{equation}\label{eq:3bbc}  
\Phi(R,{\bf \Omega})=\frac{1}{R^2}\sin\left(s_0\ln\frac{R}{R_t}\right)\frac{\phi_{is_0}({\bf \Omega})}{\sqrt{\langle \phi_{is_0}|\phi_{is_0}\rangle}}.
\end{equation}
Here, $R_t$ is related to the three-body parameter $\kappa_*$ as $R_t=\sqrt{2}\exp(\Im\ln[\Gamma(1+is_0)]/s_0)/\kappa_*$ where $\Gamma(x)$ is the Gamma function and $s_0=1.006$. The hyperangular function describing $s$-wave pairwise scatterings is \footnote{The normalization constant is given by
$\langle\phi_s|\phi_s\rangle\equiv \int d{\bf \Omega}|\phi_{s}({\bf \Omega})|^2=-\frac{12\pi}{s}\sin\left(\frac{s^\ast\pi}{2}\right)\left[\cos\left(\frac{s\pi}{2}\right)-\frac{s\pi}{2}\sin\left(\frac{s\pi}{2}\right)-\frac{4\pi}{3\sqrt{3}}\cos\left(\frac{s\pi}{6}\right)\right]$.} $\phi_{s_0}({\bf \Omega})=(1+\hat{P}_{13}+\hat{P}_{23})\varphi_{s_0}(\alpha)/\sin(2\alpha)\sqrt{4\pi}$ with $\varphi_s(\alpha)=\sin(s(\pi/2-\alpha))$ where $\hat{P}_{ij}$ swaps particles $i$ and $j$.
When this occurs, the many-body wave function $|\Psi\rangle$ takes the form
\begin{equation}\label{eq:three_bc}
\Psi({\bf r}_1,{\bf r}_2,{\bf r}_3,\dots,{\bf r}_\mathrm{N})\approx\Phi(R,{\bf \Omega})\mathcal{B}({\bf c}_\mathrm{123},{\bf r}_3\dots,{\bf r}_\mathrm{N}),
\end{equation}
where $c_{123}=({\bf r}_1+{\bf r}_2+{\bf r}_3)/3$ is the three-body center of mass, and $\mathcal{B}$ is the three-body regular part of the many-body wave function.  The microscopic behaviors of the many-body wave function (Eqs.~\eqref{eq:bp} and \eqref{eq:three_bc}, respectively) can be used to derive a set of important relationships between system properties extending the two-body contact relations discussed in Sec.~\ref{sec:c2doublet} to when the Efimov effect arises
\begin{align}
&Vn_{\bf k}\to\frac{1}{k^4}C_2+\frac{F(k)}{k^5}C_3\label{eq:nk_triplet},\\
&\mathcal{C}_2=\frac{m^2g^2}{\hbar^4}\langle \hat{d}^\dagger \hat{d}\rangle-\frac{4m^3g^3}{\Lambda^2\hbar^6}\left(H+\frac{J}{\pi}+\frac{J}{2a\Lambda}\right)\langle \hat{t}^\dagger \hat{t}\rangle\label{eq:c2_triplet},\\
&\mathcal{C}_3=-\frac{m^2g^2}{2\hbar^4\Lambda^2}\left(H'+\frac{J'}{a\Lambda}\right)\langle \hat{t}^\dagger \hat{t}\rangle,\label{eq:c3_triplet}
\end{align}
where $\hat{d}=\hat{\psi}({\bf 0})\hat{\psi}({\bf 0})$ and $\hat{t}=\hat{\psi}({\bf 0})\hat{\psi}({\bf 0})\hat{\psi}({\bf 0})$ \cite{PhysRevA.86.053633,PhysRevLett.106.153005}.  Here, the probability to measure such Efimovian triplets is quantified by the (total) three-body contact $C_3$ and three-body contact density $\mathcal{C}_3$ related as $V\mathcal{C}_3=C_3$.  The quantities $F$, $H$, and $J$ are log-periodic functions of $k$ and $\Lambda$, given by
\begin{align}
&F(k)=A\sin(2s_0\ln(k/\kappa_*)+2\phi),\\
&H(\ln(\Lambda/\Lambda_*))=h_0\frac{C-s_0S}{C+s_0S},\\
&J(\ln(\Lambda/\Lambda_*))=\frac{j_0+j_1(2SC)+j_2(C^2-S^2)}{(C+s_0S)^2},
\end{align}
where $C=\cos(s_0\ln(\Lambda/\Lambda_*))$ and $S=\sin(s_0\ln(\Lambda/\Lambda_*))$ and with universal constants $A=89.262$, $\phi=-0.669$, $h_0=0.879$, $j_0=-0.148$, $j_1=-0.892$, $j_2=-0.087$ and renormalization scale $s_0\ln(\Lambda_*/\kappa_*)=0.971$ mod $\pi$.  The $'$ notation indicates a partial derivative with respect to $\ln(\Lambda/\Lambda_*)$.  The log-periodic dependency in Eqs.~\eqref{eq:nk_triplet}-\eqref{eq:c3_triplet} on the discrete scaling $e^{\pi/s_0}\approx22.7$, reflects the infinite number of Efimov trimers which form at unitarity with binding energies scaling in the zero-range limit as $E_\mathrm{3b}^{(n)}=-e^{-2n\pi/s_0}\hbar^2\kappa_*^2/m$ for any integer $n$.  For finite-range potentials, the Efimov trimer spectrum is bounded from below $n\geq0$, and the three-body parameter $\kappa_*$ sets the wavenumber of the ground Efimov trimer $E_\mathrm{3b}^{(0)}=-\hbar^2\kappa_*^2/m$ and, importantly, introduces a nonuniversal, finite length scale \cite{BRAATEN2006259,Naidon_2017,D_Incao_2018}.   For the pairwise potential considered in this work, the three-body parameter is $\kappa_*r_\mathrm{vdW}=0.211$ (see discussion in Appendix~\ref{app:few}), which is in fair agreement with the universal result $\kappa_*r_\mathrm{vdW}\approx 0.226$ near the broad Feshbach resonances used experimentally \cite{RevModPhys.82.1225,PhysRevLett.108.263001,PhysRevA.90.022106}.

Although the formal caveats in Sec.~\ref{sec:c2doublet} were cautionary, Eq.~\eqref{eq:nk_triplet} clearly fails when generalized to the triplet model of the quenched unitary Bose gas. It is necessary then to revisit the assumptions underlying Eqs.~\eqref{eq:nk_triplet}--\eqref{eq:c3_triplet}.  Formally, Eqs.~\eqref{eq:c2_triplet}--\eqref{eq:c3_triplet} follow directly from the forms of two and three-body microscopic wave functions $\phi(r)$ and $\Phi(R,{\bf \Omega})$, respectively, that both hold locally, regardless of whether the many-body system is in equilibrium or not.  One can then define the dynamical three-body contact density by integrating over the three-body regular part, $\mathcal{B}$, of the many-body wave function in Eq.~\eqref{eq:three_bc} to obtain the relation \cite{PhysRevA.86.053633}
\begin{eqnarray}
g^{(3)}({\bf 0},{\bf r},{\bf r}',t)&\equiv&\frac{\langle\hat{\psi}^\dagger({\bf r})\hat{\psi}^\dagger({\bf r}')\hat{\psi}^\dagger({\bf 0})\hat{\psi}({\bf 0})\hat{\psi}({\bf r}')\hat{\psi}({\bf r})\rangle(t)}{n^3},\nonumber\\
&\underset{R\to0}{=}&|\Phi(R,{\bf \Omega})|^2\frac{8}{n^3 s_0^2\sqrt{3}}\mathcal{C}_3(t),\label{eq:c3g3}
\end{eqnarray}
for the functional form of the triplet correlation function in the $R\to0$ limit written here specifically for uniform systems.  To obtain Eq.~\eqref{eq:nk_triplet}, one must make additional assumptions that the short-distance divergent behaviors Eqs.~\eqref{eq:bp} and \eqref{eq:three_bc} dominate the large-$k$ limit of the Fourier transform of the many-body wave function (Eq.~\eqref{eq:fourier_nk}).  This clearly no longer holds for the dynamics of $n_{\bf k}$ in the triplet model, highlighting the nonlocal origin of the decaying exponential.  Finally, we note the additional $\langle \hat{t}^\dagger \hat{t}\rangle$ dependence in Eq.~\eqref{eq:c2_triplet}  absent in the doublet model, following the convention of Ref.~\cite{PhysRevLett.106.153005}.  This extra contribution becomes negligible, scaling as $1/\Lambda$ and has been included here for completeness.  We now discuss results for the dynamical two and three-body contacts using Eqs.~\eqref{eq:c2_triplet} and \eqref{eq:c3_triplet}, respectively.

%
\begin{figure}[t!]
\centering
\includegraphics[width=8.6cm]{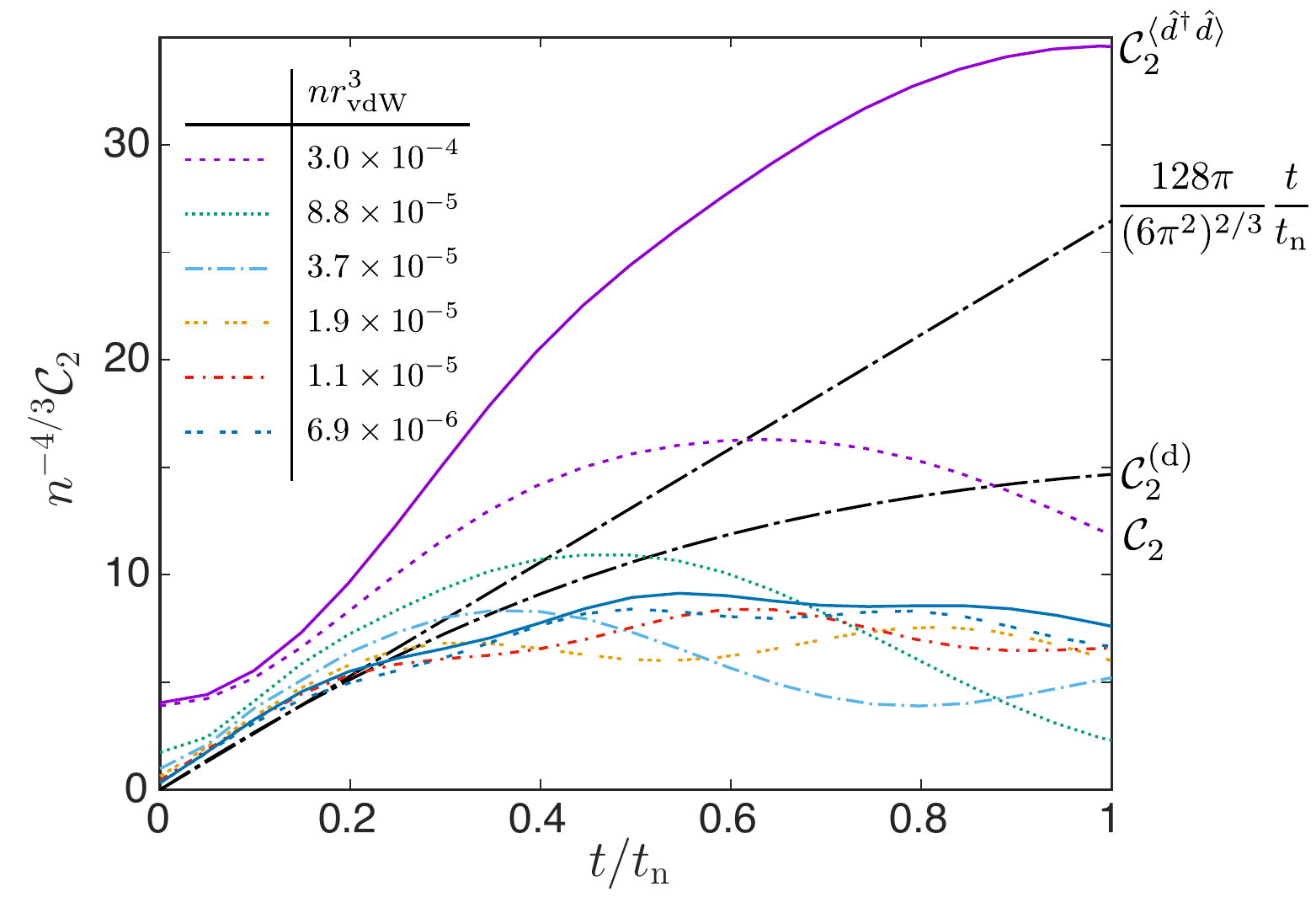}
\caption{Dynamics of the two-body contact (Eqs.~\eqref{eq:c2_triplet}) over a range of densities and times up to $1.0 t_\mathrm{n}$.  Results for the dimensionless two-body contact density $n^{-4/3}\mathcal{C}_2$ from the triplet model over the range of densities indicated in the key.  We show also the two-body contact density $\mathcal{C}_2^{\langle d^\dagger d\rangle}$ obtained from neglecting the term $\langle t^\dagger t\rangle$ in Eq.~\eqref{eq:c2_triplet}.  This is shown only for the most and least dense cases as solid curves attached to the corresponding $n^{-4/3}\mathcal{C}_2$ results in order to illustrate the diminishing of this extra contribution in the zero-range limit.  The universal doublet model results $\mathcal{C}_2^{\mathrm{(d)}}$ along with the universal early-time growth rate obtained in Ref.~\cite{PhysRevA.91.013616} are indicated by the black dashed-dotted curves.} \label{fig:c2_triplet}
\end{figure}  

\subsubsection{$\mathcal{C}_2$ dynamics}

In Fig.~\ref{fig:c2_triplet}, the numerical results for the dynamical two-body contact in the triplet model are shown over a range of densities and times up to $t=1.0t_\mathrm{n}$.  Formally, we note that the cumulant decomposition of the dominant contribution $\langle \hat{d}^\dagger\hat{d}\rangle$ to $\mathcal{C}_2$ is the same as in Sec.~\ref{sec:c2doublet} with the  addition now of the triplet $M$-cumulant.  Here we differentiate between $\mathcal{C}_2^{\langle \hat{d}^\dagger \hat{d}\rangle}(t)$ and $\mathcal{C}_2(t)$ defined without and with the $\langle\hat{t}^\dagger \hat{t}\rangle$ contribution, respectively, in Eq.~\eqref{eq:c2_triplet} to demonstrate how this term becomes negligible as $nr_\mathrm{vdW}^3$ is decreased.  Comparing against the early-time doublet results, we find that linear early-time growth rate is approached as $nr_\mathrm{vdW}^3$ is decreased.  The early-time dynamics of $\mathcal{C}_2(t)$ are therefore insensitive to the Efimov effect, consistent with Ref.~\cite{PhysRevA.99.043604}.  We note that the nonzero offset $\mathcal{C}_2(0)=g^2n^2$ is a finite-range effect scaling as $1/\Lambda^2$.  By $t\sim0.2t_\mathrm{n}$, the triplet and doublet model results for $\mathcal{C}_2(t)$ begin to depart significantly after a period of universal growth.  Echoing the findings of Ref.~\cite{PhysRevA.99.043604}, we interpret this development as the timescale when clustered pairs become sensitive to the surrounding ``few-body medium'' consisting of a third boson.  This sensitivity leads to the secondary dependence of $\mathcal{C}_2(t)$ on the Efimov effect, as its dynamics display the characteristic non-universal beating phenomenon at the frequency of an Efimov trimer.  At later times, the probability of finding pairs of atoms close together becomes less likely than in the doublet model.  In the triplet model, there is now the competition between forming pair or triplet clusters, which develop more slowly as we find from the analysis of $\mathcal{C}_3(t)$ below.  

%
\begin{figure}[ht!]
\centering
\includegraphics[width=8.6cm]{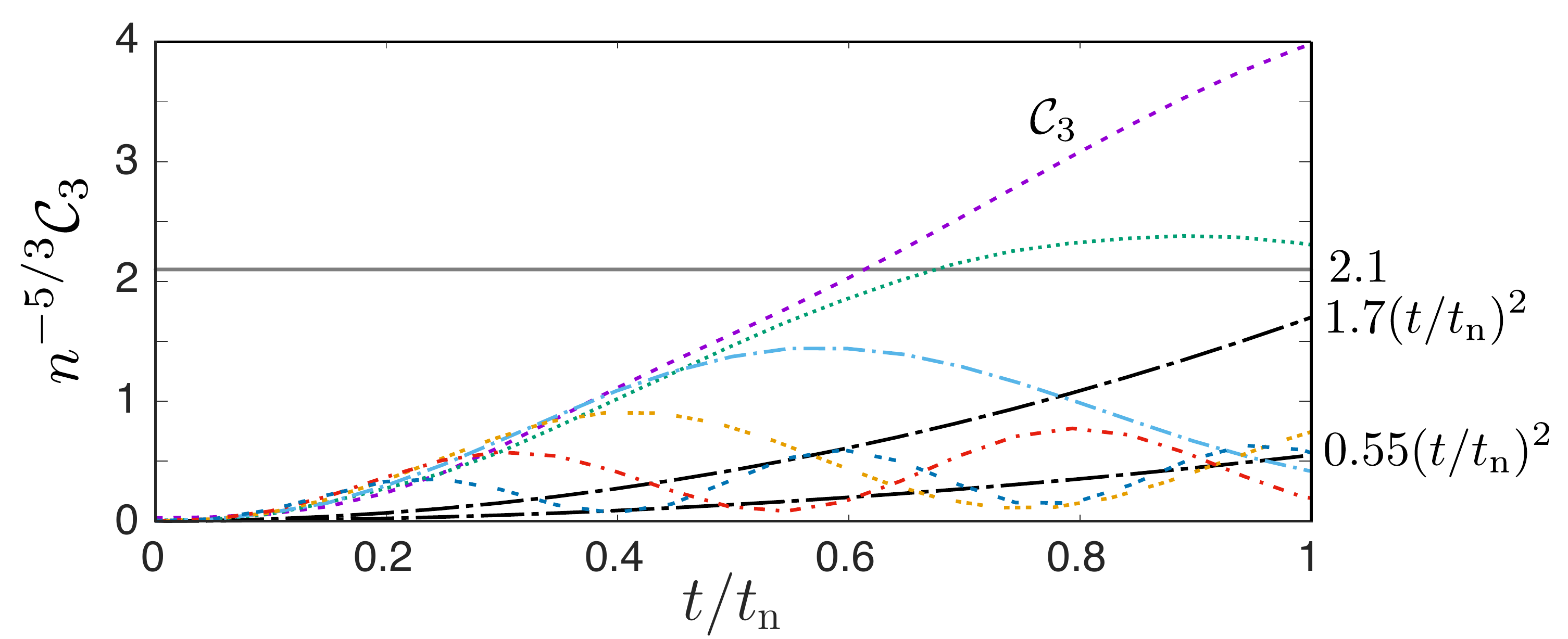}
\caption{Dynamics of the three-body contact (Eq.~\eqref{eq:c3_triplet}) over a range of densities and times up to $1.0 t_\mathrm{n}$.  Results for the dimensionless three-body contact density $n^{-5/3}\mathcal{C}_3$ (key same as Fig.~\ref{fig:c2_triplet}) from the triplet model.  For comparison, we show the range of quadratic early-time growths (dashed-dotted) found in Ref.~\cite{PhysRevLett.120.100401}.  For scale comparison, we display also the universal fit $n^{-5/3}\mathcal{C}_3\approx 2.1$ \cite{PhysRevLett.112.110402}  (grey solid) extracted from the experimental results of Ref.~\cite{makotyn2014universal} under the assumption of a locally equilibrated metastable state.} \label{fig:c3_data}
\end{figure} 
%
%
%
\begin{figure}[ht!]
\centering
\includegraphics[width=8.6cm]{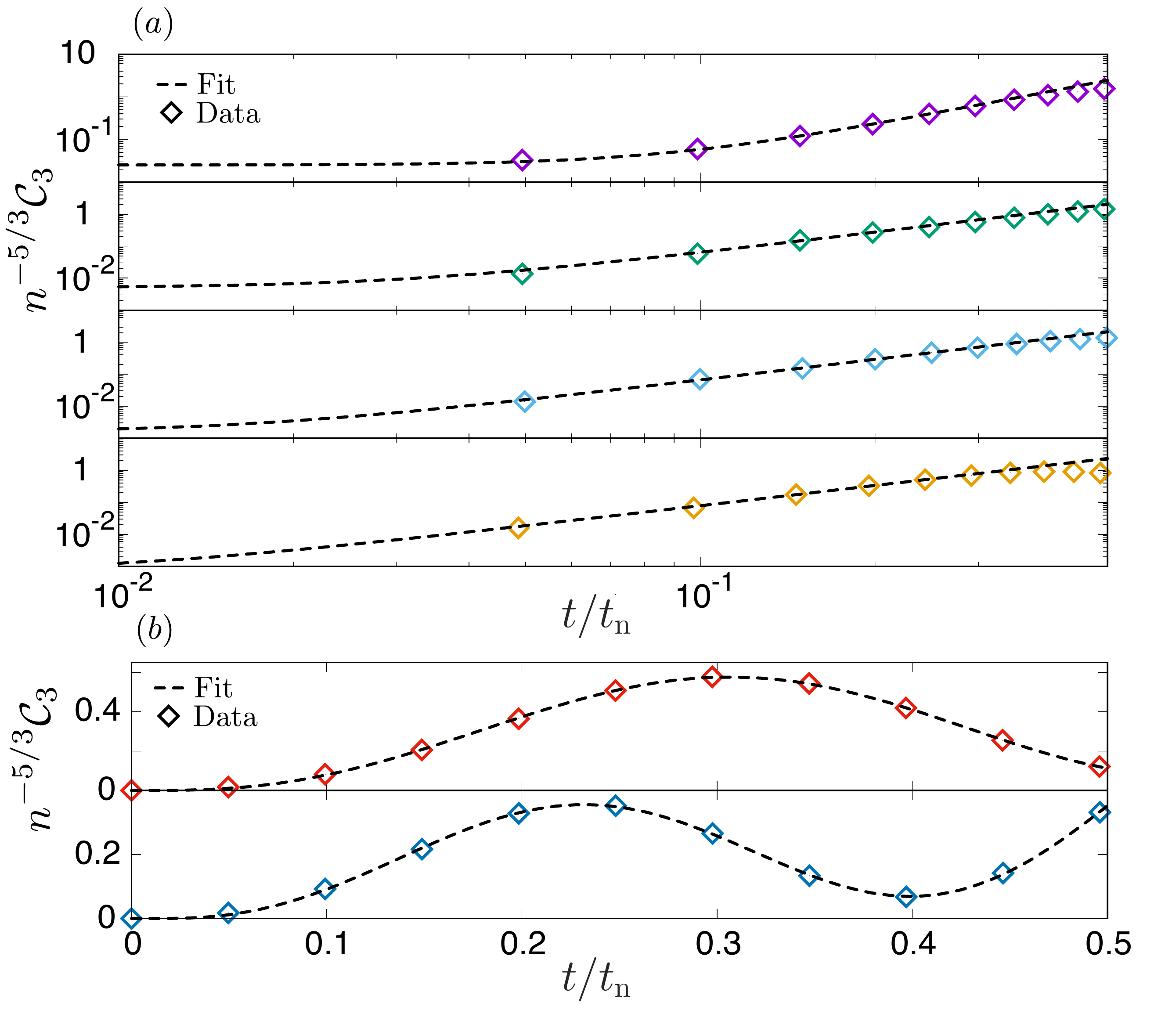}
\caption{Fits (dashed lines) of the early-time triplet model results (data points) for $n^{-5/3}\mathcal{C}_3(t)$ over a range of densities, corresponding in descending order to the key of Fig.~\ref{fig:c2_triplet}).  The results are fit to the functions (a) $f_1(t)=a_1 t^{a_2}$ and (b) $f_2(t)=b_1t^2+b_2\sqrt{t}\sin^2(b_3 t/2\hbar-b_4)$.  The fit parameters are given in Table~\ref{table:c3_fits}.} \label{fig:c3_fits}
\end{figure}
%
\begin{table}[h!]
\caption{Fits of the early-time triplet dynamics of $n^{-5/3}(\mathcal{C}_3(t)-\mathcal{C}_3(0))$ to power law $f_1(t)=a_1 t^{a_2}$ and oscillatory $f_2(t)=b_1t^2+b_2\sqrt{t}\sin^2(b_3 t/2\hbar-b_4)$ functions.  For densities where $n^{-5/3}\mathcal{C}_3(t)$ does not display a full period of oscillation in the early-time dynamics, we fit to $f_1(t)$ in the window $t\lesssim0.25t_\mathrm{n}$ to obtain the growth rates and power laws.  For densities where a full period is observable, we fit to $f_2(t)$ in the larger window $t\lesssim0.5t_\mathrm{n}$ to obtain growth rates and precise estimation of the oscillation frequency.  We note that using $f_1(t)$ in this latter regime would result in an overestimation of the growth rates.  The estimate of the ratio $|b_3/E_\mathrm{3b}^{(0)}|$ is obtained by averaging over multiple fits in the window $0.4\leq t/t_\mathrm{n}\leq 0.6$, and the uncertainty is given simply by the standard deviation.  }
\label{table:c3_fits}
\centering
\begin{ruledtabular}
\begin{tabular}{l l c c c c}
$nr_\mathrm{vdW}^3$ & $k_\mathrm{n}R^{(0)}_\mathrm{3b}$ & $a_1$ & $a_2$ \\
\noalign{\smallskip}
\hline
\noalign{\smallskip}
$3.0\times10^{-4}$ &  1.42 & $15.06$ & $2.65$ \\
\noalign{\smallskip}
$8.8\times10^{-5}$ &  0.95 & $9.31$ & $2.19$ \\
\noalign{\smallskip}
$3.7\times10^{-5}$ &  0.71 & $9.54$ & $2.17$ \\
\noalign{\smallskip}
$1.9\times10^{-5}$ &  0.57 & $10.10$ & $2.11$ \\
\noalign{\smallskip}
\hline
\hline
\noalign{\smallskip}
$nr_\mathrm{vdW}^3$ & $k_\mathrm{n}R^{(0)}_\mathrm{3b}$ & $b_1$ & $b_2$ & $|b_3/E_\mathrm{3b}^{(0)}|$ & $b_4$\\
\noalign{\smallskip}
\hline
\noalign{\smallskip}
$1.1\times10^{-5}$ & 0.48 & $0.25$ & $1.02$ & $0.97(3)$ & $0.09$ \\
\noalign{\smallskip}
$6.9\times 10^{-6}$ & 0.41 & $0.43$ & $0.71$ & $1.01(2)$ & $0.14$ \\
\end{tabular}
\end{ruledtabular}
\end{table}
\newpage
\subsubsection{$\mathcal{C}_3$ dynamics}

So far in the analysis of this section, we have assumed that the non-universal oscillations found in $\mu(t)$, $\mathcal{C}_2(t)$ and to a lesser extend $\Delta(t)$, which has no explicit dependence on triplet cumulants, are signatures of an Efimov state.  Here, we analyze the triplet model results for $\mathcal{C}_3(t)$, which directly measures the probability to measure short-range Efimovian triplets as correlations develop in the many-body system.  In Fig.~\ref{fig:c3_data}, the numerical results for the dynamical three-body contact are shown over a range of densities and times up to $t=1.0t_\mathrm{n}$.  At all times, the dynamics are density-dependent, and the non-universal oscillations in time become visible as $nr_\mathrm{vdW}^3$ is decreased.  To analyze these results, we first motivate why such oscillations should appear distinctly in $\mathcal{C}_3(t)$, then to make the discussion more quantitative, the triplet model results for $\mathcal{C}_3(t)$ are fit first for $t\leq0.25t_\mathrm{n}$ to obtain the relevant early-time scalings and then fit at later times $t<0.5t_\mathrm{n}$ to extract the oscillation frequencies, enabling a unambiguous identification of the Efimov state present in the many-body system.      

Counting powers of $\Lambda$ in the cumulant expansions in Eqs.~\eqref{eq:c2_triplet} and \eqref{eq:c3_triplet} reveals that the dynamical two and three-body contacts are dominated by the dynamics of the $c$ and $R$ cumulants, respectively.  In Sec.~\ref{sec:formal}, we discussed how the post-quench response of these cumulants (Eqs.~\eqref{eq:csol} and \eqref{eq:rsol}) is determined by the overlap between the few-body spectrum and the driving effect of the lower-order cumulants.  For example, away from resonance the dynamical two-body contact responds at the natural frequency of the universal dimer $-\hbar^2/ma^2$ as studied in Ref.~\cite{PhysRevA.91.013616,PhysRevA.99.023623}.  At unitarity, the dimer energy is at threshold, however the three-body contact can now respond at the frequency of any one of the infinity of three-body bound Efimov trimers.  In practice, the Efimov trimer whose size $R_\mathrm{3b}^{(j)}=\sqrt{2(1+s_0^2)/3}\exp(j\pi/s_0)/\kappa_*$ is comparable to the interparticle spacing $R_\mathrm{3b}^{(j)}\sim k_\mathrm{n}$ has the greatest overlap with the drive of the lower-order cumulants (see \eqref{eq:rsol}).  Consequently, the infinity of trimers accumulating at threshold play a negligible role in the dynamics of the three-body contact as was found in Refs.~\cite{PhysRevLett.120.100401,PhysRevA.99.043604}.  For densities in the regime $|E^{(j)}_\mathrm{3b}|\lesssim E_\mathrm{n}$, the $j$th Efimov state is optimally embedded in the many-body configuration, and the mode-matching $|E^{(j)}_\mathrm{3b}|\sim E_\mathrm{n}$ signals maximally-enhanced growth of the dynamical three-body contact at early times as found in Refs.~\cite{PhysRevLett.120.100401,PhysRevA.99.043604}.  When $|E^{(j)}_\mathrm{3b}|\gtrsim E_\mathrm{n}$, the Fermi and Efimovian timescales are distinct, and the frequency of the $j$th Efimov state are observable in the early-time dynamics of the three-body contact \cite{PhysRevLett.120.100401,PhysRevA.99.043604}.  Finally, when $|E^{(j)}_\mathrm{3b}|\gg E_\mathrm{n}$, there is little overlap between density scales and the $j$th Efimov state, and therefore all deeply bound Efimov states relative to $E_\mathrm{n}$ play a negligible role in the dynamics.  This behavior is then repeated over a full log-period (factor $\sim22.7^3$) in the density.

Whereas $\mathcal{C}_2(t)$ displays a universal linear growth at early-times, $\mathcal{C}_3(t)$ was shown to follow a range of gradual, density-dependent quadratic growth rates in Refs.~\cite{PhysRevLett.120.100401,PhysRevA.99.043604}.  In Fig.~\ref{fig:c3_data}, we compare the triplet model results against the predictions of these few-body models, noting the absence of inelastic three-body losses in the present work.  To quantify this comparison, we fit the early-time ($t\lesssim 0.25t_\mathrm{n}$) triplet model results for $n^{-5/3}\mathcal{C}_3(t)$ to the function $f_1(t)=a_1t^{a_2}$, and extract the growth rates $a_1$ and power laws $a_2$.  From averages the fits given in Table~\ref{table:c3_fits}, we estimate a $a_2=2.3(2)$ scaling law.  These fits are compared directly against the triplet model results in Fig.~\ref{fig:c3_fits}(a), where the breakdown of the early-time power-law growth becomes apparent by $t\sim 0.5t_\mathrm{n}$.  Even though quadratic growth is approached for decreasing density, the rates remain larger than the predictions of Refs.~\cite{PhysRevLett.120.100401,PhysRevA.99.043604} indicated by the black dashed-dotted lines in Fig.~\ref{fig:c3_data}.  At later times however, we see that this growth is overestimated as dynamics become oscillatory, which indicates that a more sophisticated fitting function should be used for the lowest densities studied in this section as we address now.

For the lowest densities studied in the triplet model, the dynamical three-body contact displays oscillations with periods visible even at early times $t<0.5t_\mathrm{n}$.  To quantify the frequency of this oscillation, we fit the dynamics of $n^{-5/3}\mathcal{C}_3(t)$ to $f_2(t)=b_1t^2+b_2\sqrt{t}\sin^2(b_3t/2\hbar-b_4)$ to obtain the growth rates $b_1$ and $b_2$, the oscillation phase $b_4$, and the oscillation frequency $b_3$ reported in units of the nearby ground-state Efimov trimer binding energy $E_\mathrm{3b}^{(0)}$ in Table~\ref{table:c3_fits}.  The non-analytic form of $f_2(t)$ was chosen as a combination of $t^2$ and $t^{5/2}$ power laws, motivated by the range of scalings found at larger densities.  This provides an excellent fit of the data in the window $t\lesssim 0.5t_\mathrm{n}$ as shown in Fig.~\ref{fig:c3_fits}(b) \footnote{To perform reliable fits over the larger window $t\leq t_\mathrm{n}$, we expect that a more involved fitting routine is required.}.  We note that the $b_4$ contribution to $f_2(t)$ adds an additional $\sqrt{t}$ scaling at early-times, which is generally negligible as the phase offset is typically small.  From Table~\ref{table:c3_fits}, we find quadratic growth rates more comparable with the findings of Refs.~\cite{PhysRevLett.120.100401,PhysRevA.99.043604} and, importantly, a {\it precise} identification of the oscillation frequency of the ground Efimov trimer to within an uncertainty of a few percent.

\subsection{Summary}
In this section, the quenched unitary Bose gas was studied within the triplet model, focusing on $(i)$ how the doublet dynamics depart from the prethermal state and on $(ii)$ signatures of the Efimov effect in the many-body observables of the system.  Although the pairing field was found to approach a (roughly) universal steady-state, the dynamics of the instantaneous chemical potential did not, displaying visible non-universal oscillations at the frequency of the ground-state Efimov trimer.  The momentum distribution $n_{\bf k}$ was found to depart from the $1/k^4$ power-law towards an exponentially-decaying tail at a momentum-dependent rate, although the violation of the total energy in the triplet model prevented any observation of the crossover to true thermalization.  The development of the exponentially-decaying tail was shown to coincide with a large buildup of kinetic energy in the large-$k$ modes and a corresponding draining of kinetic energy from the low-$k$ modes relative to the doublet model results.  The dynamics of the two-body contact were shown to depart non-universally from the doublet model results after a period of universal growth at early times, and to display the characteristic beating phenomenon at the frequency of the ground-state Efimov trimer at later times, consistent with the behavior found in the few-body studies \cite{PhysRevLett.120.100401,PhysRevA.99.043604}.  The oscillatory dynamics of the three-body contact, which quantify the probability of measuring short-range Efimovian triplet clusters, were found to match quantitatively to the frequency of the ground-state Efimov trimer in vacuum, providing an important proof of the concept of the calibrated triplet model.  We note that the sensitivity of the Efimov effect to the ultraviolet scales provides a stringent benchmark on the implementation of the numerics that are discussed further in Appendix~\ref{app:SimEqns}.  

\section{Conclusion}\label{sec:conclusion}
We have illustrated that the cumulant expansion can be used to study the sequential buildup of correlations in a degenerate ultracold Bose gas quenched to the unitary regime.  After outlining the cumulant theory of the many-body system, discussing its truncation, and identifying the few-body effects included at each level of the hierarchy, the quenched unitary Bose gas was then modeled at the doublet and triplet levels.  In the doublet model, the gas was found to reach a universal prethermalized state after a period of rapid quantum depletion of the initially pure Bose-Einstein condensate.  In this state, signatures of a universal Bogoliubov dispersion law emerge in the far-from-equilibrium dynamics of the occupation numbers.  This can be understood from the proximal universal steady-states of the chemical potential and pairing fields in the prethermalized state.  Using the dynamical two-body contact, we then analyzed the kinetic energy per particle and connected with the finite, universal kinetic temperatures measured over a restricted momentum range in Ref.~\cite{eigen2017universal}.  In the triplet model, the introduction of non-integrable three-particle processes caused the system to depart from the prethermal state at a momentum-dependent rate.  This departure manifests in the large$-k$ occupation number dynamics as a transition away from the $1/k^4$ power law towards a decaying exponential $\exp(-\alpha k)$, coinciding with a large pileup of kinetic energy.  Additionally, the many-body observables were found to display sensitivity to Efimovian length and time scales to varying degrees. By analyzing the dynamical three-body contact, we made a precise identification of this dynamical effect with Efimov states.   

The Efimov effect is predicted not only to manifest dynamically but also as log-periodic violations of the continuous scaling of system observables with the atomic density \cite{PhysRevLett.121.023401,Kokkelmans2018,PhysRevLett.120.100401,PhysRevA.99.043604}.  Such a study may shed more light on the intriguing scenario $|E_\mathrm{3b}^{(n)}|\approx E_\mathrm{n}$ when an Efimov state becomes embedded in the medium.  
 Simulating the triplet model over a factor $\sim22.7^3$ in the density however remains a practical challenge due to the $\sim \Lambda^4$ scaling of the calculation time for the numerical implementation described in Appendix~\ref{app:SimEqns}.

More generally, this study lays the groundwork for how a cumulant approach can be used to systematically include non-perturbative few-body effects in a description of strongly-correlated, far-from-equilibrium many-body systems. This method provides a flexible tool for studying quenched quantum gases, regardless of their quantum statistics, with the flexibility of including, for instance, drive and loss terms to study open systems and out-of-equilibrium phase transitions. The range of possible extensions of this method highlights the importance of developing methods for truncating the hierarchy while preserving the underlying conservation laws.  These topics however remain the subject of future study.  

\begin{acknowledgments}
We thank Murray Holland, Jos\'e D'Incao, John Corson, Paul Mestrom, Thomas Secker, Jinglun Li, and Denise Braun for fruitful discussions.  V.E.C., S.M., and S.J.J.M.F.K. acknowledge financial support by the Netherlands Organisation for Scientific Research (NWO) under Grant No. 680-47-623 and by the Foundation for Fundamental Research on Matter (FOM).  H.K., M.V.R., and M.W. acknowledge support from the Flemish Research Foundation (FWO-Vl), project FWO:G042915N.  M.W. acknowledges additional support from project FWO:G016219N.  H.K. acknowledges additional support from the  European Union's Horizon 2020 research and innovation program under the Marie Sk\l odowska-Curie grant agreement number 665501.  M.V.R. gratefully acknowledges additional support from a BAEF postdoctoral fellowship.

\end{acknowledgments}

\appendix

\section{Few-body model at unitarity}
\label{app:few}
In this section, we detail the pairwise potential used to produce the numerical data analyzed in Secs.~\ref{sec:hfb} and \ref{sec:triplet}.  Our choice of potentials is motivated by requirements to provide a good approximation of few-body scattering and bound-states on resonance while remaining computationally efficient.

\subsection{Two-body calibration}\label{sec:twobody}
The local potential introduced in Sec.~\ref{sec:hamiltonian} can always be expanded as a sum of nonlocal separable potentials
\begin{equation}\label{eq:sepexp}
\langle {\bf k}|\hat{V}|{\bf k'}\rangle=\sum_{j=1} g_j\langle{\bf k}|\zeta_j\rangle\langle\zeta_j|{\bf k'}\rangle,
\end{equation}
with form factors $|\zeta_j\rangle$ and interaction strengths $g_j$ \cite{faddeev2013quantum}.  Using the separable expansion, the Lippman Schwinger equation for the $T$-operator $\hat{T}(z)=\hat{V}+\hat{V}\hat{G}_{\mathrm{2B}}^{(0)}(z)\hat{T}(z)$, with two-body free Green's function $\hat{G}_\mathrm{2B}^{(0)}(z)=(z-2\hat{\epsilon})^{-1}$ and one-body kinetic-energy operator $\hat{\epsilon}|{\bf k}\rangle=\hbar^2k^2/2m|{\bf k}\rangle$, can be solved for a closed expression of the $T$-matrix $T({\bf k},{\bf k'},z)=\langle {\bf k}|\hat{T}(z)|{\bf k}'\rangle$ as 
\begin{equation}
T({\bf k},{\bf k'},z)=\sum_{ij} g_j\langle{\bf k}|\zeta_i\rangle\langle\zeta_j|{\bf k'}\rangle[\Xi^{-1}({\bf k'},z)]_{ij},
\end{equation}
where
\begin{equation}
\Xi_{ij}=\frac{1}{g_j}\delta_{ij}+\int \frac{d^3k}{(2\pi)^3} \frac{\langle{\bf k}|\zeta_j\rangle\langle\zeta_i|{\bf k}'\rangle}{\hbar^2k^2/m-z}.
\end{equation} 
In the limit where the binding energy of a shallow $s$-wave bound state nears threshold, referred to as a zero-energy resonance, the scattering length becomes large and the partial cross section approaches the unitarity limit \cite{newton2013scattering}.  In this case one of the $\Xi_{ii}$'s will vanish for $z\to0$, and the $T$-matrix is dominated by the corresponding simple pole
\begin{equation}\label{eq:tmatrixunitary}
T({\bf k},{\bf k'},z)= \frac{\langle{\bf k}|\zeta\rangle\langle\zeta|{\bf k'}\rangle}{\Xi(z)},
\end{equation}
known as the unitary pole approximation \cite{glockle1983}.  Within this approximation, the actual potential can be replaced by a nonlocal separable potential $\hat{V}=g|\zeta\rangle\langle\zeta|$, which reproduces Eq.~\eqref{eq:tmatrixunitary}.

Following Refs.~\cite{Kokkelmans2018,PhysRevA.100.013612}, we choose $s$-wave form factors $\langle {\bf k}|\zeta\rangle=\theta(\Lambda-|{\bf k}|)$ that are functions of the relative momentum.  The function $\theta(x)$ is the unit step function defined such that $\theta(x\geq0)=1$ and $\theta(x<0)=0$.  For a separable potential, the Lippmann-Schwinger equation for the two-body $T$ operator $\hat{T}(z)=\hat{V}+\hat{V}\hat{G}_{\mathrm{2B}}^{(0)}(z)\hat{T}(z)$ yields the closed expression
\begin{align}
\hat{T}(z)&=\frac{g|\zeta\rangle\langle \zeta|}{1-g\langle\zeta|\hat{G}_\mathrm{2B}^{(0)}(z)|\zeta\rangle},\label{eq:simplepole}\\
&=\begin{cases}
\frac{g|\zeta\rangle\langle \zeta|}{1+g\frac{m}{2\pi\hbar^2}\left[\Lambda-k\tanh^{-1}\left(\frac{\Lambda}{k}\right)+\frac{i\pi k}{2}\right]}\quad&\text{for }z=\frac{\hbar^2k^2}{m},\\
\frac{g|\zeta\rangle\langle \zeta|}{1+g\frac{m}{2\pi\hbar^2}\left[\Lambda-k\tan^{-1}\left(\frac{\Lambda}{k}\right)\right]}\quad&\text{for }z=-\frac{\hbar^2k^2}{m}.
\end{cases}\nonumber\\
\end{align}
The coupling constant $g$ is determined by matching with the low-energy limit of the on-shell $T$-matrix for $s$-wave scattering
\begin{eqnarray}
 \frac{4\pi\hbar^2}{m}a&\underset{|{\bf k}|\to0}{=}&\langle{\bf k},-{\bf k}|\hat{T}(\hbar^2k^2/m+i0)|{\bf k}',-{\bf k}'\rangle,\\
 &=&\left(\frac{1}{g}+\frac{m\Lambda}{2\pi^2\hbar^2}\right)^{-1},
 \end{eqnarray}

 which yields the expression $g=-2\pi^2\hbar^2/m\Lambda$ on resonance.  The cutoff $\Lambda$ is calibrated to reproduce finite-range corrections to the molecular binding energy $-\hbar^2/m(a-\bar{a})^2$ away from resonance, where $\bar{a}\approx0.956r_\mathrm{vdW}$ is the mean-scattering length that is set by the van der Waals length $r_\mathrm{vdW}$ for a give atomic species \cite{PhysRevA.59.1998}.  For the $^{39}$K experiments modeled in this work, we take $r_\mathrm{vdW}=64.61a_0$ \cite{D_Errico_2007,RevModPhys.82.1225}.  
 
The simple pole of the $T$ operator in Eq.~\eqref{eq:simplepole} gives the binding energy $E_\mathrm{D}=-\hbar^2\kappa^2/m$ in the limit $\tilde{\kappa}=\kappa/\Lambda\ll1$ as
 \begin{equation}
\frac{\pi\tilde{\kappa}}{2}-\tilde{\kappa}^2-\frac{\pi}{2a\Lambda}=O\left(\tilde{\kappa}^4\right).
 \end{equation}  
 Ignoring quartic and higher-order contributions and equating with the molecular binding energy with finite range corrections \cite{RevModPhys.82.1225,PhysRevA.59.1998} yields
 \begin{equation}
 \Lambda=\frac{2}{\pi\bar{a}(1-\bar{a}/a)}\approx \frac{2}{\pi\bar{a}},
 \end{equation}
 which is expanded in the small parameter $\bar{a}/a$ valid in the strongly-interacting regime $a/r_\mathrm{vdW}\gg1$.  To understand the significance of this calibration, we compare the effective range approximation of the on-shell $T$-matrix in the unitarity limit
 \begin{align}
\langle{\bf k}|\hat{T}(\hbar^2k^2/m+i0)|{\bf k}'\rangle\underset{k\to 0}{\approx}\frac{4\pi\hbar^2}{m}\frac{1}{ik-r_\mathrm{eff}k^2/2},
\end{align}
where $|{\bf k}|=|{\bf k}'|$, with the equivalent limit of Eq.~\eqref{eq:simplepole}
 \begin{align}
\langle{\bf k}|\hat{T}(\hbar^2k^2/m+i0)|{\bf k}'\rangle\underset{k\to 0}{\approx}\frac{4\pi\hbar^2}{m}\frac{1}{ik-2k^2/\pi\Lambda},
\end{align}
which yields $r_\mathrm{eff}=4\pi/\Lambda=2\bar{a}$.  We compare this with the result $r_\mathrm{eff}=3\bar{a}$ for a Lorentzian form factor \cite{Schmidt2012} and with the analytic result $r_\mathrm{eff}=\Gamma(1/4)^4\bar{a}/6\pi^2\approx2.92\bar{a}$ for the effective range of a pure $1/r^6$ van der Waals interaction at unitarity \cite{PhysRevA.59.1998}.  Taking the zero-range approximation yields the well-known $1/k^2$ scaling of the unitarity bounded partial cross section.  It is instructive to evaluate also the equivalent zero-range expression for the retarded $T$ operator \cite{newton2013scattering} in the time domain
\begin{align}
\hat{T}_+(\tau)&=\frac{1}{2\pi\hbar}\int_{-\infty}^\infty dE e^{-iE\tau/\hbar}\hat{T}(E+i0),\nonumber\\
&=-\theta(\tau)\frac{|\zeta\rangle\langle\zeta|}{\sqrt{\tau}}\sqrt{\frac{16i\pi\hbar^5}{m^3}},\label{eq:tdur}
\end{align}
which can be obtained by analytic continuation of the Gaussian integral $I(z)=\int_0^\infty dk e^{-k^2 z}$ in the half plane $\Re[z]>0$.  We contrast this gradual $\tau^{-1/2}$ decay with the sharply peaked Born approximation $\hat{T}_+(\tau)=g\delta(\tau)|\zeta\rangle\langle \zeta|$.   
\subsection{Efimov spectrum}\label{sec:threebody}
The calibration scheme for the interaction parameters yields finite range corrections to two-body binding energies and scattering amplitudes due to the long-range van der Waals interactions remaining on resonance.  On the two-body level, these corrections to the binding energy become less important near unitarity as the ratio $a/\bar{a}$ approaches infinity.  However, on the three-body level, the spectrum of three-body bound Efimov states is set by finite-range effects.  And so it is important to check that the calibration scheme produces a trimer spectrum which matches roughly what has been observed experimentally.

To solve the three-body problem in vacuum for our calibrated separable potential, we begin with the decomposition of the three-body wave function $|\Psi_\mathrm{3B}\rangle=|\Psi^{(1)}\rangle+|\Psi^{(2)}\rangle+|\Psi^{(3)}\rangle$ into Faddeev components \cite{faddeev2013quantum} satisfying the bound-state equation in momentum space
\begin{widetext}
\begin{equation}\label{eq:stm1}
\Psi^{(1)}({\bf q}_1,{\bf p}_1)=G^{(0)}_\mathrm{3B}(q_1,p_1,E)\int\frac{d^3 q'}{(2\pi)^3}\int\frac{d^3 p'}{(2\pi)^3}\langle {\bf q}_1,{\bf p}_1|\hat{T}_{23}(E)|{\bf q'},{\bf p'}\rangle\langle {\bf q'},{\bf p'}|\hat{P}_++\hat{P}_-|\Psi^{(1)}\rangle,
\end{equation}
\end{widetext}
where $\hat{T}_{23}(z)=\hat{V}_{23}+\hat{V}_{23}\hat{G}_\mathrm{3B}^{(0)}(E)\hat{T}_{23}(z)$, $E$ is the binding energy, and $\hat{G}_\mathrm{3B}^{(0)}(z)=(z-\sum_{i=1}^3 \hat{\epsilon}_i)^{-1}$ is the vacuum three-body Green's function.  In Eq.~\eqref{eq:stm1}, the three-body system with single-particle wavevectors ${\bf k}_1$, ${\bf k}_2$, and ${\bf k}_3$ is parametrized by Jacobi vectors ${\bf q}_1=({\bf k}_2-{\bf k}_3)/2$ and ${\bf p}_1=(2{\bf k}_1-{\bf k_2}-{\bf k_3})/3$.  Following the original formulation of Skorniakov and Ter-Martirosian \cite{skorniakov1957three}, we make the ansatz $|\Psi^{(1)}\rangle=N\hat{G}_\mathrm{3B}^{(0)}(E)(|\zeta\rangle\otimes|\mathcal{F}\rangle)$, where $N$ is the normalization constant, and the tensor product is defined as $\langle {\bf q}_1,{\bf p}_1|(|\zeta\rangle\otimes|\mathcal{F}\rangle)=\zeta(2q_1)\mathcal{F}(p_1)$.  Inserting this ansatz into Eq.~\eqref{eq:stm1} yields the one-dimensional integral equation
\begin{widetext}
\begin{equation}\label{eq:stm2}
\mathcal{F}(p_1)=2g\tau\left(E-\frac{3\hbar^2p_1^2}{4m}\right)\int \frac{d^3 p'}{(2\pi)^3}\ \frac{\zeta(|2{\bf p_1}+{\bf p'}|)\zeta(|2{\bf p'}+{\bf p_1}|)}{E-\frac{\hbar^2p_1^2}{m}-\frac{\hbar^2p'^2}{m}-\frac{\hbar^2{\bf p_1\cdot p'}}{m}}\mathcal{F}(p'),
\end{equation}
\end{widetext}
where $\tau(z)=1/(1-g\langle\zeta|\hat{G}^{(0)}_\mathrm{2B}(z)|\zeta\rangle$.  Nontrivial solutions of Eq.~\eqref{eq:stm2} correspond to the Efimov trimer binding energies at unitarity \cite{skorniakov1957three}.  For the calibrated separable potential introduced in Sec.~\ref{sec:twobody} and used in the many-body simulations, the resulting trimer spectrum is given in Table~\ref{table:efimov}.  The wavenumber of the ground trimer $\kappa_*\equiv\kappa^{(0)}=0.211/r_\mathrm{vdW}$ is the three-body parameter, which compares with the universal result $\kappa_*r_\mathrm{vdW}\approx0.226$ for broad, open-channel dominated Feshbach resonances \cite{RevModPhys.82.1225,PhysRevLett.108.263001,PhysRevA.90.022106}.  Additionally, the zero-range model predictions for the $22.7^2$ geometric scaling between neighboring energies is recovered for the highly-excited Efimov trimers as is expected in a finite-ranged model \cite{BRAATEN2006259,Naidon_2017,D_Incao_2018}.  Ultimately, we see that our calibrated separable potential, despite being tailored to corrections on the two-body level, captures the sensitive dependence of Efimov physics on finite-range effects on the three-body level. 
\begin{table}[ht]
\caption{Approximate values for the ground and first-excited Efimov trimer binding energies obtained from numerical solutions of Eq.~\eqref{eq:stm2} using the calibrated separable pairwise potential.  In practice, to obtain trimer binding energies from the integral equation (Eq.~\eqref{eq:stm2}), we follow the Nystrom method and convert the integral equation into a summation over a Gauss-Legendre quadrature \cite{press1989numerical}.  The resultant equation can be solved as an eigenvalue problem, with eigenvalues corresponding to trimer binding energies.}
\label{table:efimov}
\centering
\begin{ruledtabular}
\begin{tabular}{l c c c c c}
$j$ & $E^{(j)}_\mathrm{3b}/E_\mathrm{vdW}$ & $E^{(j)}_\mathrm{3b}/E^{(j+1)}_\mathrm{3b}$ & $\kappa^{(j)}r_\mathrm{vdW}$ & $\kappa^{(j)}/\Lambda$\\
\noalign{\smallskip}
\hline
\noalign{\smallskip}
0 & 0.0446 & $24.2^2$ & 0.211 & 0.317\\
\noalign{\smallskip}
1& $7.62\times 10^{-5}$ & $22.7^2$ & 0.00873 & 0.0131  \\
\end{tabular}
\end{ruledtabular}
\end{table}

\renewcommand{\vec}{\mathbf}

\section{Numerical Methods}
\label{app:SimEqns}
In this section, the cumulant equations of motion of Sec.~\ref{sec:eom} are rewritten for the nonlocal separable potential discussed in Sec.~\ref{app:few} that is used in our numerics.  We will see in Sec.~\ref{app:implementation} that the factorized form of the $s$-wave separable potential effectively reduce some integrations in the triplet cumulant equations of motions from 3D to 2D.
 
In Sec.~\ref{app:eom}, we give this `simulation form' of the cumulant equations.  Besides the different interaction potential, these cumulant equations differ from those in Sec.~\ref{sec:eom} in that we ignore all quadruplets to yield a triplet model, and we simulate only a subset of the most dominant terms in $\mathcal{M}^{H_4}$ (Eq.~\eqref{eq:MH4}) and $\mathcal{R}^{H_4}$ (Eq.~\eqref{eq:RH4}) to simplify numerics.  In Sec.~\ref{app:implementation}, we discuss our numerical methods for simulating the cumulant equations.  In Sec.~\ref{app:convergence}, the convergence of our simulation with respect to grid parameters is analyzed for completeness. 
 
 \subsection{Equations of motion}\label{app:eom}

 We begin by rewriting the many-body Hamiltonian (Eq.~\eqref{hamiltonianF}) for the nonlocal separable potential
 \begin{align}
\hat{H}=&\sum_{\bf k}\epsilon_{\bf k}\hat{a}^\dagger_{\bf k}\hat{a}_{\bf k}\nonumber\\
&+\frac{g}{2V}\sum_{{\bf p},{\bf p'},{\bf q}}\zeta_{{\bf p}-{\bf p'}+2{\bf q}}\zeta_{{\bf p}-{\bf p'}} a^\dagger_{{\bf p}+{\bf q}}a^\dagger_{{\bf p'}-{\bf q}}\hat{a}_{\bf p} \hat{a}_{\bf p'},
\end{align}
where we have used the shorthand $\langle {\bf k,k'}|\hat{V}|{\bf k'',k'''}\rangle=g\delta_{{\bf k+k'},{\bf k''+k'''}}\zeta_{\bf k-k'}\zeta_{\bf k''-k'''}$ with $\zeta_{\bf k-k'}=\theta(\Lambda-|{\bf k-k'}|/2)$ to take expectation values of the form factors in the lab frame.
From the Gross-Pitaevskii equation
\begin{align}\label{eq:psi_sep} 
i \hbar\partial_t \psi_0=&g\left(\zeta_0^2n_0+\frac{2g}{V}\sum_{\bf l}\zeta_{\bf l}^2n_{\bf l}\right)\psi_0+\frac{g\psi_0^*}{V}\sum_{\bf l}\zeta_0\zeta_{2{\bf l}}c_{\bf l}\nonumber\\
&+\frac{g}{V^{3/2}}\sum_{{\bf l},{\bf s}}\zeta_{\bf l}\zeta_{2{\bf s}-{\bf l}}M^*_{{\bf l},{\bf s}},
\end{align}
we extract the condensate phase derivative as in Eq.~\eqref{thetapoint} 
\begin{align}
\hbar\frac{\dd\theta_0}{\dd t}=&-\frac{1}{2n_0} \bb{\psi_0^*\ii\hbar \frac{\dd\psi_0}{\dd t}-\ii\hbar \frac{\dd\psi_0^*}{\dd t}\psi_0},\\
=&-\Bigg[g\zeta_0^2n_0+\frac{2g}{V}\sum_{\bf l} \zeta_{\bf l}^2 n_{\bf l}+\frac{g}{V}\sum_{\bf l}\zeta_0\zeta_{2{\bf l}}\Re c_{\bf l}\nonumber\\
&+\frac{g}{\sqrt{n_0V^3}}\sum_{{\bf l},{\bf s}}\zeta_{\bf l}\zeta_{2{\bf s}-{\bf l}}M^*_{{\bf l},{\bf s}}\Bigg].\label{eq:thetadot_sep}
\end{align}
From the Heisenberg equation for the unrotated operators $\hat{b}_{\bf k}$ (See Eq.~\eqref{dtb}), we obtain the form of the doublet equations of motion for the separable potential
\begin{widetext}
\begin{eqnarray}\label{eq:n_num}
i\hbar\partial_t n_{\bf k}&=&2i\ \text{Im}\left[\Delta_{\bf k}c_{\bf k}^*+2g\sqrt{\frac{n_0}{V}}\sum_{\bf l}\zeta_{2{\bf k}-{\bf l}}\zeta_{\bf l} M_{{\bf l},{\bf k}}+g\sqrt{\frac{n_0}{V}}\sum_{\bf l}\zeta_{\bf k}\zeta_{2{\bf l}-{\bf k}}M^*_{{\bf k},{\bf l}}\right],\\
\label{eq:n_sep}
i\hbar\partial_t c_{\bf k}&=&2E_{\bf k}c_{\bf k}+(1+2n_{\bf k})\Delta_{\bf k}+4g\sqrt{\frac{n_0}{V}}\sum_{\bf l}\zeta_{\bf l+k}\zeta_{\bf l-k} M^*_{{\bf l},{\bf k}}+2g\sqrt{\frac{n_0}{V}}\sum_{\bf l}\zeta_{\bf k}\zeta_{2{\bf l}-{\bf k}}R_{{\bf k},{\bf l}},\label{eq:c_sep}
\end{eqnarray}
\end{widetext}
 where we use the forms of the Hartree-Fock hamiltonian (Eq.~\eqref{eq:kerenorm}) and pairing field (Eq.~\eqref{eq:delta}) for a separable potential
\begin{align}\label{eq:bog_param}
&E_{\bf k}=\epsilon_{\bf k}+2g\left[\zeta_{\bf k}^2n_0+\frac{1}{V}\sum_{{\bf l}}\zeta_{{\bf k}-{\bf l}}^2n_{\bf l}\right]+\hbar\partial_t\theta_0,\\
&\Delta_{\bf k}=g\zeta_{2{\bf k}}\left[\zeta_0n_0+\frac{1}{V}\sum_{\bf l}\zeta_{2{\bf l}}c_{\bf l}\right].
\end{align}
For the triplet equations of motion, we obtain the forms of Eqs.~\eqref{eq:Mdot} and \eqref{eq:Rdot} for a separable potential
\begin{widetext}
\begin{eqnarray}
\label{eq:M_sep}
i\hbar\partial_t M_{\vec{k},\vec{q}} &=& \Big( E_\vec{k} - E_\vec{q} - E_{\vec{k}-\vec{q}}  \Big) M_{\vec{k},\vec{q}} 
-  \Delta^*_{\bf k-q}M^\ast_{\vec{q},\vec{k}} -\Delta^*_{\bf q} M^\ast_{\vec{k}-\vec{q},\vec{k}} + \Delta_{\bf k} R_{\vec{k},\vec{q}}^\ast + \mathcal{M}^{H_3}_{\bf k,q} +\mathcal{M}^{H_4}_{\bf k,q},\\
\label{eq:R_sep}
i\hbar\partial_t R_{\vec{k},\vec{q}} &=& \Big( E_\vec{k} + E_\vec{q} + E_{\vec{k}-\vec{q}} \Big) R_{\vec{k},\vec{q}} + \Delta_{\bf k} M^\ast_{\vec{k},\vec{q}} + \Delta_{\bf q}M^\ast_{\vec{q},\vec{k}} +\Delta_{\bf k-q} M^\ast_{\vec{k}-\vec{q},\vec{k}} + \mathcal{R}^{H_3}_{\bf k,q} +\mathcal{R}^{H_4}_{\bf k,q},
\end{eqnarray}
\end{widetext}
where the doublet sources are the forms of Eqs.~\eqref{eq:MH3} and \eqref{eq:RH3} for a separable potential
\begin{widetext}
\begin{eqnarray}
\label{eq:MH3_sep}
 \frac{\mathcal{M}^{H_3}_{\bf k,q}}{\sqrt{n_0/V} }&=&2g\left(\zeta_{\bf 2k-q}\zeta_{\bf q}c^*_{\vec{k}-\vec{q}}n_\vec{q} + \zeta_{\bf k+q}\zeta_{\bf k-q}n_{\vec{k}-\vec{q}}c^*_\vec{q} - n_\vec{k}(\zeta_{\bf k+q}\zeta_{\bf k-q}c^*_\vec{q}+\zeta_{\bf q}\zeta_{\bf 2k-q}c^*_{\vec{k}-\vec{q}})\right)\nonumber\\
&&+ 2g\Big(\zeta_{\bf 2q-k}\zeta_{\bf k}n_{\vec{k}-\vec{q}}n_\vec{q}-\zeta_{\bf 2q-k}\zeta_{\bf k}n_\vec{k}(1+n_\vec{q} +n_{\vec{k}-\vec{q}})  - c_\vec{k}(\zeta_{\bf 2k-q}\zeta_{\bf q}c^*_\vec{q}+\zeta_{\bf k+q}\zeta_{\bf k-q}c^*_{\vec{k}-\vec{q}})\Big), \\
\label{eq:RH3_sep}
 \frac{\mathcal{R}^{H_3}_{\bf k,q}}{\sqrt{n_0/V} } &=&2g\Big(\zeta_{\bf 2q-k}\zeta_{\bf k}c_\vec{k}(1+n_\vec{q}+n_{\vec{k}-\vec{q}}) + \zeta_{\bf 2k-q}\zeta_{\bf q}c_\vec{q}(1+n_\vec{k}+n_{\vec{k}-\vec{q}}) +\zeta_{\bf k+q}\zeta_{\bf k-q} c_{\vec{k}-\vec{q}}(1+n_\vec{k}+n_{\vec{q}})\Big)\nonumber\\
&&+2g\Big(\zeta_{\bf k-q}\zeta_{\bf  k+q}c_\vec{k} c_\vec{q} + \zeta_{\bf q}\zeta_{\bf 2k-q}c_\vec{k} c_{\vec{k}-\vec{q}} + \zeta_{\bf k}\zeta_{\bf 2q-k}c_\vec{q} c_{\vec{k}-\vec{q}} \Big).
\end{eqnarray}
\end{widetext}
We approximate the triplet terms $\mathcal{M}^{H_4}_{\bf k,q}$ and $\mathcal{R}^{H_4}_{\bf k,q}$ by calculating only their most dominant contributions
\begin{widetext}
\begin{eqnarray}
\label{eq:MH4_sep}
\mathcal{M}^{H_4}_{\bf k,q}&\approx& 
 -\frac{g}{V} \sum_\vec{l} \zeta_{\bf 2q-k}\zeta_{\bf 2l-k}M_{\vec{k},\vec{l}} \big( n_{\vec{k}-\vec{q}} +n_\vec{q} +1 \big),\\
\label{eq:RH4_sep}
\mathcal{R}^{H_4}_{\bf k,q}&\approx& \frac{g}{V}\sum_\vec{l}  \Bigg(\zeta_{\bf 2q-k}\zeta_{\bf 2l-k}R_{\vec{l},\vec{k}} \big( n_\vec{q}+ n_{\vec{k}-\vec{q}} + 1 \big) + \zeta_{\bf 2k-q}\zeta_{\bf 2l-q} R_{\vec{l},\vec{q}} \big( n_\vec{k} + n_{\vec{k}-\vec{q}}+ 1\big) \nonumber\\
&&+ \zeta_{\bf k+q}\zeta_{\bf k-q+2l}R_{\vec{l},\vec{k}-\vec{q}} \big(n_\vec{k} + n_\vec{q} +1 \big) \Bigg).
\end{eqnarray}
\end{widetext}
The `$+1$' terms make the most dominant contributions to $\mathcal{M}^{H_4}$ and $\mathcal{R}^{H_4}$ at early times before quantum depletion becomes appreciable.  This can be understood by simply counting the number of operator products, but is also something that we confirmed numerically.  From Sec.~\ref{sec:formal}, we know also that these terms are required to produce the correct form of the interacting few-body Hamiltonian.  In addition to the these terms we have included the sub-dominant Bose-enhancement factors of the form `$1+n+n$' so that scattering is described at the level of the many-body $T$-matrix, consistent with the equation of motion for the $c$ cumulant (Eq.~\eqref{eq:c_sep}).  We emphasize however that due to the restriction of our analysis of the triplet simulation to $t\lesssim t_\mathrm{n}$ before quantum depletion becomes significant, the difference between vacuum and many-body $T$-matrices is minimal (c.f. the discussion in Ref.~\cite{Kokkelmans2018}).  Finally, the approximations in Eq~\eqref{eq:MH4} and \eqref{eq:RH4} also have a practical purpose in reducing significantly the computational burden which is addressed in the following subsection on implementation.

 \subsection{Implementation}\label{app:implementation}
Because we simulate a uniform Bose gas at rest in three dimensions, the doublets $n_\vec{k}$ and $c_\vec{k}$ are spherically symmetric and can be represented as a vector with index $k_i=|\vec{k}|_i$. For the triplets $M_{\vec{k},\vec{q}}$ and $R_{\vec{k},\vec{q}}$ the situation is a little bit more complicated. We have that they are encoded by two 3D momentum vectors and should therefore depend on six parameters. However, we first have an overall rotation symmetry of $\vec{k}$ and $\vec{q}$ simultaneously, which excludes already two angles, and then also a rotation symmetry of $\vec{q}$ with respect to $\vec{k}$, which excludes another rotation angle. Therefore we are left with three independent parameters and we can parametrize $M_{\vec{k},\vec{q}}\equiv M(k,q,\cos{\theta_{\vec{k},\vec{q}}})$ as a 3D array on a grid $(k_i,q_i,\cos\theta|_i)$, where $k_i$ are $q_i$ are the vector norms and $ \cos{\theta}_i$ is the discretized cosine of the polar angle between $\vec{k}$ and $\vec{q}$. Many operations in the cumulant equations of motion in Sec.~\ref{app:eom} are pointwise and can be evaluated directly within this parametrization.

For the implementation, we also need to evaluate objects with swapped and/or shifted indices, like the doublet $n_{\vec{k}-\vec{q}}$ or the triplets $M_{\vec{q},\vec{k}}$ or $M_{\vec{k}-\vec{q},\vec{k}}$. For the doublets we can simply evaluate the vector norm $|\vec{k}-\vec{q}| = \sqrt{k^2+q^2-2kq\cos{\theta_{\vec{k},\vec{q}}}}$ and project the result to the nearest $k_i$ in our predefined grid, so that now $n_{\vec{k}-\vec{q}}$ becomes a 3D array after interpolation.
If we have to swap two indices $\vec{k}$ and $\vec{q}$ of a triplet, we can simply swap the first two momentum indices of the 3D array, since $\cos{\theta_{\vec{k},\vec{q}}}$ is invariant under the exchange of the two momenta, i.e. $M_{\vec{q},\vec{k}}\equiv M(q,k,\cos{\theta_{\vec{k},\vec{q}}})$. To evaluate $M_{\vec{k}-\vec{q},\vec{k}}\equiv M(|\vec{k}-\vec{q}|,k,\cos{\theta_{\vec{k}-\vec{q},\vec{k}}})$, we also have to evaluate $\cos{\theta_{\vec{k}-\vec{q},\vec{k}}}=(k-q\cos{\theta_{\vec{k},\vec{q}}})/|\vec{k}-\vec{q}|$, if we align $\vec{k}$ along the $z$-axis, after which we can apply a zeroth-order interpolation in 3D, i.e. we select the 3D index $(k_i,q_i,\cos\theta|_i)$ that is closest to the point $(|\vec{k}-\vec{q}|,k,\cos{\theta_{\vec{k}-\vec{q},\vec{k}}})$.
We have numerically compared the zeroth-order interpolation with first-order and even spline methods \cite{press1989numerical}, but the result is indistinguishable when the grid spacing is chosen finely enough. Note that zeroth-order interpolation is essentially a map of indices and can be precomputed, making it much more efficient than higher-order interpolation schemes.  The form factors $\zeta_\vec{k}$,  $\zeta_{\bf 2q-k}$ etc, can also be precomputed for our 3D grid $(k_i,q_i,\cos\theta|_i)$ and stored as logical 3D arrays for later use in the equations of motion.

Furthermore we have to evaluate the summations in spherical coordinates. For example, in (\ref{eq:bog_param}) we encounter a spherically symmetric summation, which can be evaluated as follows
\begin{equation}
\sum_\vec{l} \zeta_{2\vec{l}} c_\vec{l} \equiv \frac{V}{2\pi^2} \int_0^{k_\text{max}} l^2dl\, \zeta(2l) c(l) \approx \frac{V\Delta k}{2\pi^2} \sum_{i} k_i^2 \zeta_{2i} c_i.
\end{equation}
 Here $\Delta k$ is the difference between two consecutive elements in the vector $k_i$ (if uniform) and $k_\text{max}$ is the numerical grid cutoff and we do a simple form of Riemann integration, where we use the spherically symmetric vectors $\zeta_i\equiv \zeta(k_i)$ and $c_i\equiv c(k_i)$. In principle more involved algorithms can be implemented (like trapezoidal or Simpson's rule) but with a fine enough grid this turns out to be satisfactory. 
 
 Similarly, we also have integrals over a momentum index of a $(\vec{k},\vec{q})$-object. Also in (\ref{eq:bog_param}) we find the summation
 \begin{widetext}
\begin{equation}
\sum_\vec{q} \zeta^2_{\vec{k}-\vec{q}} n_\vec{q} \equiv \frac{V}{4\pi^2} \int_0^{k_\text{max}} q^2dq \int_{-1}^1 d\cos{\theta} \; \zeta_{\vec{k}-\vec{q}}(k,q,\cos\theta) n(q) \approx \frac{V\Delta k\, \Delta c}{4\pi^2} \sum_{j,m} k_j^2 \zeta_{\vec{k}-\vec{q};i,j,m} n_j \equiv \mathcal{H}_i,
\end{equation}
\end{widetext}
where $\Delta c$ is the differential element of the $\cos{\theta}|_i$ and we have defined the 3D array $\zeta_{\vec{k}-\vec{q}}$ with indexing $\zeta_{\vec{k}-\vec{q};i,j,m}\equiv \zeta_{\vec{k}-\vec{q}}(k_i,q_j,\cos{\theta}|_m)$. The result $\mathcal{H}_i$ is again a spherically symmetric array. To summarize, we implement the summation over an index of a $(\vec{k},\vec{q})$-object as a summation over two of the three indices of the corresponding 3D array, with the correct differential element for the Riemann integration.  We note that using the seperable potential, any summation in (\ref{eq:n_num})-(\ref{eq:RH4_sep}) can be evaluated with one of the two ways described above, after construction of the right vector or 3D array as the integrand.
 
 \subsection{Convergence}\label{app:convergence}

In this section, we provide details related to the numerical convergence of the triplet simulation.  We discuss the convergence of various simulation quantities with respect to the angular and momentum grid parameters.  We confirm also that the violation of the total energy in the triplet simulation agrees with analytics, providing an additional convergence test.  Finally, we detail the computing resources used to simulate the triplet equations of motion.

We choose a uniformly spaced momentum grid ${\bf k}=\{k_i\}_{i=1,\dots \mathrm{nk}}$ extending from $k_i=\Delta k$ to $k_\mathrm{nk}=k_\mathrm{max}$, where the system volume $V$ determines the grid-spacing $\Delta k$ through the usual relation $V=(2\pi/\Delta k)^3$.  The numerical cutoff $k_\mathrm{max}$, which is a truncation in the single-particle plane-wave basis, is distinct from the form factor cutoff $\Lambda$, which places an upper bound on both incoming and outgoing relative momentum involved in pairwise interactions.  Therefore, the pairwise generation of excitations with zero center of mass momentum described by the $c$ cumulant is inherently limited to single-particle momenta $k\leq\Lambda$.  Setting $k_\mathrm{max}=\Lambda$ is therefore justified for the HFB simulation, and we have numerically confirmed that $n_k$ vanishes for $k>\Lambda$.  In the triplet simulation, however, the $R$ and $M$ cumulants describe interactions where the center of mass momentum of an interacting pair does not vanish.  In fact, if this pairwise center of mass momentum does vanish, then the $R$ and $M$ cumulants are zero by construction.  For the $R$ cumulant the nonzero center of mass momentum of the interacting pair is offset by the third spectator atom.  For the $M$ cumulant the nonzero center of mass momentum of the interacting pair corresponds to the momentum of the incoming atom, which has been defined in the rest frame of the condensate.  Therefore, we have taken $k_\mathrm{max}>\Lambda$ in the triplet simulation to allow for the complete description of these processes.  

\begin{figure}[t!]
\centering
\includegraphics[width=8.6cm]{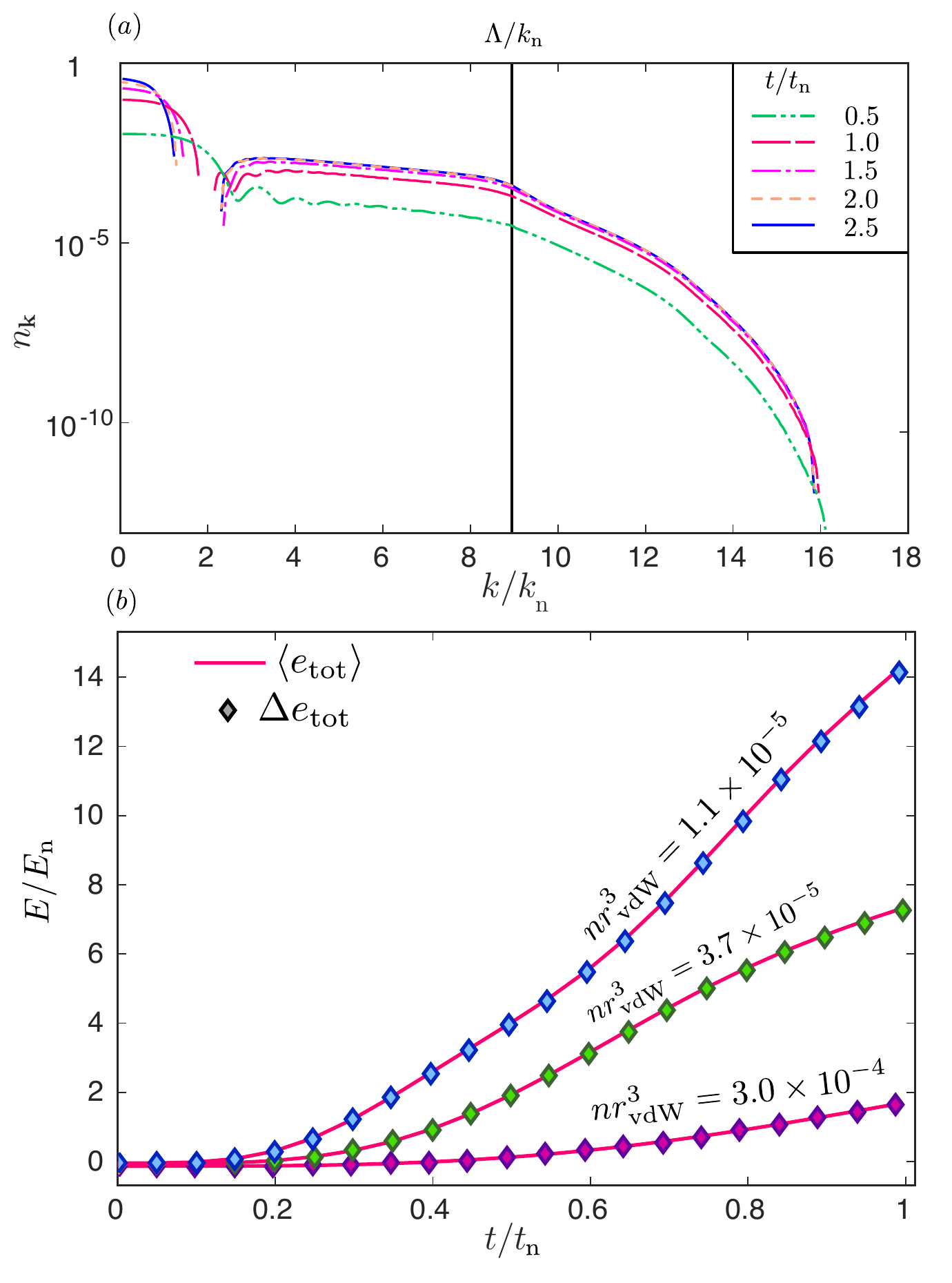}
\caption{(a) Time evolution of the momentum distribution for $nr_\mathrm{vdW}^3=6.9\times10^{-6}$ up to time $t=2.5t_\mathrm{n}$.   The cutoff scale $\Lambda/k_\mathrm{n}\approx9$ is indicated by the solid vertical line, and by $k/\Lambda= 2$ the momentum distribution remains vanishingly small for all times, demonstrating convergence with respect to the numerical cutoff $k_\mathrm{max}$.  The development and subsequent growth of the regime of negative $n_k$ near $k/k_\mathrm{n}$ can be seen for $t/t_\mathrm{n}\gtrsim 1$.  (b) Dynamics of $\Delta e_\mathrm{tot}$ (diamonds) versus $\langle e_\mathrm{tot}\rangle$ (solid red) for three different densities.  The simulation parameters used to produce the data in (a) and (b) are $\mathrm{nc}=150$, $\mathrm{nk}=5k_\mathrm{max}/n^{1/3}$, $k_\mathrm{max}=2\Lambda$, $\Delta t=m/2\hbar k_\mathrm{max}^2$, following the convergence guidelines in Sec.~\ref{app:convergence}.}\label{fig:nk_etot_conv}
\end{figure}

The natural question then is what to choose for $k_\mathrm{max}$ in the triplet simulation given that the processes described by $R$ and $M$ can involve single-particle momenta larger than $\Lambda$.  For example, consider the process described by $M_{\bf k,q}$ where the incoming excitation with momentum ${\bf k}$ decays into two excitations.  The form factor for the incoming scattering will be of the form $\zeta_{\bf k}=\theta(\Lambda-|{\bf k}|/2)$, which is restricted to momentum ${\bf k}\leq 2\Lambda$.  Conversely, this applies to the outgoing excitation of the process described by $M_{\bf k,q}^*$.  The population of single-particle modes in the triplet model is described by $\dot{n}_{\bf k}$.  Inspection of Eq.~\eqref{eq:n_sep} reveals the form factors $\zeta_{\bf l}$ and $\zeta_{\bf k}$, in the second and third terms, respectively, which act to restrict scattering into single-particle modes beyond $2\Lambda$.   In Fig.~\ref{fig:nk_etot_conv}(a), we show how $n_k$ remains nonzero for $1<k/\Lambda<2$ in the triplet model and confirm that $n_{\bf k}$ is numerically zero by $k=2\Lambda$.  We note that the choice of $k_\mathrm{max}$ also has consequences for the spectrum of bound states in the simulation because of the ultraviolet sensitivity of the three-body parameter discussed in Sec.~\ref{sec:threebody}.  The choice $k_\mathrm{max}=2\Lambda$ of the numerical cutoff was used to produce the results of Sec.~\ref{sec:triplet}, which also matches the expected frequency of the ground Efimov trimer in vacuum, serving as an additional consistency check.  

\begin{figure}[t!]
\centering
\includegraphics[width=8.6cm]{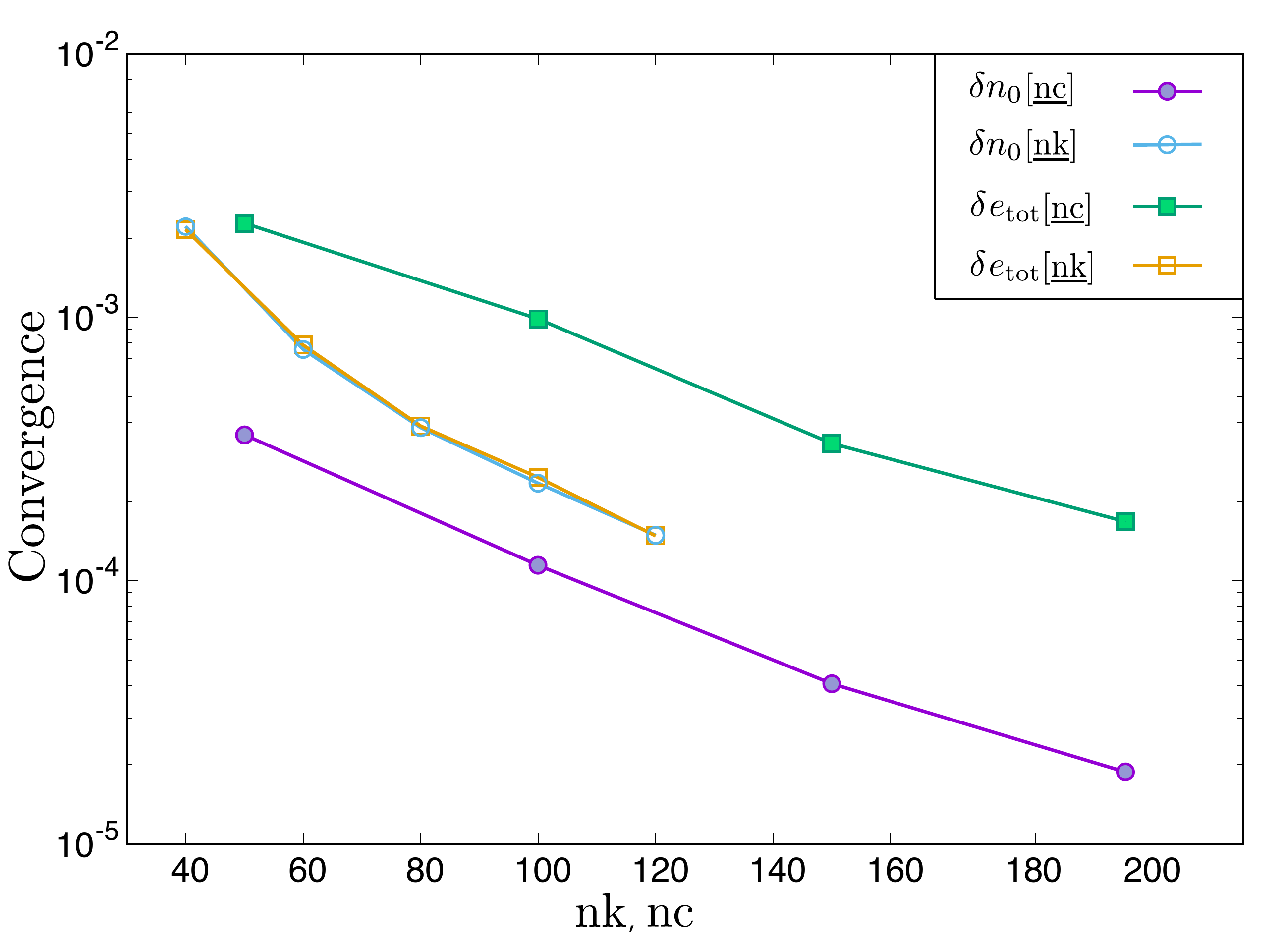}
\caption{Percent variation of the condensate fraction $\delta n_0$ (Eq.~\eqref{eq:delta_nzero}) and total energy $\delta e_\mathrm{tot}$ (Eq.~\eqref{eq:delta_E_tot}) as a function of momentum space and angular grid vectors $\underline{\mathrm{nc}}=\{25,50,100,150,200\}$  and $\underline{\mathrm{nk}}=\{j\times (k_\mathrm{max}/n^{1/3})\}_{j=1}^{6}$, respectively.  The simulation parameters that remained fixed to produce this data at density $nr_\mathrm{vdW}^3=3.0\times10^{-4}$ and time $t=t_\mathrm{n}$ are $k_\mathrm{max}=2\Lambda$ and $\Delta t=m/2\hbar k_\mathrm{max}^2$, following the convergence guidelines in Sec.~\ref{app:convergence}.}\label{fig:grid_conv}
\end{figure} 
As a general rule, the chosen simulation time step, $\Delta t$, must be at least as fast as the frequency set by the largest energy in the simulation.  In practice, we choose $\Delta t=m/2\hbar k_\mathrm{max}^2$ when the cutoff of the single-particle momentum sets the largest frequency in the simulation, which is generally the case.  The simulation is then run up to $t\sim t_\mathrm{n}$, beyond which the positivity of $n_k$ becomes violated typically for momentum in the vicinity of $k/k_\mathrm{n}\sim 2k_\mathrm{n}$ (see Fig.~\ref{fig:nk_etot_conv}(a)), which becomes a persistent feature at later times.  Because this violation is not physical, in Sec.~\ref{sec:triplet} we have restricted our analysis to results before this behavior occurs.  

This violation is a symptom of the non-conservation of the total energy, which is inherent in the triplet cumulant theory as discussed in Sec.~\ref{sec:conslaws}.  Analytically, one can predict the extent to which the total energy $E_\mathrm{tot}=\langle \hat{H}\rangle$ will change by calculating its time derivative from the restricted source term (Eq.~\eqref{eq:MH4_sep})
\begin{equation}\label{eq:dE}
\hbar\frac{dE_\mathrm{tot}}{dt}=\frac{2g^2}{V^{3/2}}\Im\left[\psi_0^*\sum_{\bf q,k,l} \zeta_{2{\bf q}-{\bf k}}\zeta_{2{\bf l}-{\bf k}}(1+n_{\bf q})M^*_{\bf k,l}\right].
\end{equation} 
In Fig.~\ref{fig:nk_etot_conv}(b), we compare the total energy per particle $\langle e_\mathrm{tot}\rangle=E_\mathrm{tot}/N$ with the quantity, $\Delta e_\mathrm{tot}=\int dt\ (d\langle e_\mathrm{tot}\rangle/dt)$, which is the result of simulating Eq.~\eqref{eq:dE} as an independent equation of motion supplied with an initial condition $\Delta e_\mathrm{tot}(t=0)=\langle e_\mathrm{tot}(t=0)\rangle$.  The excellent agreement between $\langle e_\mathrm{tot}\rangle$ and $\Delta e_\mathrm{tot}$ indicates that the observed violation of the total energy is inherent in the theory and {\it not} due to technical issues within the simulation itself.  Although not shown, we find in general that the total number is conserved at all times within the triplet simulation as expected.  

We now discuss the choice of the number of momentum and angular grid points and consequences for the convergence of simulation results.  To study the convergence, we track the total energy per particle and the condensate fraction as a function of $\mathrm{nk}$ and $\mathrm{nc}$ at the latest times $t\sim t_\mathrm{n}$ analyzed in Sec.~\ref{sec:triplet}.  We fix $nr_\mathrm{vdW}^3 = 3.0\times10^{-4}$, $k_\mathrm{max}=2\Lambda$ and $\Delta t=m/2\hbar k_\mathrm{max}^2$ and define the normalized variation of the slope
\begin{align}
&\delta e_\mathrm{tot}[x_i,x_{i-1}]=\left|\frac{1-\langle e_\mathrm{tot}\rangle[x_{i-1}]/\langle e_\mathrm{tot}\rangle[x_i]}{x_i-x_{i-1}}\right|,\label{eq:delta_E_tot}\\
&\delta n_0[x_i,x_{i-1}]=\left|\frac{1-n_0[x_{i-1}]/n_0[x_i]}{x_i-x_{i-1}}\right|,\label{eq:delta_nzero}
\end{align}
in terms of a vector of grid parameters $\underline{x}=\{x_1\dots x_\mathrm{f}\}$ as our measure of convergence.  In Fig.~\ref{fig:grid_conv}, we have evaluated $\delta e_\mathrm{tot}$ and $\delta n_0$ using to different grid vectors 
$\underline{\mathrm{nc}}=\{25,50,100,150,200\}$ and $\underline{\mathrm{nk}}=\{j\times (k_\mathrm{max}/n^{1/3})\}_{j=1}^{6}$ rounded to the nearest integer.  When evaluating convergence with respect to $\underline{\mathrm{nc}}$, we fix $\mathrm{nk}=5k_\mathrm{max}/n^{1/3}$, and when evaluating with respect to $\underline{\mathrm{nk}}$, we fix $\mathrm{nc}=150$.  We find that the simulation results for $\langle e_\mathrm{tot}\rangle$ and $n_0$ are converged to the level of a percent or less when the normalized slope variations measured by $\delta e_\mathrm{tot}$ and $\delta n_0$, respectively, are on the order of $10^{-4}$.  Therefore, we have taken $\mathrm{nc}=150$ and $\mathrm{nk}=5k_\mathrm{max}/n^{1/3}$ as the standard grid parameters used to produce the data of Sec.~\ref{sec:triplet}.  Although we have not discussed the convergence of the HFB simulation, we follow the same guidelines for the grid parameters.

Finally, we note that the HFB and triplet simulations were run on an NVIDIA Tesla P100 GPGPU card which has 16GB of memory and 3584 cores.  Although GPGPUs greatly speed up point-wise arithmetic in the simulation, the limited amount of memory means that triplet simulations cannot be taken to large values of the numerical cutoff (beyond $k_\mathrm{max}/n^{1/3}\gtrsim70$ in our case) while simultaneously fixing $\mathrm{nc}=150$ and $\mathrm{nk}=5k_\mathrm{max}/n^{1/3}$ in order to achieve convergence.  This hardware restriction places a practical limit on the range of results presented in this work.  Additionally, we note that the calculation time for the numerical implementation scheme described in this section scales roughly as $\sim \Lambda^4$.  In practice, the triplet simulations can be taken to larger values of $k_\mathrm{max}$ on workstations with a large number of CPUs and memory, although the slowdown compared to a GPGPU becomes significant.

\section{Quadruplet Cumulants}\label{app:quad}
To numerically simulate each of the quadruplets requires storing a six-dimensional complex array, which requires an enormous computational capacity and is beyond the present work.  Motivations of completeness aside, it is illustrative to discuss the explicit equations of motion for the quadruplets and to discuss their structure and formal solution in the early-time limit as we do in this section.  
\subsection{Equations of motion}
Here, we give explicit expressions of the cubic $\meanv{[\hat a \hat b \hat c \hat d,\hat H_3]}$ and quartic $\meanv{[\hat a \hat b \hat c \hat d,\hat H_4^{\rm eff}]}$ contributions to the equations of motion for the quadruplets Eqs.~\eqref{Q}, \eqref{P}, \eqref{T}. 
The contributions of the cubic hamiltonian $\hat H_3$ are contractions of $5$-body operators, hence products of doublets and triplets:
\begin{widetext}
\begin{multline}
\frac{\mathcal{Q}^{H_3}_{\alpha,\beta;\gamma}}{\sqrt{n_0/V}}=\mathcal{A}_{\{(\alpha,\beta),(\gamma,\delta)\}}\mathcal{S}_{\{\alpha,\beta\}}{\mathcal{S}_{\{\gamma,\delta\}}}\bbcrol{\vphantom{\frac{1}{2}}\frac{V_\alpha}{2}(1+n_\alpha+n_\beta)M_{\gamma+\delta,\gamma}+{\frac{V_\alpha+V_{\alpha+\beta}}{2} c_\alpha M_{\gamma+\delta,\gamma}}} \\\bbcror{+\frac{V_\beta+V_{\alpha+\beta}}{2}c_\alpha R_{\gamma,-\delta}^*+M_{\delta,\beta}^*\bbaco{(V_\gamma+V_{\alpha-\gamma})(n_\gamma-n_\alpha)+c_\gamma^*(V_\alpha+V_\gamma)}\vphantom{\frac{1}{2}}}, \label{F5Q}
\end{multline}
\begin{multline}
\frac{\mathcal{P}^{H_3}_{\alpha,\beta,\gamma}}{\sqrt{n_0/V}}=\mathcal{S}_{\{\alpha,\beta,\gamma\}}\bbcrol{\vphantom{\frac{1}{2}} V_\beta (1+n_\beta+n_\gamma) M_{\delta',\alpha}^*+\frac{V_\alpha+V_{\alpha-\delta'}}{2}(n_{\delta'}-n_\alpha)M_{\beta+\gamma,\beta}^*+\frac{V_{\delta'}+V_{\alpha-\delta'}}{2}(n_{\delta'}-n_\alpha)R_{\beta,-\gamma}}\\
\bbcror{+\frac{V_\alpha+V_{\delta'}}{2}(c_{\delta'}^* R_{\beta,-\gamma} -c_\alpha M_{\beta+\gamma,\beta}^* )+{(V_\alpha +V_{\alpha+\beta})c_\alpha M_{\delta',\gamma}^*+(V_\beta +V_{\alpha+\beta})c_\alpha M_{\gamma,\delta'}}\vphantom{\frac{1}{2}} } \label{F5P},
\end{multline}
\begin{multline}
\frac{\mathcal{T}^{H_3}_{\alpha,\beta,\gamma}}{\sqrt{n_0/V}}=\frac{\mathcal{S}_{\{\alpha,\beta,\gamma,\delta''\}}}{2}\bbcro{\vphantom{\frac{1}{2}} (V_\alpha+V_{\alpha+\beta}) c_\alpha R_{\gamma,-\delta''}+(V_\alpha+V_{\beta}) \bb{\frac{1}{2}+n_\alpha} R_{\gamma,-\delta''}+(V_\beta+V_{\alpha+\beta}) c_\alpha M_{\gamma+\delta'',\gamma}^*} \label{F5T}.
\end{multline}
\end{widetext}
The contributions of the quartic hamiltonian $\hat H_4^{\rm eff}$
are the most difficult. Since they are contractions of $6$-body operators,
they contain $(i)$ products of (2 or 3) doublets, $(ii)$ products of 2 triplets, and $(iii)$ quadruplets eventually multiplied by a doublet.
Separating those three contributions, we have
\begin{widetext}
\bea
\mathcal{Q}^{H_4,\rm doublets}_{\alpha,\beta;\gamma}&=&\frac{1}{V}\mathcal{A}_{\{(\alpha,\beta),(\gamma,\delta)\}}\mathcal{S}_{\{\alpha,\beta\}}\mathcal{S}_{\{\gamma,\delta\}}\bbcro{{V_{\gamma-\alpha}}\bb{\frac{1}{2}+n_\beta} n_\gamma n_\delta+(V_{\alpha+\beta}+V_{\alpha-\delta})c_\alpha c_\delta^*n_\gamma} \label{F6dQ}, \\
\mathcal{P}^{H_4,\rm doublets}_{\alpha,\beta,\gamma}&=&\frac{1}{V}\mathcal{S}_{\{\alpha,\beta,\gamma\}}\bbcro{V_{\alpha+\gamma}(1+n_\alpha+n_\beta)c_\gamma n_{\delta'}+V_{\alpha+\beta}c_\alpha c_\gamma c_{\delta'}^*-V_{\alpha+\beta}c_\alpha n_\beta n_\gamma} \label{F6dP},\\
\mathcal{T}^{H_4,\rm doublets}_{\alpha,\beta,\gamma} &=&\frac{1}{V}\mathcal{S}_{\{\alpha,\beta,\gamma,\delta''\}}\bbcro{V_{\alpha+\delta''} \bb{\frac{1}{2}+n_\beta} c_\gamma c_{\delta''} } \label{F6dT},
\eea
\end{widetext}
\begin{widetext}
\bea
\notag \mathcal{Q}^{H_4,\rm triplets}_{\alpha,\beta;\gamma}&=&\frac{1}{V}\mathcal{A}_{\{(\alpha,\beta),(\gamma,\delta)\}}\mathcal{S}_{\{\alpha,\beta\}}\mathcal{S}_{\{\gamma,\delta\}}\sum_\qq\bbcrol{\frac{R_{\gamma,-\delta}^*}{2} V_\qq R_{\alpha+\qq,-\beta} +\frac{M_{\gamma+\delta,\gamma}}{2}{ (V_\beta+V_{\qq-\alpha})M_{\beta+\qq,\beta}^*}} \\
&&\bbcror{+M_{\beta,\gamma} V_\qq M_{\delta,\alpha+\qq}^*+M_{\gamma,\beta}^* (V_\delta+V_{\qq+\delta-\alpha})M_{\delta+\qq,\delta}\vphantom{\frac{1}{2}}} \label{F6tQ}, \\
\notag \mathcal{P}^{H_4,\rm triplets}_{\alpha,\beta,\gamma}&=& \frac{1}{V}\mathcal{S}_{\{\alpha,\beta,\gamma\}}\sum_\qq \bbcrol{\frac{R_{\beta,-\gamma}}{2} (V_{\delta'}+V_{\qq+\beta+\gamma})M_{\delta'+\qq,\delta'}+M_{\delta',\beta}^*(V_\alpha+V_{\qq-\gamma})M_{\alpha+\qq,\alpha}^*+M_{\beta+\gamma,\beta}^*\frac{V_\qq}{2}M_{\delta',\alpha-\qq}^*}\\
&&\bbcror{+M_{\beta,\delta'} V_\qq R_{\alpha,-\gamma-\qq}-R_{\beta,-\gamma}\frac{V_\qq}{2} M_{\alpha,\delta'-\qq}-M_{\beta+\gamma,\beta}^*\frac{V_\alpha+V_{\qq-\beta-\gamma}}{2}M_{\alpha+\qq,\alpha}^*\vphantom{\frac{1}{2}}} \label{F6tP},\\
\mathcal{T}^{H_4,\rm triplets}_{\alpha,\beta,\gamma}&=&\frac{1}{2V}\mathcal{S}_{\{\alpha,\beta,\gamma,\delta''\}}\sum_\qq \bbcro{M_{\gamma+\delta'',\gamma}^* V_\qq R_{\alpha+\qq,-\beta}+R_{\gamma,-\delta''} (V_\beta+V_{\qq-\alpha}) M_{\beta+\qq,\beta}^*} \label{F6tT},
\eea
\end{widetext}
\begin{widetext}
\bea
\notag \mathcal{Q}^{H_4,\rm quadruplets}_{\alpha,\beta,\gamma}&=&\frac{1}{V}\mathcal{A}_{\{(\alpha,\beta),(\gamma,\delta)\}}\mathcal{S}_{\{\alpha,\beta\}}\mathcal{S}_{\{\gamma,\delta\}}\sum_\qq \bbcrol{V_\qq \frac{1+n_\alpha+n_\beta}{4}Q_{\alpha+\qq,\beta-\qq;\gamma}+(V_{\alpha-\gamma}+V_{\alpha-\qq})\frac{n_\gamma-n_\alpha}{2}Q_{\beta,\qq;\delta}}\\
&&\bbcror{+V_\qq c_\gamma^* P_{\alpha+\qq,\beta,-\gamma-\qq} + \frac{V_{\alpha+\beta}+V_{\qq+\beta}}{2}c_\beta P^*_{\gamma,\delta,\qq}\vphantom{\frac{1}{2}}} \label{F6qQ},\\
\notag \mathcal{P}^{H_4,\rm quadruplets}_{\alpha,\beta,\gamma}&=&\frac{1}{2V}\mathcal{S}_{\{\alpha,\beta,\gamma\}}\sum_\qq \bbcrol{V_\qq (1+n_\alpha+n_\beta)P_{\alpha+\qq,\beta-\qq,\gamma}+(V_{\alpha-\delta'}+V_{\alpha-\qq})(n_{\delta'}-n_\alpha)P_{\beta,\gamma,\qq}}\\
&&\bbcror{+V_\qq {c_{\delta'}}^* T_{\alpha+\qq,\beta,\gamma}+ 2(V_{\alpha+\beta}+V_{\qq-\beta})c_\alpha Q_{\gamma,\qq;\delta'}-V_\qq {c_\alpha}Q_{\beta,\gamma;\delta'-\qq}\vphantom{\frac{1}{2}}}  \label{F6qP},\\
\mathcal{T}^{H_4,\rm quadruplets}_{\alpha,\beta,\gamma}&=&\frac{1}{2V}\mathcal{S}_{\{\alpha,\beta,\gamma,\delta''\}}\sum_\qq \bbcro{V_\qq \bb{n_\beta+\frac{1}{2}}T_{\alpha+\qq,\gamma,\delta''}+(V_{\gamma+\delta''}+V_{\alpha-\qq})c_\beta P_{\qq,\gamma,\delta''}}  \label{F6qT}.
\eea
\end{widetext}
Here the replacements of $\delta\to\alpha+\beta-\gamma$,
$\delta'\to\alpha+\beta+\gamma$ and $\delta''\to-\alpha-\beta-\gamma$ should be done after acting with the symmetrizer and antisymetrizer.
\subsection{Solution}
The length expressions in Eqs.~\eqref{F6dQ}--\eqref{F6qT} hide the underlying structure of the quadruplet equations of motion as  coupled few-body Schr\"odinger equations with nonlinear and drive terms.  In this section, we follow Sec.~\ref{sec:formal} and reduce these equations to their early-time form to make this structure explicit and to illustrate how one, two, three, and four-body physics are encoded in the formal solutions.

First, we begin by reduced the equation of motion for the $Q$ cumulant (Eq.~\eqref{Q}) to the early-time form
\begin{widetext}
\begin{align}\label{eq:qdot}
i\hbar\partial_t|Q_t,Q_t\rangle\langle Q_t,Q_t|&=\hat{H}_{12}(t)|Q_t,Q_t\rangle\langle Q_t,Q_t|-|Q_t,Q_t\rangle\langle Q_t,Q_t|\hat{H}_{12}(t)\nonumber\\
&+(1+\hat{P}_{12})\hat{V}\left[|\psi_t,M_t\rangle\langle M_t,M_t|+|n_{1,t},n_{2,t}\rangle\langle n_{2,t},n_{1,t}|\right]\nonumber\\
&-\left[|M_t,M_t\rangle\langle M_t,\psi_{0,t}|+|n_{1,t},n_{2,t}\rangle\langle n_{2,t},n_{1,t}|\right]\hat{V}(1+\hat{P}_{12}),
\end{align}
\end{widetext}
where we have defined the rank (2,2) tensors $\langle \alpha,\beta |Q_t,Q_t\rangle\langle Q_t,Q_t|\gamma,\delta\rangle=VQ_{\alpha\beta;\gamma}(t)\delta_{\alpha+\beta,-\delta-\gamma}$ and 
\begin{equation}
\langle \alpha,\beta|n_{1,t},n_{2,t}\rangle\langle n_{2,t},n_{1,t}|\delta,\gamma\rangle=n_\alpha(t)n_\beta(t)\delta_{\alpha,\gamma}\delta_{\beta,\delta}.
\end{equation}
Equation~\eqref{eq:qdot} can be solved formally as
\begin{widetext}
\begin{align}
|Q_t,Q_t\rangle\langle Q_t,Q_t|=&\hat{\mathcal{U}}_{12}(t-t_0)|Q_{t_0},Q_{t_0}\rangle\langle Q_{t_0},Q_{t_0}|\hat{\mathcal{U}}_{12}(t_0-t)\nonumber\\
&+\frac{1}{i\hbar}\int_{t_0}^t d\tau \hat{\mathcal{U}}_{12}(t-\tau)\hat{V}(1+\hat{P}_{12})\left[|\psi_{0,\tau},M_\tau\rangle\langle M_\tau,M_\tau|+|n_{1,\tau},n_{2,\tau}\rangle\langle n_{2,\tau},n_{1,\tau}|\right](\hat{\mathcal{U}}_{12}(\tau-t)\nonumber\\
&-\frac{1}{i\hbar}\int_{t_0}^t d\tau \hat{\mathcal{U}}_{12}(t-\tau)\left[|M_\tau,M_\tau\rangle\langle \psi_{0,\tau},M_\tau|+|n_{1,\tau},n_{2,\tau}\rangle\langle n_{2,\tau},n_{1,\tau}|\right](1+\hat{P}_{12})\hat{V}\hat{\mathcal{U}}_{12}(\tau-t).\label{eq:qsol}
\end{align}
\end{widetext}
Here, we see that the $|n_{1,\tau},n_{2,\tau}\rangle\langle n_{2,\tau},n_{1,\tau}|$ parts of the memory kernel describes forward and backward Boltzmannian scattering in a classical dilute gas \cite{huang1987statistical}.  Including this contribution of $Q$ in $\dot{n}_k$ it is possible to retrieve the Boltzmann equation describing two-body scattering as the level of the $T$-matrix (c.f. Ref.~\cite{KREMP1997320}).    

Next, we reduce the equations of motion for the $P$ cumulant (Eq.~\eqref{P}) to the early time form
\begin{widetext}
\begin{align}\label{eq:psol}
i\hbar\partial_t |P_t,P_t,P_t\rangle\langle P_t|=&\hat{H}_{123}(t)|P_t,P_t,P_t\rangle\langle P_t|-|P_t,P_t,P_t\rangle\langle P_t|\hat{H}_1(t)\nonumber\\
&+(1+\hat{P}_++\hat{P}_-)(\hat{V}_{12}+\hat{V}_{13})\left[|n_t,c_t,c_t\rangle\langle n_t|+|\psi_{0,t},M_t,M_t\rangle\langle M_t|\right],
\end{align}
\end{widetext} 
where we have defined the rank (1,3) tensor $\langle \alpha,\beta,\gamma|P_t,P_t,P_t\rangle\langle P_t|\delta\rangle=VP_{\alpha,\beta,\gamma}(t)\delta_{\alpha+\beta+\gamma,\delta}$.  Equation~\eqref{eq:psol} can be formally solved as
\begin{widetext}
\begin{align}\label{eq:psol1}
 |P_t,P_t,P_t\rangle\langle P_t|=&\hat{\mathcal{U}}_\mathrm{123}(t-t_0) |P_{t_0},P_{t_0},P_{t_0}\rangle\langle P_{t_0}|\hat{\mathcal{U}}_1(t_0-t)\nonumber\\
 &+\frac{1}{i\hbar}\int_{t_0}^t d\tau \hat{\mathcal{U}}_\mathrm{123}(t-\tau)(1+\hat{P}_++\hat{P}_-)(\hat{V}_{12}+\hat{V}_{13})\left[|n_\tau,c_\tau,c_\tau\rangle\langle n_\tau|+|\psi_{0,\tau},M_\tau,M_\tau\rangle\langle M_\tau|\right]\hat{\mathcal{U}}_1(\tau-t).
\end{align}
\end{widetext}
Finally, the $T$ cumulant equation of motion reduces to the early-time form
\begin{widetext}
\begin{align}
i\hbar\partial_t |T_t,T_t,T_t,T_t\rangle=&\hat{H}_\mathrm{1234}(t)|T_t,T_t,T_t,T_t\rangle\nonumber\\
&+(1+\hat{P}_{1234}+\hat{P}_{1324}+\hat{P}_{1423})\left(\hat{V}_{12}+\hat{V}_{13}+\hat{V}_{14}\right)|\psi_{0,t},R_t,R_t,R_t\rangle\nonumber\\
&+\left(1+\hat{P}_{234}+\hat{P}_{243}\right)\left(\hat{V}_{13}+\hat{V}_{14}+\hat{V}_{23}+\hat{V}_{24}\right)|c_{1,t},c_{1,t},c_{2,t},c_{2,t}\rangle\label{eq:tsol},
\end{align}
\end{widetext}
with $\hat{P}_{1234}|\alpha,\beta,\gamma,\delta\rangle=|\delta,\alpha,\beta,\gamma\rangle$, and where we have defined the range (0,4) tensor $\langle \alpha,\beta,\gamma,\delta|T_t,T_t,T_t,T_t\rangle=VT_{\alpha,\beta,\gamma}(t)\delta_{\alpha+\beta,-\gamma-\delta}$ and $\hat{H}_{1234}=\sum^4_{i<j}\hat{H}_{ij}(t)$ as the vacuum four-body Hamiltonian in the rotating frame of the condensate.  Equation~\eqref{eq:tsol} can be formally solved as
\begin{widetext}
\begin{align}
|T_t,T_t,T_t,T_t\rangle=&\hat{\mathcal{U}}_{1234}(t-t_0)|T_{t_0},T_{t_0},T_{t_0},T_{t_0}\rangle\nonumber\\
&+\frac{1}{i\hbar}\int_{t_0}^t d\tau\hat{\mathcal{U}}_{1234}(t-\tau)(1+\hat{P}_{1234}+\hat{P}_{1324}+\hat{P}_{1423})\left(\hat{V}_{12}+\hat{V}_{13}+\hat{V}_{14}\right)|\psi_{0,\tau},R_\tau,R_\tau,R_\tau\rangle\nonumber\\
&+\frac{1}{i\hbar}\int_{t_0}^t d\tau\hat{\mathcal{U}}_{1234}(t-\tau)\left(1+\hat{P}_{234}+\hat{P}_{243}\right)\left(\hat{V}_{13}+\hat{V}_{14}+\hat{V}_{23}+\hat{V}_{24}\right)|c_{1,\tau},c_{1,\tau},c_{2,\tau},c_{2,\tau}\rangle\label{eq:tsol1},
\end{align}
\end{widetext}
where $\hat{\mathcal{U}}_\mathrm{1234}(t)=\exp\left[-i\int_{t_0}^td\tau\hat{H}_{1234}(\tau)/\hbar\right]$ is the four-body evolution operator in the rotating frame of the condensate.  Analogous the connection between Eq.~\eqref{eq:rsol} and the Faddeev equations (see Sec.~\ref{sec:formal} and Refs.~\cite{kohler2002,Kokkelmans2018}), Eq.~\eqref{eq:tsol}) yields generalized four-body $T$-matrices satisfying the Yakubovsky equations \cite{faddeev2013quantum}.  To include the physics of the four-body bound states tied to Efimov states \cite{PhysRevA.97.033621,BRAATEN2006259,Naidon_2017,D_Incao_2018} in the cumulant theory of the unitary Bose gas, the hierarchy must be taken then to at least the quadruplet level.

\section{Relation To Alternative Approaches}\label{app:connect}
To construct the cumulant theory used in this work, we make two main approximations.  First, we describe Bose-condensation in the $U(1)$-symmetry-breaking approach, and second, we truncate the cumulant hierarchy to consider cumulants only up to some finite order.  In this section, we first connect with the number-conserving description of Bose-condensation \cite{PhysRevA.57.3008,PhysRevA.56.1414} in Sec.~\ref{app:NC}.  Second, we group many equivalent models of the quenched unitary Bose gas found in the literature \cite{PhysRevA.100.013612,Kokkelmans2018,PhysRevA.89.021601,PhysRevA.91.013616,PhysRevLett.124.040403,PhysRevA.99.023623} under the umbrella of the doublet model presented in the present work in Sec.~\ref{app:nsj}.  Already in Sec.~\ref{sec:eom}, the connection between the HFB theory and the doublet model was established.  We note that the Popov and bath theories of Refs.~\cite{PhysRevA.88.063611,PhysRevA.93.033653,PhysRevA.90.021602} both set the effective interaction strength $g$ in an ad hoc fashion and ignored the $c$-cumulant dynamics.  These works are not consistent with the unitarity limit of the $s$-wave cross section $\sigma\propto1/k^2$ and therefore do not describe resonant scattering processes.  We remark that the Hyperbolic Bloch equations, derived in Ref.~\cite{KIRA2015185} (see Eqs.~\eqref{n} and \eqref{c}), have not been simulated to date as they require handling of the resource-intensive quadruplets.  The triplet model studied in Sec.~\ref{sec:triplet} represents the state-of-the-art in this regard.  Although these are mostly formal remarks, making distinctions and connections between approaches is instructive both for understanding the limitations of the present work and for uniting equivalent lines of research on the quenched unitary Bose gas.  

\subsection{Number-conserving approach}
\label{app:NC}

In the number conserving approach, one performs a {quantum} modulus-phase decomposition of the condensate operator $\hat{a}_0$:
\be
\hat{a}_0=\eee^{\ii\hat{\theta}_0}\sqrt{\hat{N}_0}.
\ee
The phase $\hat{\theta}_0$ and population ${\hat{N}_0}$ of the condensate are canonically conjugated
\be
\left[\hat{\theta}_0,\hat{N}_0\right] =-\ii.
\ee
They inherit this relation from the bosonic nature of $\hat{a}_0$. Note that this phase-modulus decomposition is possible only in the approximation that the condensate is never empty.

We then introduce the excitation field for $\kk\neq0$
\be
\hat\Lambda_\kk=\eee^{-\ii\hat{\theta}_0} \hat{a}_\kk.
\ee
Conceptually, the advantage of using $\hat\Lambda_\kk$ rather than $\hat{b}_\kk$ as in the main text is that $\hat\Lambda_\kk$ conserves the number of particles (it transfers one particle from the non-condensed fraction to the condensate). Thus, one can still have nonzero anomalous averages $\meanv{\hat\Lambda \hat\Lambda}\neq0$ even in states with a fixed number of particles. In terms of $\hat\Lambda$, $\hat{\Lambda}^\dagger$ and $\hat{N}_0$, the Hamiltonian reads \footnote{To avoid restrictions on the summations, we use the convention $\hat \Lambda_\zero=0$.}
\begin{widetext}
\begin{multline}
\hat{H}^{\rm (NC)}=\frac{V_\zero N^2}{2V} + \sum_{\kk} \bb{\bbcro{\epsilon_\kk+\frac{V_\kk\hat N_0}{V}}\hat\Lambda_\kk^\dagger\hat\Lambda_\kk + \frac{V_\kk\sqrt{\hat N_0(\hat N_0-1)}}{2V}[\hat\Lambda_{-\kk}\hat\Lambda_\kk +\hat\Lambda_\kk^\dagger\hat\Lambda_{-\kk}^\dagger]} \\
+\frac{\sqrt{\hat N_0}}{V}\sum_{\kk,\qq} V_\qq \bb{\hat\Lambda_{\kk+\qq}^\dagger\hat\Lambda_{\kk}\hat\Lambda_{\qq}+\mbox{h.c.}} + \frac{1}{2V}\sum_{\kk,\kk',\qq} V_\qq \Lambda_{\kk'+\qq}^\dagger \hat{\Lambda}_{\kk-\qq}^\dagger \hat{\Lambda}_\kk \hat{\Lambda}_{\kk'}.
\label{HNC}
\end{multline}
\end{widetext}
To avoid cumbersome restrictions in the sums over $\kk, \kk'$ and $\qq$, we set $\hat\Lambda_\zero=0$ by convention. Here, we collected the terms proportional to $V_\zero$ in the constant first term using conservation of the total number of particles $N=\hat N_0+\sum_{\kk}  \hat\Lambda_\kk^\dagger\hat\Lambda_\kk$. Although not done in the main text (Sec.~\ref{sec:hamiltonian}), we note that such simplification is also possible in the symmetry-breaking picture. In Eq.~\eqref{HNC}, we have kept the $O(1/\meanv{\hat N_0})$ corrections that come from the non-commutation of $\hat{\theta}_0$ and $\hat{N}_0$. In the thermodynamic limit, these corrections are negligible (as long as the condensate is macroscopically occupied).

We use the approach of Ref.~\cite{kira2014excitation} to identify all the terms that become negligible in the thermodynamic limit. We define $N_0(t)\equiv\meanv{\hat N_0}$ and write $\hat N_0=N_0(t)+\delta\hat N_0$, and similarly for the macroscopic sums of the non-condensed field, for example $\sum_{\kk}  \hat\Lambda_\kk^\dagger\hat\Lambda_\kk=\sum_{\kk}  \meanv{\hat\Lambda_\kk^\dagger\hat\Lambda_\kk}+\sum_{\kk}  \delta(\hat\Lambda_\kk^\dagger\hat\Lambda_\kk)$. The product of fluctuations is of order $O(1/\sqrt{N})$ smaller than the leading (non-scalar) terms in the Hamiltonian so it can be neglected. We then obtain
\begin{widetext}
\begin{multline}
\hat{H}^{\rm  (NC)}\simeq \frac{V_\zero Nn}{2}+ \sum_{\kk} \bb{\bbcro{\epsilon_\kk+V_\kk n_0(t)}\hat\Lambda_\kk^\dagger\hat\Lambda_\kk + \frac{V_\kk{ n_0(t)}}{2}[\hat\Lambda_{-\kk}\hat\Lambda_\kk +\hat\Lambda_\kk^\dagger\hat\Lambda_{-\kk}^\dagger]} \\
+\sqrt{\frac{n_0(t)}{V}}\sum_{\kk,\qq} V_\qq \bb{\hat\Lambda_{\kk+\qq}^\dagger\hat\Lambda_{\kk}\hat\Lambda_{\qq}+\mbox{h.c.}} + \frac{1}{2V}\sum_{\kk,\kk',\qq} V_\qq \Lambda_{\kk'+\qq}^\dagger \hat{\Lambda}_{\kk-\qq}^\dagger \hat{\Lambda}_\kk \hat{\Lambda}_{\kk'} - (\hat N_0 - N_0(t)) \meanvlr{\hbar\frac{\dd\hat\theta_0}{\dd t}}.
\label{HNCapprox}
\end{multline}
\end{widetext}
This Hamiltonian is the same as $\hat H_b$ (Eq.~\eqref{dtb}), the Hamiltonian found in the symmetry-breaking approach (up to the replacement $\hat b \to \hat \Lambda$, and the collection of the terms containing $V_\zero$ discussed above). To show the complete equivalence of the two theories, we calculate the phase derivative in the number-conserving approach from the commutator of $\hat N_0$ with the exact expression \eqref{HNC} of the Hamiltonian:
\begin{align}
\meanvlr{\hbar\frac{\dd\hat \theta_0}{\dd t}}=&-\frac{1}{V}\sum_\kk \bbcro{V_\kk \meanv{\hat \Lambda_\kk^\dagger \hat \Lambda_\kk} +\frac{V_\kk}{2} \bb{\meanv{\hat \Lambda_{-\kk} \hat \Lambda_\kk}+\mbox{cc.}}}\nonumber\\
&-\frac{1}{2\sqrt{n_0V^3}} \sum_{\kk\qq} V_\qq\bb{\meanv{\hat \Lambda_{\kk+\qq}^\dagger \hat \Lambda_\kk \hat \Lambda_\qq} +\mbox{cc.}},
\end{align}
which is the same as \eqref{thetapoint}, up to the constant rotation velocity $V_\zero n$.
\subsection{Nozi\`eres-Saint James approach}\label{app:nsj}
The time-dependent generalization of the Nozi\`eres-Saint James approach (NSJ) \cite{James1982,PhysRevA.89.021601,PhysRevA.91.013616,PhysRevA.99.023623,PhysRevLett.124.040403} is based on the variational ansatz for the ground state wave function
\begin{equation}
 |\Psi_\mathrm{NSJ}(t)\rangle = \frac{1}{\mathcal{N}}\,\mathrm{expt}\left[\sqrt{V}\alpha_0(t) \hat{a}^\dagger_0 + \frac{1}{2}\sum_{{\bf k}} \beta_{\bf k}(t) \hat{a}^\dagger_{\bf-k} \hat{a}^\dagger_{\bf k}\right] |0\rangle,
 \label{eq:Nozieres-SaintJames}
\end{equation}
with normalization factor $\mathcal{N}$, and variational parameters $\alpha_0(t)$ and $\beta_{\bf k}(t)$, and factor of $1/2$ in the summation to account for double counting of pairs $({\bf k,-k})$.  The NSJ variational parameters are connected to the single and doublet cumulants as $\psi_0=\alpha_0$, ${\tilde c}_{\bf k}=\beta_{\bf k}/(1 - |\beta_{\bf k}|^2)$, and $n_{\bf k}= |\beta_{\bf k}|^2/(1 - |\beta_{\bf k}|^2)$ with total number $N= |\alpha_0|^2 + \sum_{\bf k} |\beta_{\bf k}|^2/(1-|\beta_{\bf k}|^2)$.
The equation of motion for $\alpha_0(t)$ corresponding to the Hamiltonian (Eq.~\eqref{hamiltonianF}) is
\begin{align}\label{eq:alphadot}
 i\hbar \frac{d \alpha_0}{dt} = &  n V_{\bf 0} \alpha_0 +  \alpha_0\frac{1}{V} \sum_{\bf k} V_{\bf k}\frac{|\beta_{\bf k}|^2 }{1 - |\beta_{\bf k}|^2} \nonumber\\
 &+ \alpha_0^*\frac{1}{V} \sum_{\bf k} V_{\bf k} \frac{ \beta_{\bf k}}{1 - |\beta_{\bf k}|^2},
 \end{align}
 where the term by term equivalence with the GPE (Eq.~\eqref{GPE}) (for vanishing triplet contributions) is apparent.    The corresponding equation of motion for $\beta_{\bf k}(t)$ is 
\begin{align}  \label{eq:betadot}
  i\hbar \frac{d \beta_{\bf k}}{dt} =& 2(\epsilon_{\bf k} + V_{\bf 0}n )\beta_{\bf k} + V_{\bf k}\left(\alpha_0^2+ (\alpha_0^*)^2 \beta_{\bf k}^2 + 2 |\alpha_0|^2 \beta_{\bf k} \right)\nonumber\\
  &+\frac{1}{V} \sum_{\bf q} V_{\bf k-q}\frac{2|\beta_{\bf q}|^2 \beta_{\bf k}+\beta_{\bf q}+\beta_{\bf q}^*\beta_{\bf k}^2}{1- |\beta_{\bf q}|^2}.
\end{align}
To evaluate the equation of motion for the $c$-cumulant, we consider the corresponding expression in the NSJ approach 
\begin{widetext}
\begin{align}
  i\hbar \frac{d}{dt}\left( \frac{\beta_{\bf k}}{1 - |\beta_{\bf k}|^2} \right) =& i\hbar \frac{d \beta_{\bf k}}{dt} \frac{1}{(1 - |\beta_{\bf k}|^2)^2} - \left(-i \hbar \frac{d \beta_{\bf k}^*}{dt}\right)\frac{\beta_{\bf k}^2}{(1 - |\beta_{\bf k}|^2)^2}, \\  
=&   \left[2(\epsilon_{\bf k} + V_{\bf 0}n)c_{\bf k}+2\left(V_{\bf k}n_0+\frac{1}{V}\sum_{\bf q}V_{\bf k-q}n_{\bf q}\right)c_{\bf k}\right]\left[\frac{1}{1-|\beta_{\bf k}|^2}-\frac{|\beta_{\bf k}|^2}{1-|\beta_{\bf k}|^2}\right]\nonumber\\
&+\left[V_{\bf k}\psi_0^2+\frac{1}{V}\sum_{\bf q}V_{\bf k-q}c_{\bf q}\right]\left[\frac{1}{(1-|\beta_{\bf k}|^2)^2}-\frac{|\beta_{\bf k}|^4}{(1-|\beta_{\bf k}|^2)^2}\right],\label{eq:cdotnsj}
 \end{align}
 \end{widetext}
 where
 \begin{equation}
 \left[\frac{1}{(1-|\beta_{\bf k}|^2)^2}-\frac{|\beta_{\bf k}|^4}{(1-|\beta_{\bf k}|^2)^2}\right]=1+2n_{\bf k}.
 \end{equation}
From the relation $|c_{\bf k}|^2=n_{\bf k}(n_{\bf k}+1)$, which is clear from the definitions of ${\tilde c}_{\bf k}$ and $n_{\bf k}$ in terms of the variational parameters, the equation of motion for $\dot{n}$ follows immediately.  Setting finally $c_\kk=e^{-2i\theta_0}\tilde c_\kk$ to switch to the rotating frame of the condensate, we see then that the NSJ approach yields equations of motion that are identical to the doublet model (Eqs.~\eqref{GPE}, \eqref{n}, and \eqref{c}) considered in this work. Therefore, both the NSJ and HFB approaches are equivalent to each other as also suggested in Ref.~\cite{PhysRevA.91.013616} and all correspond a truncation of the cumulant hierarchy at the doublet level.  

\bibliographystyle{apsrev4-1}
\bibliography{HFB}

\begin{thebibliography}{94}%
\makeatletter
\providecommand \@ifxundefined [1]{%
 \@ifx{#1\undefined}
}%
\providecommand \@ifnum [1]{%
 \ifnum #1\expandafter \@firstoftwo
 \else \expandafter \@secondoftwo
 \fi
}%
\providecommand \@ifx [1]{%
 \ifx #1\expandafter \@firstoftwo
 \else \expandafter \@secondoftwo
 \fi
}%
\providecommand \natexlab [1]{#1}%
\providecommand \enquote  [1]{``#1''}%
\providecommand \bibnamefont  [1]{#1}%
\providecommand \bibfnamefont [1]{#1}%
\providecommand \citenamefont [1]{#1}%
\providecommand \href@noop [0]{\@secondoftwo}%
\providecommand \href [0]{\begingroup \@sanitize@url \@href}%
\providecommand \@href[1]{\@@startlink{#1}\@@href}%
\providecommand \@@href[1]{\endgroup#1\@@endlink}%
\providecommand \@sanitize@url [0]{\catcode `\\12\catcode `\$12\catcode
  `\&12\catcode `\#12\catcode `\^12\catcode `\_12\catcode `\%12\relax}%
\providecommand \@@startlink[1]{}%
\providecommand \@@endlink[0]{}%
\providecommand \url  [0]{\begingroup\@sanitize@url \@url }%
\providecommand \@url [1]{\endgroup\@href {#1}{\urlprefix }}%
\providecommand \urlprefix  [0]{URL }%
\providecommand \Eprint [0]{\href }%
\providecommand \doibase [0]{http://dx.doi.org/}%
\providecommand \selectlanguage [0]{\@gobble}%
\providecommand \bibinfo  [0]{\@secondoftwo}%
\providecommand \bibfield  [0]{\@secondoftwo}%
\providecommand \translation [1]{[#1]}%
\providecommand \BibitemOpen [0]{}%
\providecommand \bibitemStop [0]{}%
\providecommand \bibitemNoStop [0]{.\EOS\space}%
\providecommand \EOS [0]{\spacefactor3000\relax}%
\providecommand \BibitemShut  [1]{\csname bibitem#1\endcsname}%
\let\auto@bib@innerbib\@empty
\bibitem [{\citenamefont {Pitaevskii}\ and\ \citenamefont
  {Stringari}(2016)}]{pitaevskii2016bose}%
  \BibitemOpen
  \bibfield  {author} {\bibinfo {author} {\bibfnamefont {L.}~\bibnamefont
  {Pitaevskii}}\ and\ \bibinfo {author} {\bibfnamefont {S.}~\bibnamefont
  {Stringari}},\ }\href@noop {} {\emph {\bibinfo {title} {Bose-Einstein
  Condensation and Superfluidity}}},\ Vol.\ \bibinfo {volume} {164}\ (\bibinfo
  {publisher} {Oxford University Press},\ \bibinfo {year} {2016})\BibitemShut
  {NoStop}%
\bibitem [{\citenamefont {Greiner}\ \emph {et~al.}(2003)\citenamefont
  {Greiner}, \citenamefont {Regal},\ and\ \citenamefont {Jin}}]{Jin2003}%
  \BibitemOpen
  \bibfield  {author} {\bibinfo {author} {\bibfnamefont {M.}~\bibnamefont
  {Greiner}}, \bibinfo {author} {\bibfnamefont {C.~A.}\ \bibnamefont {Regal}},
  \ and\ \bibinfo {author} {\bibfnamefont {D.~S.}\ \bibnamefont {Jin}},\ }\href
  {http://dx.doi.org/10.1038/nature02199} {\bibfield  {journal} {\bibinfo
  {journal} {Nature}\ }\textbf {\bibinfo {volume} {426}},\ \bibinfo {pages}
  {537} (\bibinfo {year} {2003})}\BibitemShut {NoStop}%
\bibitem [{\citenamefont {Zwierlein}\ \emph {et~al.}(2003)\citenamefont
  {Zwierlein}, \citenamefont {Stan}, \citenamefont {Schunck}, \citenamefont
  {Raupach}, \citenamefont {Gupta}, \citenamefont {Hadzibabic},\ and\
  \citenamefont {Ketterle}}]{Ketterle2003}%
  \BibitemOpen
  \bibfield  {author} {\bibinfo {author} {\bibfnamefont {M.~W.}\ \bibnamefont
  {Zwierlein}}, \bibinfo {author} {\bibfnamefont {C.~A.}\ \bibnamefont {Stan}},
  \bibinfo {author} {\bibfnamefont {C.~H.}\ \bibnamefont {Schunck}}, \bibinfo
  {author} {\bibfnamefont {S.~M.~F.}\ \bibnamefont {Raupach}}, \bibinfo
  {author} {\bibfnamefont {S.}~\bibnamefont {Gupta}}, \bibinfo {author}
  {\bibfnamefont {Z.}~\bibnamefont {Hadzibabic}}, \ and\ \bibinfo {author}
  {\bibfnamefont {W.}~\bibnamefont {Ketterle}},\ }\href {\doibase
  10.1103/PhysRevLett.91.250401} {\bibfield  {journal} {\bibinfo  {journal}
  {Phys. Rev. Lett.}\ }\textbf {\bibinfo {volume} {91}},\ \bibinfo {pages}
  {250401} (\bibinfo {year} {2003})}\BibitemShut {NoStop}%
\bibitem [{\citenamefont {Jochim}\ \emph {et~al.}(2003)\citenamefont {Jochim},
  \citenamefont {Bartenstein}, \citenamefont {Altmeyer}, \citenamefont {Hendl},
  \citenamefont {Riedl}, \citenamefont {Chin}, \citenamefont
  {Hecker~Denschlag},\ and\ \citenamefont {Grimm}}]{Grimm2003}%
  \BibitemOpen
  \bibfield  {author} {\bibinfo {author} {\bibfnamefont {S.}~\bibnamefont
  {Jochim}}, \bibinfo {author} {\bibfnamefont {M.}~\bibnamefont {Bartenstein}},
  \bibinfo {author} {\bibfnamefont {A.}~\bibnamefont {Altmeyer}}, \bibinfo
  {author} {\bibfnamefont {G.}~\bibnamefont {Hendl}}, \bibinfo {author}
  {\bibfnamefont {S.}~\bibnamefont {Riedl}}, \bibinfo {author} {\bibfnamefont
  {C.}~\bibnamefont {Chin}}, \bibinfo {author} {\bibfnamefont {J.}~\bibnamefont
  {Hecker~Denschlag}}, \ and\ \bibinfo {author} {\bibfnamefont
  {R.}~\bibnamefont {Grimm}},\ }\href {\doibase 10.1126/science.1093280}
  {\bibfield  {journal} {\bibinfo  {journal} {Science}\ }\textbf {\bibinfo
  {volume} {302}},\ \bibinfo {pages} {2101} (\bibinfo {year}
  {2003})}\BibitemShut {NoStop}%
\bibitem [{\citenamefont {Navon}\ \emph {et~al.}(2010)\citenamefont {Navon},
  \citenamefont {Nascimb{\`e}ne}, \citenamefont {Chevy},\ and\ \citenamefont
  {Salomon}}]{Navon729}%
  \BibitemOpen
  \bibfield  {author} {\bibinfo {author} {\bibfnamefont {N.}~\bibnamefont
  {Navon}}, \bibinfo {author} {\bibfnamefont {S.}~\bibnamefont
  {Nascimb{\`e}ne}}, \bibinfo {author} {\bibfnamefont {F.}~\bibnamefont
  {Chevy}}, \ and\ \bibinfo {author} {\bibfnamefont {C.}~\bibnamefont
  {Salomon}},\ }\href {\doibase 10.1126/science.1187582} {\bibfield  {journal}
  {\bibinfo  {journal} {Science}\ }\textbf {\bibinfo {volume} {328}},\ \bibinfo
  {pages} {729} (\bibinfo {year} {2010})}\BibitemShut {NoStop}%
\bibitem [{\citenamefont {Horikoshi}\ \emph {et~al.}(2010)\citenamefont
  {Horikoshi}, \citenamefont {Nakajima}, \citenamefont {Ueda},\ and\
  \citenamefont {Mukaiyama}}]{Horikoshi442}%
  \BibitemOpen
  \bibfield  {author} {\bibinfo {author} {\bibfnamefont {M.}~\bibnamefont
  {Horikoshi}}, \bibinfo {author} {\bibfnamefont {S.}~\bibnamefont {Nakajima}},
  \bibinfo {author} {\bibfnamefont {M.}~\bibnamefont {Ueda}}, \ and\ \bibinfo
  {author} {\bibfnamefont {T.}~\bibnamefont {Mukaiyama}},\ }\href {\doibase
  10.1126/science.1183012} {\bibfield  {journal} {\bibinfo  {journal}
  {Science}\ }\textbf {\bibinfo {volume} {327}},\ \bibinfo {pages} {442}
  (\bibinfo {year} {2010})}\BibitemShut {NoStop}%
\bibitem [{\citenamefont {Ku}\ \emph {et~al.}(2012)\citenamefont {Ku},
  \citenamefont {Sommer}, \citenamefont {Cheuk},\ and\ \citenamefont
  {Zwierlein}}]{Ku563}%
  \BibitemOpen
  \bibfield  {author} {\bibinfo {author} {\bibfnamefont {M.~J.~H.}\
  \bibnamefont {Ku}}, \bibinfo {author} {\bibfnamefont {A.~T.}\ \bibnamefont
  {Sommer}}, \bibinfo {author} {\bibfnamefont {L.~W.}\ \bibnamefont {Cheuk}}, \
  and\ \bibinfo {author} {\bibfnamefont {M.~W.}\ \bibnamefont {Zwierlein}},\
  }\href {\doibase 10.1126/science.1214987} {\bibfield  {journal} {\bibinfo
  {journal} {Science}\ }\textbf {\bibinfo {volume} {335}},\ \bibinfo {pages}
  {563} (\bibinfo {year} {2012})}\BibitemShut {NoStop}%
\bibitem [{\citenamefont {Zwerger}(2011)}]{zwerger2011bcs}%
  \BibitemOpen
  \bibfield  {author} {\bibinfo {author} {\bibfnamefont {W.}~\bibnamefont
  {Zwerger}},\ }\href@noop {} {\emph {\bibinfo {title} {The BCS-BEC Crossover
  and the Unitary Fermi Gas}}},\ Vol.\ \bibinfo {volume} {836}\ (\bibinfo
  {publisher} {Springer Science \& Business Media},\ \bibinfo {year}
  {2011})\BibitemShut {NoStop}%
\bibitem [{\citenamefont {Son}(2007)}]{PhysRevLett.98.020604}%
  \BibitemOpen
  \bibfield  {author} {\bibinfo {author} {\bibfnamefont {D.~T.}\ \bibnamefont
  {Son}},\ }\href {\doibase 10.1103/PhysRevLett.98.020604} {\bibfield
  {journal} {\bibinfo  {journal} {Phys. Rev. Lett.}\ }\textbf {\bibinfo
  {volume} {98}},\ \bibinfo {pages} {020604} (\bibinfo {year}
  {2007})}\BibitemShut {NoStop}%
\bibitem [{\citenamefont {Cao}\ \emph {et~al.}(2011)\citenamefont {Cao},
  \citenamefont {Elliott}, \citenamefont {Joseph}, \citenamefont {Wu},
  \citenamefont {Petricka}, \citenamefont {Sch{\"a}fer},\ and\ \citenamefont
  {Thomas}}]{Cao58}%
  \BibitemOpen
  \bibfield  {author} {\bibinfo {author} {\bibfnamefont {C.}~\bibnamefont
  {Cao}}, \bibinfo {author} {\bibfnamefont {E.}~\bibnamefont {Elliott}},
  \bibinfo {author} {\bibfnamefont {J.}~\bibnamefont {Joseph}}, \bibinfo
  {author} {\bibfnamefont {H.}~\bibnamefont {Wu}}, \bibinfo {author}
  {\bibfnamefont {J.}~\bibnamefont {Petricka}}, \bibinfo {author}
  {\bibfnamefont {T.}~\bibnamefont {Sch{\"a}fer}}, \ and\ \bibinfo {author}
  {\bibfnamefont {J.~E.}\ \bibnamefont {Thomas}},\ }\href {\doibase
  10.1126/science.1195219} {\bibfield  {journal} {\bibinfo  {journal}
  {Science}\ }\textbf {\bibinfo {volume} {331}},\ \bibinfo {pages} {58}
  (\bibinfo {year} {2011})}\BibitemShut {NoStop}%
\bibitem [{\citenamefont {Ho}\ and\ \citenamefont
  {Mueller}(2004)}]{PhysRevLett.92.160404}%
  \BibitemOpen
  \bibfield  {author} {\bibinfo {author} {\bibfnamefont {T.-L.}\ \bibnamefont
  {Ho}}\ and\ \bibinfo {author} {\bibfnamefont {E.~J.}\ \bibnamefont
  {Mueller}},\ }\href {\doibase 10.1103/PhysRevLett.92.160404} {\bibfield
  {journal} {\bibinfo  {journal} {Phys. Rev. Lett.}\ }\textbf {\bibinfo
  {volume} {92}},\ \bibinfo {pages} {160404} (\bibinfo {year}
  {2004})}\BibitemShut {NoStop}%
\bibitem [{\citenamefont {Chiofalo}\ \emph {et~al.}(2002)\citenamefont
  {Chiofalo}, \citenamefont {Kokkelmans}, \citenamefont {Milstein},\ and\
  \citenamefont {Holland}}]{PhysRevLett.88.090402}%
  \BibitemOpen
  \bibfield  {author} {\bibinfo {author} {\bibfnamefont {M.~L.}\ \bibnamefont
  {Chiofalo}}, \bibinfo {author} {\bibfnamefont {S.~J. J. M.~F.}\ \bibnamefont
  {Kokkelmans}}, \bibinfo {author} {\bibfnamefont {J.~N.}\ \bibnamefont
  {Milstein}}, \ and\ \bibinfo {author} {\bibfnamefont {M.~J.}\ \bibnamefont
  {Holland}},\ }\href {\doibase 10.1103/PhysRevLett.88.090402} {\bibfield
  {journal} {\bibinfo  {journal} {Phys. Rev. Lett.}\ }\textbf {\bibinfo
  {volume} {88}},\ \bibinfo {pages} {090402} (\bibinfo {year}
  {2002})}\BibitemShut {NoStop}%
\bibitem [{\citenamefont {Holland}\ \emph {et~al.}(2001)\citenamefont
  {Holland}, \citenamefont {Kokkelmans}, \citenamefont {Chiofalo},\ and\
  \citenamefont {Walser}}]{PhysRevLett.87.120406}%
  \BibitemOpen
  \bibfield  {author} {\bibinfo {author} {\bibfnamefont {M.}~\bibnamefont
  {Holland}}, \bibinfo {author} {\bibfnamefont {S.~J. J. M.~F.}\ \bibnamefont
  {Kokkelmans}}, \bibinfo {author} {\bibfnamefont {M.~L.}\ \bibnamefont
  {Chiofalo}}, \ and\ \bibinfo {author} {\bibfnamefont {R.}~\bibnamefont
  {Walser}},\ }\href {\doibase 10.1103/PhysRevLett.87.120406} {\bibfield
  {journal} {\bibinfo  {journal} {Phys. Rev. Lett.}\ }\textbf {\bibinfo
  {volume} {87}},\ \bibinfo {pages} {120406} (\bibinfo {year}
  {2001})}\BibitemShut {NoStop}%
\bibitem [{\citenamefont {Van~Houcke}\ \emph {et~al.}(2012)\citenamefont
  {Van~Houcke}, \citenamefont {Werner}, \citenamefont {Kozik}, \citenamefont
  {Prokof'ev}, \citenamefont {Svistunov}, \citenamefont {Ku}, \citenamefont
  {Sommer}, \citenamefont {Cheuk}, \citenamefont {Schirotzek},\ and\
  \citenamefont {Zwierlein}}]{houcke2012}%
  \BibitemOpen
  \bibfield  {author} {\bibinfo {author} {\bibfnamefont {K.}~\bibnamefont
  {Van~Houcke}}, \bibinfo {author} {\bibfnamefont {F.}~\bibnamefont {Werner}},
  \bibinfo {author} {\bibfnamefont {E.}~\bibnamefont {Kozik}}, \bibinfo
  {author} {\bibfnamefont {N.}~\bibnamefont {Prokof'ev}}, \bibinfo {author}
  {\bibfnamefont {B.}~\bibnamefont {Svistunov}}, \bibinfo {author}
  {\bibfnamefont {M.~J.~H.}\ \bibnamefont {Ku}}, \bibinfo {author}
  {\bibfnamefont {A.~T.}\ \bibnamefont {Sommer}}, \bibinfo {author}
  {\bibfnamefont {L.~W.}\ \bibnamefont {Cheuk}}, \bibinfo {author}
  {\bibfnamefont {A.}~\bibnamefont {Schirotzek}}, \ and\ \bibinfo {author}
  {\bibfnamefont {M.~W.}\ \bibnamefont {Zwierlein}},\ }\href {\doibase
  10.1038/nphys2273} {\bibfield  {journal} {\bibinfo  {journal} {Nature
  Physics}\ }\textbf {\bibinfo {volume} {8}},\ \bibinfo {pages} {366} (\bibinfo
  {year} {2012})}\BibitemShut {NoStop}%
\bibitem [{\citenamefont {Levinsen}\ \emph {et~al.}(2017)\citenamefont
  {Levinsen}, \citenamefont {Massignan}, \citenamefont {Endo},\ and\
  \citenamefont {Parish}}]{Levinsen_2017}%
  \BibitemOpen
  \bibfield  {author} {\bibinfo {author} {\bibfnamefont {J.}~\bibnamefont
  {Levinsen}}, \bibinfo {author} {\bibfnamefont {P.}~\bibnamefont {Massignan}},
  \bibinfo {author} {\bibfnamefont {S.}~\bibnamefont {Endo}}, \ and\ \bibinfo
  {author} {\bibfnamefont {M.~M.}\ \bibnamefont {Parish}},\ }\href {\doibase
  10.1088/1361-6455/aa5a1e} {\bibfield  {journal} {\bibinfo  {journal} {Journal
  of Physics B: Atomic, Molecular and Optical Physics}\ }\textbf {\bibinfo
  {volume} {50}},\ \bibinfo {pages} {072001} (\bibinfo {year}
  {2017})}\BibitemShut {NoStop}%
\bibitem [{\citenamefont {Efimov}(1971)}]{efimov1971weakly}%
  \BibitemOpen
  \bibfield  {author} {\bibinfo {author} {\bibfnamefont {V.}~\bibnamefont
  {Efimov}},\ }\href@noop {} {\bibfield  {journal} {\bibinfo  {journal} {Sov.
  J. Nucl. Phys}\ }\textbf {\bibinfo {volume} {12}},\ \bibinfo {pages} {101}
  (\bibinfo {year} {1971})}\BibitemShut {NoStop}%
\bibitem [{\citenamefont {Efimov}(1979)}]{efimov1979low}%
  \BibitemOpen
  \bibfield  {author} {\bibinfo {author} {\bibfnamefont {V.}~\bibnamefont
  {Efimov}},\ }\href@noop {} {\bibfield  {journal} {\bibinfo  {journal} {Sov.
  J. Nucl. Phys.}\ }\textbf {\bibinfo {volume} {29}},\ \bibinfo {pages} {546}
  (\bibinfo {year} {1979})}\BibitemShut {NoStop}%
\bibitem [{\citenamefont {Braaten}\ and\ \citenamefont
  {Hammer}(2006)}]{BRAATEN2006259}%
  \BibitemOpen
  \bibfield  {author} {\bibinfo {author} {\bibfnamefont {E.}~\bibnamefont
  {Braaten}}\ and\ \bibinfo {author} {\bibfnamefont {H.-W.}\ \bibnamefont
  {Hammer}},\ }\href {\doibase https://doi.org/10.1016/j.physrep.2006.03.001}
  {\bibfield  {journal} {\bibinfo  {journal} {Physics Reports}\ }\textbf
  {\bibinfo {volume} {428}},\ \bibinfo {pages} {259 } (\bibinfo {year}
  {2006})}\BibitemShut {NoStop}%
\bibitem [{\citenamefont {D'Incao}(2018)}]{D_Incao_2018}%
  \BibitemOpen
  \bibfield  {author} {\bibinfo {author} {\bibfnamefont {J.~P.}\ \bibnamefont
  {D'Incao}},\ }\href {\doibase 10.1088/1361-6455/aaa116} {\bibfield  {journal}
  {\bibinfo  {journal} {Journal of Physics B: Atomic, Molecular and Optical
  Physics}\ }\textbf {\bibinfo {volume} {51}},\ \bibinfo {pages} {043001}
  (\bibinfo {year} {2018})}\BibitemShut {NoStop}%
\bibitem [{\citenamefont {Naidon}\ and\ \citenamefont
  {Endo}(2017)}]{Naidon_2017}%
  \BibitemOpen
  \bibfield  {author} {\bibinfo {author} {\bibfnamefont {P.}~\bibnamefont
  {Naidon}}\ and\ \bibinfo {author} {\bibfnamefont {S.}~\bibnamefont {Endo}},\
  }\href {\doibase 10.1088/1361-6633/aa50e8} {\bibfield  {journal} {\bibinfo
  {journal} {Reports on Progress in Physics}\ }\textbf {\bibinfo {volume}
  {80}},\ \bibinfo {pages} {056001} (\bibinfo {year} {2017})}\BibitemShut
  {NoStop}%
\bibitem [{\citenamefont {Greene}\ \emph {et~al.}(2017)\citenamefont {Greene},
  \citenamefont {Giannakeas},\ and\ \citenamefont
  {P\'erez-R\'{\i}os}}]{RevModPhys.89.035006}%
  \BibitemOpen
  \bibfield  {author} {\bibinfo {author} {\bibfnamefont {C.~H.}\ \bibnamefont
  {Greene}}, \bibinfo {author} {\bibfnamefont {P.}~\bibnamefont {Giannakeas}},
  \ and\ \bibinfo {author} {\bibfnamefont {J.}~\bibnamefont
  {P\'erez-R\'{\i}os}},\ }\href {\doibase 10.1103/RevModPhys.89.035006}
  {\bibfield  {journal} {\bibinfo  {journal} {Rev. Mod. Phys.}\ }\textbf
  {\bibinfo {volume} {89}},\ \bibinfo {pages} {035006} (\bibinfo {year}
  {2017})}\BibitemShut {NoStop}%
\bibitem [{\citenamefont {Nishida}(2012)}]{Nishida2012}%
  \BibitemOpen
  \bibfield  {author} {\bibinfo {author} {\bibfnamefont {Y.}~\bibnamefont
  {Nishida}},\ }\href {\doibase 10.1103/PhysRevLett.109.240401} {\bibfield
  {journal} {\bibinfo  {journal} {Phys. Rev. Lett.}\ }\textbf {\bibinfo
  {volume} {109}},\ \bibinfo {pages} {240401} (\bibinfo {year}
  {2012})}\BibitemShut {NoStop}%
\bibitem [{\citenamefont {Tajima}\ and\ \citenamefont
  {Naidon}(2019)}]{Naidon2019}%
  \BibitemOpen
  \bibfield  {author} {\bibinfo {author} {\bibfnamefont {H.}~\bibnamefont
  {Tajima}}\ and\ \bibinfo {author} {\bibfnamefont {P.}~\bibnamefont
  {Naidon}},\ }\href {\doibase 10.1088/1367-2630/ab306b} {\bibfield  {journal}
  {\bibinfo  {journal} {New Journal of Physics}\ }\textbf {\bibinfo {volume}
  {21}},\ \bibinfo {pages} {073051} (\bibinfo {year} {2019})}\BibitemShut
  {NoStop}%
\bibitem [{\citenamefont {Makotyn}\ \emph {et~al.}(2014)\citenamefont
  {Makotyn}, \citenamefont {Klauss}, \citenamefont {Goldberger}, \citenamefont
  {Cornell},\ and\ \citenamefont {Jin}}]{makotyn2014universal}%
  \BibitemOpen
  \bibfield  {author} {\bibinfo {author} {\bibfnamefont {P.}~\bibnamefont
  {Makotyn}}, \bibinfo {author} {\bibfnamefont {C.~E.}\ \bibnamefont {Klauss}},
  \bibinfo {author} {\bibfnamefont {D.~L.}\ \bibnamefont {Goldberger}},
  \bibinfo {author} {\bibfnamefont {E.}~\bibnamefont {Cornell}}, \ and\
  \bibinfo {author} {\bibfnamefont {D.~S.}\ \bibnamefont {Jin}},\ }\href@noop
  {} {\bibfield  {journal} {\bibinfo  {journal} {Nat. Phys.}\ }\textbf
  {\bibinfo {volume} {10}},\ \bibinfo {pages} {116} (\bibinfo {year}
  {2014})}\BibitemShut {NoStop}%
\bibitem [{\citenamefont {Klauss}\ \emph {et~al.}(2017)\citenamefont {Klauss},
  \citenamefont {Xie}, \citenamefont {Lopez-Abadia}, \citenamefont {D'Incao},
  \citenamefont {Hadzibabic}, \citenamefont {Jin},\ and\ \citenamefont
  {Cornell}}]{klauss2017observation}%
  \BibitemOpen
  \bibfield  {author} {\bibinfo {author} {\bibfnamefont {C.~E.}\ \bibnamefont
  {Klauss}}, \bibinfo {author} {\bibfnamefont {X.}~\bibnamefont {Xie}},
  \bibinfo {author} {\bibfnamefont {C.}~\bibnamefont {Lopez-Abadia}}, \bibinfo
  {author} {\bibfnamefont {J.~P.}\ \bibnamefont {D'Incao}}, \bibinfo {author}
  {\bibfnamefont {Z.}~\bibnamefont {Hadzibabic}}, \bibinfo {author}
  {\bibfnamefont {D.~S.}\ \bibnamefont {Jin}}, \ and\ \bibinfo {author}
  {\bibfnamefont {E.~A.}\ \bibnamefont {Cornell}},\ }\href {\doibase
  10.1103/PhysRevLett.119.143401} {\bibfield  {journal} {\bibinfo  {journal}
  {Phys. Rev. Lett.}\ }\textbf {\bibinfo {volume} {119}},\ \bibinfo {pages}
  {143401} (\bibinfo {year} {2017})}\BibitemShut {NoStop}%
\bibitem [{\citenamefont {Eigen}\ \emph
  {et~al.}(2017{\natexlab{a}})\citenamefont {Eigen}, \citenamefont {Glidden},
  \citenamefont {Lopes}, \citenamefont {Navon}, \citenamefont {Hadzibabic},\
  and\ \citenamefont {Smith}}]{eigen2017universal}%
  \BibitemOpen
  \bibfield  {author} {\bibinfo {author} {\bibfnamefont {C.}~\bibnamefont
  {Eigen}}, \bibinfo {author} {\bibfnamefont {J.~A.~P.}\ \bibnamefont
  {Glidden}}, \bibinfo {author} {\bibfnamefont {R.}~\bibnamefont {Lopes}},
  \bibinfo {author} {\bibfnamefont {N.}~\bibnamefont {Navon}}, \bibinfo
  {author} {\bibfnamefont {Z.}~\bibnamefont {Hadzibabic}}, \ and\ \bibinfo
  {author} {\bibfnamefont {R.~P.}\ \bibnamefont {Smith}},\ }\href {\doibase
  10.1103/PhysRevLett.119.250404} {\bibfield  {journal} {\bibinfo  {journal}
  {Phys. Rev. Lett.}\ }\textbf {\bibinfo {volume} {119}},\ \bibinfo {pages}
  {250404} (\bibinfo {year} {2017}{\natexlab{a}})}\BibitemShut {NoStop}%
\bibitem [{\citenamefont {Eigen}\ \emph {et~al.}(2018)\citenamefont {Eigen},
  \citenamefont {Glidden}, \citenamefont {Lopes}, \citenamefont {Cornell},
  \citenamefont {Smith},\ and\ \citenamefont
  {Hadzibabic}}]{eigen2018prethermal}%
  \BibitemOpen
  \bibfield  {author} {\bibinfo {author} {\bibfnamefont {C.}~\bibnamefont
  {Eigen}}, \bibinfo {author} {\bibfnamefont {J.~A.~P.}\ \bibnamefont
  {Glidden}}, \bibinfo {author} {\bibfnamefont {R.}~\bibnamefont {Lopes}},
  \bibinfo {author} {\bibfnamefont {E.~A.}\ \bibnamefont {Cornell}}, \bibinfo
  {author} {\bibfnamefont {R.~P.}\ \bibnamefont {Smith}}, \ and\ \bibinfo
  {author} {\bibfnamefont {Z.}~\bibnamefont {Hadzibabic}},\ }\href@noop {}
  {\bibfield  {journal} {\bibinfo  {journal} {Nature}\ }\textbf {\bibinfo
  {volume} {556}},\ \bibinfo {pages} {221} (\bibinfo {year}
  {2018})}\BibitemShut {NoStop}%
\bibitem [{\citenamefont {D'Incao}\ \emph {et~al.}(2018)\citenamefont
  {D'Incao}, \citenamefont {Wang},\ and\ \citenamefont
  {Colussi}}]{PhysRevLett.121.023401}%
  \BibitemOpen
  \bibfield  {author} {\bibinfo {author} {\bibfnamefont {J.~P.}\ \bibnamefont
  {D'Incao}}, \bibinfo {author} {\bibfnamefont {J.}~\bibnamefont {Wang}}, \
  and\ \bibinfo {author} {\bibfnamefont {V.~E.}\ \bibnamefont {Colussi}},\
  }\href {\doibase 10.1103/PhysRevLett.121.023401} {\bibfield  {journal}
  {\bibinfo  {journal} {Phys. Rev. Lett.}\ }\textbf {\bibinfo {volume} {121}},\
  \bibinfo {pages} {023401} (\bibinfo {year} {2018})}\BibitemShut {NoStop}%
\bibitem [{\citenamefont {Colussi}\ \emph
  {et~al.}(2018{\natexlab{a}})\citenamefont {Colussi}, \citenamefont {Corson},\
  and\ \citenamefont {D'Incao}}]{PhysRevLett.120.100401}%
  \BibitemOpen
  \bibfield  {author} {\bibinfo {author} {\bibfnamefont {V.~E.}\ \bibnamefont
  {Colussi}}, \bibinfo {author} {\bibfnamefont {J.~P.}\ \bibnamefont {Corson}},
  \ and\ \bibinfo {author} {\bibfnamefont {J.~P.}\ \bibnamefont {D'Incao}},\
  }\href {\doibase 10.1103/PhysRevLett.120.100401} {\bibfield  {journal}
  {\bibinfo  {journal} {Phys. Rev. Lett.}\ }\textbf {\bibinfo {volume} {120}},\
  \bibinfo {pages} {100401} (\bibinfo {year} {2018}{\natexlab{a}})}\BibitemShut
  {NoStop}%
\bibitem [{\citenamefont {Colussi}\ \emph {et~al.}(2019)\citenamefont
  {Colussi}, \citenamefont {van Zwol}, \citenamefont {D'Incao},\ and\
  \citenamefont {Kokkelmans}}]{PhysRevA.99.043604}%
  \BibitemOpen
  \bibfield  {author} {\bibinfo {author} {\bibfnamefont {V.~E.}\ \bibnamefont
  {Colussi}}, \bibinfo {author} {\bibfnamefont {B.~E.}\ \bibnamefont {van
  Zwol}}, \bibinfo {author} {\bibfnamefont {J.~P.}\ \bibnamefont {D'Incao}}, \
  and\ \bibinfo {author} {\bibfnamefont {S.~J. J. M.~F.}\ \bibnamefont
  {Kokkelmans}},\ }\href {\doibase 10.1103/PhysRevA.99.043604} {\bibfield
  {journal} {\bibinfo  {journal} {Phys. Rev. A}\ }\textbf {\bibinfo {volume}
  {99}},\ \bibinfo {pages} {043604} (\bibinfo {year} {2019})}\BibitemShut
  {NoStop}%
\bibitem [{\citenamefont {Berges}\ \emph {et~al.}(2004)\citenamefont {Berges},
  \citenamefont {Bors\'anyi},\ and\ \citenamefont
  {Wetterich}}]{PhysRevLett.93.142002}%
  \BibitemOpen
  \bibfield  {author} {\bibinfo {author} {\bibfnamefont {J.}~\bibnamefont
  {Berges}}, \bibinfo {author} {\bibfnamefont {S.}~\bibnamefont {Bors\'anyi}},
  \ and\ \bibinfo {author} {\bibfnamefont {C.}~\bibnamefont {Wetterich}},\
  }\href {\doibase 10.1103/PhysRevLett.93.142002} {\bibfield  {journal}
  {\bibinfo  {journal} {Phys. Rev. Lett.}\ }\textbf {\bibinfo {volume} {93}},\
  \bibinfo {pages} {142002} (\bibinfo {year} {2004})}\BibitemShut {NoStop}%
\bibitem [{\citenamefont {Van~Regemortel}\ \emph {et~al.}(2018)\citenamefont
  {Van~Regemortel}, \citenamefont {Kurkjian}, \citenamefont {Wouters},\ and\
  \citenamefont {Carusotto}}]{PhysRevA.98.053612}%
  \BibitemOpen
  \bibfield  {author} {\bibinfo {author} {\bibfnamefont {M.}~\bibnamefont
  {Van~Regemortel}}, \bibinfo {author} {\bibfnamefont {H.}~\bibnamefont
  {Kurkjian}}, \bibinfo {author} {\bibfnamefont {M.}~\bibnamefont {Wouters}}, \
  and\ \bibinfo {author} {\bibfnamefont {I.}~\bibnamefont {Carusotto}},\ }\href
  {\doibase 10.1103/PhysRevA.98.053612} {\bibfield  {journal} {\bibinfo
  {journal} {Phys. Rev. A}\ }\textbf {\bibinfo {volume} {98}},\ \bibinfo
  {pages} {053612} (\bibinfo {year} {2018})}\BibitemShut {NoStop}%
\bibitem [{\citenamefont {Sykes}\ \emph {et~al.}(2014)\citenamefont {Sykes},
  \citenamefont {Corson}, \citenamefont {D'Incao}, \citenamefont {Koller},
  \citenamefont {Greene}, \citenamefont {Rey}, \citenamefont {Hazzard},\ and\
  \citenamefont {Bohn}}]{PhysRevA.89.021601}%
  \BibitemOpen
  \bibfield  {author} {\bibinfo {author} {\bibfnamefont {A.~G.}\ \bibnamefont
  {Sykes}}, \bibinfo {author} {\bibfnamefont {J.~P.}\ \bibnamefont {Corson}},
  \bibinfo {author} {\bibfnamefont {J.~P.}\ \bibnamefont {D'Incao}}, \bibinfo
  {author} {\bibfnamefont {A.~P.}\ \bibnamefont {Koller}}, \bibinfo {author}
  {\bibfnamefont {C.~H.}\ \bibnamefont {Greene}}, \bibinfo {author}
  {\bibfnamefont {A.~M.}\ \bibnamefont {Rey}}, \bibinfo {author} {\bibfnamefont
  {K.~R.~A.}\ \bibnamefont {Hazzard}}, \ and\ \bibinfo {author} {\bibfnamefont
  {J.~L.}\ \bibnamefont {Bohn}},\ }\href {\doibase 10.1103/PhysRevA.89.021601}
  {\bibfield  {journal} {\bibinfo  {journal} {Phys. Rev. A}\ }\textbf {\bibinfo
  {volume} {89}},\ \bibinfo {pages} {021601} (\bibinfo {year}
  {2014})}\BibitemShut {NoStop}%
\bibitem [{\citenamefont {Gao}\ \emph {et~al.}(2020)\citenamefont {Gao},
  \citenamefont {Sun}, \citenamefont {Zhang},\ and\ \citenamefont
  {Zhai}}]{PhysRevLett.124.040403}%
  \BibitemOpen
  \bibfield  {author} {\bibinfo {author} {\bibfnamefont {C.}~\bibnamefont
  {Gao}}, \bibinfo {author} {\bibfnamefont {M.}~\bibnamefont {Sun}}, \bibinfo
  {author} {\bibfnamefont {P.}~\bibnamefont {Zhang}}, \ and\ \bibinfo {author}
  {\bibfnamefont {H.}~\bibnamefont {Zhai}},\ }\href {\doibase
  10.1103/PhysRevLett.124.040403} {\bibfield  {journal} {\bibinfo  {journal}
  {Phys. Rev. Lett.}\ }\textbf {\bibinfo {volume} {124}},\ \bibinfo {pages}
  {040403} (\bibinfo {year} {2020})}\BibitemShut {NoStop}%
\bibitem [{\citenamefont {Mu\~noz de~las Heras}\ \emph
  {et~al.}(2019)\citenamefont {Mu\~noz de~las Heras}, \citenamefont {Parish},\
  and\ \citenamefont {Marchetti}}]{PhysRevA.99.023623}%
  \BibitemOpen
  \bibfield  {author} {\bibinfo {author} {\bibfnamefont {A.}~\bibnamefont
  {Mu\~noz de~las Heras}}, \bibinfo {author} {\bibfnamefont {M.~M.}\
  \bibnamefont {Parish}}, \ and\ \bibinfo {author} {\bibfnamefont {F.~M.}\
  \bibnamefont {Marchetti}},\ }\href {\doibase 10.1103/PhysRevA.99.023623}
  {\bibfield  {journal} {\bibinfo  {journal} {Phys. Rev. A}\ }\textbf {\bibinfo
  {volume} {99}},\ \bibinfo {pages} {023623} (\bibinfo {year}
  {2019})}\BibitemShut {NoStop}%
\bibitem [{\citenamefont {Yuzbashyan}\ \emph {et~al.}(2015)\citenamefont
  {Yuzbashyan}, \citenamefont {Dzero}, \citenamefont {Gurarie},\ and\
  \citenamefont {Foster}}]{Foster2015}%
  \BibitemOpen
  \bibfield  {author} {\bibinfo {author} {\bibfnamefont {E.~A.}\ \bibnamefont
  {Yuzbashyan}}, \bibinfo {author} {\bibfnamefont {M.}~\bibnamefont {Dzero}},
  \bibinfo {author} {\bibfnamefont {V.}~\bibnamefont {Gurarie}}, \ and\
  \bibinfo {author} {\bibfnamefont {M.~S.}\ \bibnamefont {Foster}},\ }\href
  {\doibase 10.1103/PhysRevA.91.033628} {\bibfield  {journal} {\bibinfo
  {journal} {Phys. Rev. A}\ }\textbf {\bibinfo {volume} {91}},\ \bibinfo
  {pages} {033628} (\bibinfo {year} {2015})}\BibitemShut {NoStop}%
\bibitem [{\citenamefont {Huang}(1987)}]{huang1987statistical}%
  \BibitemOpen
  \bibfield  {author} {\bibinfo {author} {\bibfnamefont {K.}~\bibnamefont
  {Huang}},\ }\href@noop {} {\emph {\bibinfo {title} {Statistical Mechanics}}}\
  (\bibinfo  {publisher} {Wiley},\ \bibinfo {year} {1987})\BibitemShut
  {NoStop}%
\bibitem [{\citenamefont {Fricke}(1996)}]{fricke1996transport}%
  \BibitemOpen
  \bibfield  {author} {\bibinfo {author} {\bibfnamefont {J.}~\bibnamefont
  {Fricke}},\ }\href@noop {} {\bibfield  {journal} {\bibinfo  {journal} {Annals
  of Physics}\ }\textbf {\bibinfo {volume} {252}},\ \bibinfo {pages} {479}
  (\bibinfo {year} {1996})}\BibitemShut {NoStop}%
\bibitem [{\citenamefont {K{\"o}hler}\ and\ \citenamefont
  {Burnett}(2002)}]{burnett2002}%
  \BibitemOpen
  \bibfield  {author} {\bibinfo {author} {\bibfnamefont {T.}~\bibnamefont
  {K{\"o}hler}}\ and\ \bibinfo {author} {\bibfnamefont {K.}~\bibnamefont
  {Burnett}},\ }\href@noop {} {\bibfield  {journal} {\bibinfo  {journal} {Phys.
  Rev. A}\ }\textbf {\bibinfo {volume} {65}},\ \bibinfo {pages} {033601}
  (\bibinfo {year} {2002})}\BibitemShut {NoStop}%
\bibitem [{\citenamefont {Kira}\ and\ \citenamefont
  {Koch}(2011)}]{kira2011semiconductor}%
  \BibitemOpen
  \bibfield  {author} {\bibinfo {author} {\bibfnamefont {M.}~\bibnamefont
  {Kira}}\ and\ \bibinfo {author} {\bibfnamefont {S.~W.}\ \bibnamefont
  {Koch}},\ }\href@noop {} {\emph {\bibinfo {title} {Semiconductor Quantum
  Optics}}}\ (\bibinfo  {publisher} {Cambridge University Press},\ \bibinfo
  {year} {2011})\BibitemShut {NoStop}%
\bibitem [{\citenamefont {Kira}(2014)}]{kira2014excitation}%
  \BibitemOpen
  \bibfield  {author} {\bibinfo {author} {\bibfnamefont {M.}~\bibnamefont
  {Kira}},\ }\href@noop {} {\bibfield  {journal} {\bibinfo  {journal} {Annals
  of Physics}\ }\textbf {\bibinfo {volume} {351}},\ \bibinfo {pages} {200}
  (\bibinfo {year} {2014})}\BibitemShut {NoStop}%
\bibitem [{\citenamefont {Kira}(2015{\natexlab{a}})}]{KIRA2015185}%
  \BibitemOpen
  \bibfield  {author} {\bibinfo {author} {\bibfnamefont {M.}~\bibnamefont
  {Kira}},\ }\href {\doibase https://doi.org/10.1016/j.aop.2015.02.030}
  {\bibfield  {journal} {\bibinfo  {journal} {Annals of Physics}\ }\textbf
  {\bibinfo {volume} {356}},\ \bibinfo {pages} {185 } (\bibinfo {year}
  {2015}{\natexlab{a}})}\BibitemShut {NoStop}%
\bibitem [{\citenamefont {Kira}(2015{\natexlab{b}})}]{kirancomm}%
  \BibitemOpen
  \bibfield  {author} {\bibinfo {author} {\bibfnamefont {M.}~\bibnamefont
  {Kira}},\ }\href@noop {} {\bibfield  {journal} {\bibinfo  {journal} {Nat.
  Commun.}\ }\textbf {\bibinfo {volume} {6}},\ \bibinfo {pages} {6624}
  (\bibinfo {year} {2015}{\natexlab{b}})}\BibitemShut {NoStop}%
\bibitem [{\citenamefont {Colussi}\ \emph
  {et~al.}(2018{\natexlab{b}})\citenamefont {Colussi}, \citenamefont
  {Musolino},\ and\ \citenamefont {Kokkelmans}}]{Kokkelmans2018}%
  \BibitemOpen
  \bibfield  {author} {\bibinfo {author} {\bibfnamefont {V.~E.}\ \bibnamefont
  {Colussi}}, \bibinfo {author} {\bibfnamefont {S.}~\bibnamefont {Musolino}}, \
  and\ \bibinfo {author} {\bibfnamefont {S.~J. J. M.~F.}\ \bibnamefont
  {Kokkelmans}},\ }\href {\doibase 10.1103/PhysRevA.98.051601} {\bibfield
  {journal} {\bibinfo  {journal} {Phys. Rev. A}\ }\textbf {\bibinfo {volume}
  {98}},\ \bibinfo {pages} {051601} (\bibinfo {year}
  {2018}{\natexlab{b}})}\BibitemShut {NoStop}%
\bibitem [{\citenamefont {Musolino}\ \emph {et~al.}(2019)\citenamefont
  {Musolino}, \citenamefont {Colussi},\ and\ \citenamefont
  {Kokkelmans}}]{PhysRevA.100.013612}%
  \BibitemOpen
  \bibfield  {author} {\bibinfo {author} {\bibfnamefont {S.}~\bibnamefont
  {Musolino}}, \bibinfo {author} {\bibfnamefont {V.~E.}\ \bibnamefont
  {Colussi}}, \ and\ \bibinfo {author} {\bibfnamefont {S.~J. J. M.~F.}\
  \bibnamefont {Kokkelmans}},\ }\href {\doibase 10.1103/PhysRevA.100.013612}
  {\bibfield  {journal} {\bibinfo  {journal} {Phys. Rev. A}\ }\textbf {\bibinfo
  {volume} {100}},\ \bibinfo {pages} {013612} (\bibinfo {year}
  {2019})}\BibitemShut {NoStop}%
\bibitem [{\citenamefont {Corson}\ and\ \citenamefont
  {Bohn}(2015)}]{PhysRevA.91.013616}%
  \BibitemOpen
  \bibfield  {author} {\bibinfo {author} {\bibfnamefont {J.~P.}\ \bibnamefont
  {Corson}}\ and\ \bibinfo {author} {\bibfnamefont {J.~L.}\ \bibnamefont
  {Bohn}},\ }\href {\doibase 10.1103/PhysRevA.91.013616} {\bibfield  {journal}
  {\bibinfo  {journal} {Phys. Rev. A}\ }\textbf {\bibinfo {volume} {91}},\
  \bibinfo {pages} {013616} (\bibinfo {year} {2015})}\BibitemShut {NoStop}%
\bibitem [{\citenamefont {Chin}\ \emph {et~al.}(2010)\citenamefont {Chin},
  \citenamefont {Grimm}, \citenamefont {Julienne},\ and\ \citenamefont
  {Tiesinga}}]{RevModPhys.82.1225}%
  \BibitemOpen
  \bibfield  {author} {\bibinfo {author} {\bibfnamefont {C.}~\bibnamefont
  {Chin}}, \bibinfo {author} {\bibfnamefont {R.}~\bibnamefont {Grimm}},
  \bibinfo {author} {\bibfnamefont {P.}~\bibnamefont {Julienne}}, \ and\
  \bibinfo {author} {\bibfnamefont {E.}~\bibnamefont {Tiesinga}},\ }\href
  {\doibase 10.1103/RevModPhys.82.1225} {\bibfield  {journal} {\bibinfo
  {journal} {Rev. Mod. Phys.}\ }\textbf {\bibinfo {volume} {82}},\ \bibinfo
  {pages} {1225} (\bibinfo {year} {2010})}\BibitemShut {NoStop}%
\bibitem [{\citenamefont {Taylor}(2006)}]{taylor2006scattering}%
  \BibitemOpen
  \bibfield  {author} {\bibinfo {author} {\bibfnamefont {J.~R.}\ \bibnamefont
  {Taylor}},\ }\href@noop {} {\emph {\bibinfo {title} {Scattering Theory: The
  Quantum Theory of Nonrelativistic Collisions}}}\ (\bibinfo  {publisher}
  {Courier Corporation},\ \bibinfo {year} {2006})\BibitemShut {NoStop}%
\bibitem [{\citenamefont {K\"ohler}\ \emph {et~al.}(2003)\citenamefont
  {K\"ohler}, \citenamefont {Gasenzer},\ and\ \citenamefont
  {Burnett}}]{burnett2003}%
  \BibitemOpen
  \bibfield  {author} {\bibinfo {author} {\bibfnamefont {T.}~\bibnamefont
  {K\"ohler}}, \bibinfo {author} {\bibfnamefont {T.}~\bibnamefont {Gasenzer}},
  \ and\ \bibinfo {author} {\bibfnamefont {K.}~\bibnamefont {Burnett}},\ }\href
  {\doibase 10.1103/PhysRevA.67.013601} {\bibfield  {journal} {\bibinfo
  {journal} {Phys. Rev. A}\ }\textbf {\bibinfo {volume} {67}},\ \bibinfo
  {pages} {013601} (\bibinfo {year} {2003})}\BibitemShut {NoStop}%
\bibitem [{\citenamefont {Flambaum}\ \emph {et~al.}(1999)\citenamefont
  {Flambaum}, \citenamefont {Gribakin},\ and\ \citenamefont
  {Harabati}}]{PhysRevA.59.1998}%
  \BibitemOpen
  \bibfield  {author} {\bibinfo {author} {\bibfnamefont {V.~V.}\ \bibnamefont
  {Flambaum}}, \bibinfo {author} {\bibfnamefont {G.~F.}\ \bibnamefont
  {Gribakin}}, \ and\ \bibinfo {author} {\bibfnamefont {C.}~\bibnamefont
  {Harabati}},\ }\href {\doibase 10.1103/PhysRevA.59.1998} {\bibfield
  {journal} {\bibinfo  {journal} {Phys. Rev. A}\ }\textbf {\bibinfo {volume}
  {59}},\ \bibinfo {pages} {1998} (\bibinfo {year} {1999})}\BibitemShut
  {NoStop}%
\bibitem [{\citenamefont {Sinatra}\ \emph {et~al.}(2009)\citenamefont
  {Sinatra}, \citenamefont {Castin},\ and\ \citenamefont
  {Witkowska}}]{Witkowska2009}%
  \BibitemOpen
  \bibfield  {author} {\bibinfo {author} {\bibfnamefont {A.}~\bibnamefont
  {Sinatra}}, \bibinfo {author} {\bibfnamefont {Y.}~\bibnamefont {Castin}}, \
  and\ \bibinfo {author} {\bibfnamefont {E.}~\bibnamefont {Witkowska}},\ }\href
  {\doibase 10.1103/PhysRevA.80.033614} {\bibfield  {journal} {\bibinfo
  {journal} {Phys. Rev. A}\ }\textbf {\bibinfo {volume} {80}},\ \bibinfo
  {pages} {033614} (\bibinfo {year} {2009})}\BibitemShut {NoStop}%
\bibitem [{\citenamefont {Kurkjian}\ \emph {et~al.}(2016)\citenamefont
  {Kurkjian}, \citenamefont {Castin},\ and\ \citenamefont
  {Sinatra}}]{Sinatra2016}%
  \BibitemOpen
  \bibfield  {author} {\bibinfo {author} {\bibfnamefont {H.}~\bibnamefont
  {Kurkjian}}, \bibinfo {author} {\bibfnamefont {Y.}~\bibnamefont {Castin}}, \
  and\ \bibinfo {author} {\bibfnamefont {A.}~\bibnamefont {Sinatra}},\ }\href
  {\doibase http://dx.doi.org/10.1016/j.crhy.2016.02.005} {\bibfield  {journal}
  {\bibinfo  {journal} {{Comptes Rendus Physique}}\ }\textbf {\bibinfo {volume}
  {17}},\ \bibinfo {pages} {789 } (\bibinfo {year} {2016})}\BibitemShut
  {NoStop}%
\bibitem [{\citenamefont {Blaizot}\ and\ \citenamefont
  {Ripka}(1985)}]{Ripka1985}%
  \BibitemOpen
  \bibfield  {author} {\bibinfo {author} {\bibfnamefont {J.-P.}\ \bibnamefont
  {Blaizot}}\ and\ \bibinfo {author} {\bibfnamefont {G.}~\bibnamefont
  {Ripka}},\ }\href@noop {} {\emph {\bibinfo {title} {Quantum {T}heory of
  {F}inite {S}ystems}}}\ (\bibinfo  {publisher} {MIT Press},\ \bibinfo
  {address} {Cambridge, Massachusetts},\ \bibinfo {year} {1985})\BibitemShut
  {NoStop}%
\bibitem [{\citenamefont {Glyde}\ \emph {et~al.}(2000)\citenamefont {Glyde},
  \citenamefont {Azuah},\ and\ \citenamefont {Stirling}}]{Stirling2000}%
  \BibitemOpen
  \bibfield  {author} {\bibinfo {author} {\bibfnamefont {H.~R.}\ \bibnamefont
  {Glyde}}, \bibinfo {author} {\bibfnamefont {R.~T.}\ \bibnamefont {Azuah}}, \
  and\ \bibinfo {author} {\bibfnamefont {W.~G.}\ \bibnamefont {Stirling}},\
  }\href {\doibase 10.1103/PhysRevB.62.14337} {\bibfield  {journal} {\bibinfo
  {journal} {Phys. Rev. B}\ }\textbf {\bibinfo {volume} {62}},\ \bibinfo
  {pages} {14337} (\bibinfo {year} {2000})}\BibitemShut {NoStop}%
\bibitem [{rep()}]{repository}%
  \BibitemOpen
  \href {\doibase 10.4121/uuid:720f48bc-e64e-47e1-8910-09f6832a09bb} {}\bibinfo
  {note} {This program uses the SNEG package \cite{Zitko2011} for basic second
  quantization operations such as commutation relations and is available online
  at this address
  \url{https://doi.org/10.4121/uuid:720f48bc-e64e-47e1-8910-09f6832a09bb}.}\BibitemShut
  {Stop}%
\bibitem [{\citenamefont {Proukakis}\ and\ \citenamefont
  {Burnett}(1996)}]{proukakis1996generalized}%
  \BibitemOpen
  \bibfield  {author} {\bibinfo {author} {\bibfnamefont {N.}~\bibnamefont
  {Proukakis}}\ and\ \bibinfo {author} {\bibfnamefont {K.}~\bibnamefont
  {Burnett}},\ }\href@noop {} {\bibfield  {journal} {\bibinfo  {journal} {J.
  Res. Natl. Inst. Stand. Technol.}\ }\textbf {\bibinfo {volume} {101}},\
  \bibinfo {pages} {457} (\bibinfo {year} {1996})}\BibitemShut {NoStop}%
\bibitem [{\citenamefont {Newton}(2013)}]{newton2013scattering}%
  \BibitemOpen
  \bibfield  {author} {\bibinfo {author} {\bibfnamefont {R.~G.}\ \bibnamefont
  {Newton}},\ }\href@noop {} {\emph {\bibinfo {title} {Scattering Theory of
  Waves and Particles}}}\ (\bibinfo  {publisher} {Springer Science \& Business
  Media},\ \bibinfo {year} {2013})\BibitemShut {NoStop}%
\bibitem [{\citenamefont {K\"ohler}(2002)}]{kohler2002}%
  \BibitemOpen
  \bibfield  {author} {\bibinfo {author} {\bibfnamefont {T.}~\bibnamefont
  {K\"ohler}},\ }\href {\doibase 10.1103/PhysRevLett.89.210404} {\bibfield
  {journal} {\bibinfo  {journal} {Phys. Rev. Lett.}\ }\textbf {\bibinfo
  {volume} {89}},\ \bibinfo {pages} {210404} (\bibinfo {year}
  {2002})}\BibitemShut {NoStop}%
\bibitem [{\citenamefont {Mestrom}\ \emph {et~al.}(2019)\citenamefont
  {Mestrom}, \citenamefont {Colussi}, \citenamefont {Secker},\ and\
  \citenamefont {Kokkelmans}}]{PhysRevA.100.050702}%
  \BibitemOpen
  \bibfield  {author} {\bibinfo {author} {\bibfnamefont {P.~M.~A.}\
  \bibnamefont {Mestrom}}, \bibinfo {author} {\bibfnamefont {V.~E.}\
  \bibnamefont {Colussi}}, \bibinfo {author} {\bibfnamefont {T.}~\bibnamefont
  {Secker}}, \ and\ \bibinfo {author} {\bibfnamefont {S.~J. J. M.~F.}\
  \bibnamefont {Kokkelmans}},\ }\href {\doibase 10.1103/PhysRevA.100.050702}
  {\bibfield  {journal} {\bibinfo  {journal} {Phys. Rev. A}\ }\textbf {\bibinfo
  {volume} {100}},\ \bibinfo {pages} {050702} (\bibinfo {year}
  {2019})}\BibitemShut {NoStop}%
\bibitem [{\citenamefont {Mestrom}\ \emph {et~al.}(2020)\citenamefont
  {Mestrom}, \citenamefont {Colussi}, \citenamefont {Secker}, \citenamefont
  {Groeneveld},\ and\ \citenamefont {Kokkelmans}}]{mestrom2019van}%
  \BibitemOpen
  \bibfield  {author} {\bibinfo {author} {\bibfnamefont {P.~M.~A.}\
  \bibnamefont {Mestrom}}, \bibinfo {author} {\bibfnamefont {V.~E.}\
  \bibnamefont {Colussi}}, \bibinfo {author} {\bibfnamefont {T.}~\bibnamefont
  {Secker}}, \bibinfo {author} {\bibfnamefont {G.~P.}\ \bibnamefont
  {Groeneveld}}, \ and\ \bibinfo {author} {\bibfnamefont {S.~J. J. M.~F.}\
  \bibnamefont {Kokkelmans}},\ }\href {\doibase 10.1103/PhysRevLett.124.143401}
  {\bibfield  {journal} {\bibinfo  {journal} {Phys. Rev. Lett.}\ }\textbf
  {\bibinfo {volume} {124}},\ \bibinfo {pages} {143401} (\bibinfo {year}
  {2020})}\BibitemShut {NoStop}%
\bibitem [{\citenamefont {Tan}(2008{\natexlab{a}})}]{PhysRevA.78.013636}%
  \BibitemOpen
  \bibfield  {author} {\bibinfo {author} {\bibfnamefont {S.}~\bibnamefont
  {Tan}},\ }\href {\doibase 10.1103/PhysRevA.78.013636} {\bibfield  {journal}
  {\bibinfo  {journal} {Phys. Rev. A}\ }\textbf {\bibinfo {volume} {78}},\
  \bibinfo {pages} {013636} (\bibinfo {year} {2008}{\natexlab{a}})}\BibitemShut
  {NoStop}%
\bibitem [{\citenamefont {Werner}\ and\ \citenamefont
  {Castin}(2006)}]{PhysRevLett.97.150401}%
  \BibitemOpen
  \bibfield  {author} {\bibinfo {author} {\bibfnamefont {F.}~\bibnamefont
  {Werner}}\ and\ \bibinfo {author} {\bibfnamefont {Y.}~\bibnamefont
  {Castin}},\ }\href {\doibase 10.1103/PhysRevLett.97.150401} {\bibfield
  {journal} {\bibinfo  {journal} {Phys. Rev. Lett.}\ }\textbf {\bibinfo
  {volume} {97}},\ \bibinfo {pages} {150401} (\bibinfo {year}
  {2006})}\BibitemShut {NoStop}%
\bibitem [{\citenamefont {Hadzibabic}\ \emph {et~al.}(2018)\citenamefont
  {Hadzibabic}, \citenamefont {Eigen}, \citenamefont {Glidden}, \citenamefont
  {Lopes}, \citenamefont {Smith},\ and\ \citenamefont
  {Cornell}}]{eigen_2018_data}%
  \BibitemOpen
  \bibfield  {author} {\bibinfo {author} {\bibfnamefont {Z.}~\bibnamefont
  {Hadzibabic}}, \bibinfo {author} {\bibfnamefont {C.}~\bibnamefont {Eigen}},
  \bibinfo {author} {\bibfnamefont {J.}~\bibnamefont {Glidden}}, \bibinfo
  {author} {\bibfnamefont {R.}~\bibnamefont {Lopes}}, \bibinfo {author}
  {\bibfnamefont {R.}~\bibnamefont {Smith}}, \ and\ \bibinfo {author}
  {\bibfnamefont {E.}~\bibnamefont {Cornell}},\ }\href {\doibase
  https://doi.org/10.17863/CAM.30242} {\  (\bibinfo {year} {2018}),\
  https://doi.org/10.17863/CAM.30242}\BibitemShut {NoStop}%
\bibitem [{\citenamefont {Sun}\ \emph {et~al.}(2020)\citenamefont {Sun},
  \citenamefont {Zhang},\ and\ \citenamefont {Zhai}}]{sun2020high}%
  \BibitemOpen
  \bibfield  {author} {\bibinfo {author} {\bibfnamefont {M.}~\bibnamefont
  {Sun}}, \bibinfo {author} {\bibfnamefont {P.}~\bibnamefont {Zhang}}, \ and\
  \bibinfo {author} {\bibfnamefont {H.}~\bibnamefont {Zhai}},\ }\href@noop {}
  {\bibfield  {journal} {\bibinfo  {journal} {arXiv:2006.07766v1
  [cond-mat.quant-gas]}\ } (\bibinfo {year} {2020})}\BibitemShut {NoStop}%
\bibitem [{\citenamefont {Griffin}(1996)}]{Griffin1996}%
  \BibitemOpen
  \bibfield  {author} {\bibinfo {author} {\bibfnamefont {A.}~\bibnamefont
  {Griffin}},\ }\href {\doibase 10.1103/PhysRevB.53.9341} {\bibfield  {journal}
  {\bibinfo  {journal} {Phys. Rev. B}\ }\textbf {\bibinfo {volume} {53}},\
  \bibinfo {pages} {9341} (\bibinfo {year} {1996})}\BibitemShut {NoStop}%
\bibitem [{\citenamefont {{Nozi\`eres, P.}}\ and\ \citenamefont {{Saint James,
  D.}}(1982)}]{James1982}%
  \BibitemOpen
  \bibfield  {author} {\bibinfo {author} {\bibnamefont {{Nozi\`eres, P.}}}\
  and\ \bibinfo {author} {\bibnamefont {{Saint James, D.}}},\ }\href {\doibase
  10.1051/jphys:019820043070113300} {\bibfield  {journal} {\bibinfo  {journal}
  {J. Phys. France}\ }\textbf {\bibinfo {volume} {43}},\ \bibinfo {pages}
  {1133} (\bibinfo {year} {1982})}\BibitemShut {NoStop}%
\bibitem [{\citenamefont {Yin}\ and\ \citenamefont
  {Radzihovsky}(2016)}]{PhysRevA.93.033653}%
  \BibitemOpen
  \bibfield  {author} {\bibinfo {author} {\bibfnamefont {X.}~\bibnamefont
  {Yin}}\ and\ \bibinfo {author} {\bibfnamefont {L.}~\bibnamefont
  {Radzihovsky}},\ }\href {\doibase 10.1103/PhysRevA.93.033653} {\bibfield
  {journal} {\bibinfo  {journal} {Phys. Rev. A}\ }\textbf {\bibinfo {volume}
  {93}},\ \bibinfo {pages} {033653} (\bibinfo {year} {2016})}\BibitemShut
  {NoStop}%
\bibitem [{\citenamefont {Eigen}\ \emph
  {et~al.}(2017{\natexlab{b}})\citenamefont {Eigen}, \citenamefont {Glidden},
  \citenamefont {Lopes}, \citenamefont {Navon}, \citenamefont {Hadzibabic},\
  and\ \citenamefont {Smith}}]{eigen_2017_data}%
  \BibitemOpen
  \bibfield  {author} {\bibinfo {author} {\bibfnamefont {C.}~\bibnamefont
  {Eigen}}, \bibinfo {author} {\bibfnamefont {J.}~\bibnamefont {Glidden}},
  \bibinfo {author} {\bibfnamefont {R.}~\bibnamefont {Lopes}}, \bibinfo
  {author} {\bibfnamefont {N.}~\bibnamefont {Navon}}, \bibinfo {author}
  {\bibfnamefont {Z.}~\bibnamefont {Hadzibabic}}, \ and\ \bibinfo {author}
  {\bibfnamefont {R.}~\bibnamefont {Smith}},\ }\href {\doibase
  https://doi.org/10.17863/CAM.16741} {\  (\bibinfo {year}
  {2017}{\natexlab{b}}),\ https://doi.org/10.17863/CAM.16741}\BibitemShut
  {NoStop}%
\bibitem [{\citenamefont {Tan}(2008{\natexlab{b}})}]{TAN20082952}%
  \BibitemOpen
  \bibfield  {author} {\bibinfo {author} {\bibfnamefont {S.}~\bibnamefont
  {Tan}},\ }\href {\doibase http://dx.doi.org/10.1016/j.aop.2008.03.004}
  {\bibfield  {journal} {\bibinfo  {journal} {Annals of Physics}\ }\textbf
  {\bibinfo {volume} {323}},\ \bibinfo {pages} {2952 } (\bibinfo {year}
  {2008}{\natexlab{b}})}\BibitemShut {NoStop}%
\bibitem [{\citenamefont {Tan}(2008{\natexlab{c}})}]{TAN20082971}%
  \BibitemOpen
  \bibfield  {author} {\bibinfo {author} {\bibfnamefont {S.}~\bibnamefont
  {Tan}},\ }\href {\doibase http://dx.doi.org/10.1016/j.aop.2008.03.005}
  {\bibfield  {journal} {\bibinfo  {journal} {Annals of Physics}\ }\textbf
  {\bibinfo {volume} {323}},\ \bibinfo {pages} {2971 } (\bibinfo {year}
  {2008}{\natexlab{c}})}\BibitemShut {NoStop}%
\bibitem [{\citenamefont {Tan}(2008{\natexlab{d}})}]{TAN20082987}%
  \BibitemOpen
  \bibfield  {author} {\bibinfo {author} {\bibfnamefont {S.}~\bibnamefont
  {Tan}},\ }\href {\doibase http://dx.doi.org/10.1016/j.aop.2008.03.003}
  {\bibfield  {journal} {\bibinfo  {journal} {Annals of Physics}\ }\textbf
  {\bibinfo {volume} {323}},\ \bibinfo {pages} {2987 } (\bibinfo {year}
  {2008}{\natexlab{d}})}\BibitemShut {NoStop}%
\bibitem [{\citenamefont {Werner}\ and\ \citenamefont
  {Castin}(2012{\natexlab{a}})}]{PhysRevA.86.053633}%
  \BibitemOpen
  \bibfield  {author} {\bibinfo {author} {\bibfnamefont {F.}~\bibnamefont
  {Werner}}\ and\ \bibinfo {author} {\bibfnamefont {Y.}~\bibnamefont
  {Castin}},\ }\href {\doibase 10.1103/PhysRevA.86.053633} {\bibfield
  {journal} {\bibinfo  {journal} {Phys. Rev. A}\ }\textbf {\bibinfo {volume}
  {86}},\ \bibinfo {pages} {053633} (\bibinfo {year}
  {2012}{\natexlab{a}})}\BibitemShut {NoStop}%
\bibitem [{\citenamefont {Braaten}\ \emph {et~al.}(2011)\citenamefont
  {Braaten}, \citenamefont {Kang},\ and\ \citenamefont
  {Platter}}]{PhysRevLett.106.153005}%
  \BibitemOpen
  \bibfield  {author} {\bibinfo {author} {\bibfnamefont {E.}~\bibnamefont
  {Braaten}}, \bibinfo {author} {\bibfnamefont {D.}~\bibnamefont {Kang}}, \
  and\ \bibinfo {author} {\bibfnamefont {L.}~\bibnamefont {Platter}},\ }\href
  {\doibase 10.1103/PhysRevLett.106.153005} {\bibfield  {journal} {\bibinfo
  {journal} {Phys. Rev. Lett.}\ }\textbf {\bibinfo {volume} {106}},\ \bibinfo
  {pages} {153005} (\bibinfo {year} {2011})}\BibitemShut {NoStop}%
\bibitem [{\citenamefont {Corson}\ and\ \citenamefont
  {Bohn}(2016)}]{PhysRevA.94.023604}%
  \BibitemOpen
  \bibfield  {author} {\bibinfo {author} {\bibfnamefont {J.~P.}\ \bibnamefont
  {Corson}}\ and\ \bibinfo {author} {\bibfnamefont {J.~L.}\ \bibnamefont
  {Bohn}},\ }\href {\doibase 10.1103/PhysRevA.94.023604} {\bibfield  {journal}
  {\bibinfo  {journal} {Phys. Rev. A}\ }\textbf {\bibinfo {volume} {94}},\
  \bibinfo {pages} {023604} (\bibinfo {year} {2016})}\BibitemShut {NoStop}%
\bibitem [{\citenamefont {Werner}\ and\ \citenamefont
  {Castin}(2012{\natexlab{b}})}]{PhysRevA.86.013626}%
  \BibitemOpen
  \bibfield  {author} {\bibinfo {author} {\bibfnamefont {F.}~\bibnamefont
  {Werner}}\ and\ \bibinfo {author} {\bibfnamefont {Y.}~\bibnamefont
  {Castin}},\ }\href {\doibase 10.1103/PhysRevA.86.013626} {\bibfield
  {journal} {\bibinfo  {journal} {Phys. Rev. A}\ }\textbf {\bibinfo {volume}
  {86}},\ \bibinfo {pages} {013626} (\bibinfo {year}
  {2012}{\natexlab{b}})}\BibitemShut {NoStop}%
\bibitem [{\citenamefont {Olshanii}\ and\ \citenamefont
  {Dunjko}(2003)}]{PhysRevLett.91.090401}%
  \BibitemOpen
  \bibfield  {author} {\bibinfo {author} {\bibfnamefont {M.}~\bibnamefont
  {Olshanii}}\ and\ \bibinfo {author} {\bibfnamefont {V.}~\bibnamefont
  {Dunjko}},\ }\href {\doibase 10.1103/PhysRevLett.91.090401} {\bibfield
  {journal} {\bibinfo  {journal} {Phys. Rev. Lett.}\ }\textbf {\bibinfo
  {volume} {91}},\ \bibinfo {pages} {090401} (\bibinfo {year}
  {2003})}\BibitemShut {NoStop}%
\bibitem [{\citenamefont {Rem}\ \emph {et~al.}(2013)\citenamefont {Rem},
  \citenamefont {Grier}, \citenamefont {Ferrier-Barbut}, \citenamefont
  {Eismann}, \citenamefont {Langen}, \citenamefont {Navon}, \citenamefont
  {Khaykovich}, \citenamefont {Werner}, \citenamefont {Petrov}, \citenamefont
  {Chevy},\ and\ \citenamefont {Salomon}}]{PhysRevLett.110.163202}%
  \BibitemOpen
  \bibfield  {author} {\bibinfo {author} {\bibfnamefont {B.~S.}\ \bibnamefont
  {Rem}}, \bibinfo {author} {\bibfnamefont {A.~T.}\ \bibnamefont {Grier}},
  \bibinfo {author} {\bibfnamefont {I.}~\bibnamefont {Ferrier-Barbut}},
  \bibinfo {author} {\bibfnamefont {U.}~\bibnamefont {Eismann}}, \bibinfo
  {author} {\bibfnamefont {T.}~\bibnamefont {Langen}}, \bibinfo {author}
  {\bibfnamefont {N.}~\bibnamefont {Navon}}, \bibinfo {author} {\bibfnamefont
  {L.}~\bibnamefont {Khaykovich}}, \bibinfo {author} {\bibfnamefont
  {F.}~\bibnamefont {Werner}}, \bibinfo {author} {\bibfnamefont {D.~S.}\
  \bibnamefont {Petrov}}, \bibinfo {author} {\bibfnamefont {F.}~\bibnamefont
  {Chevy}}, \ and\ \bibinfo {author} {\bibfnamefont {C.}~\bibnamefont
  {Salomon}},\ }\href {\doibase 10.1103/PhysRevLett.110.163202} {\bibfield
  {journal} {\bibinfo  {journal} {Phys. Rev. Lett.}\ }\textbf {\bibinfo
  {volume} {110}},\ \bibinfo {pages} {163202} (\bibinfo {year}
  {2013})}\BibitemShut {NoStop}%
\bibitem [{kin()}]{kineticsTBP}%
  \BibitemOpen
  \href@noop {} {}\bibinfo {note} {V. E. Colussi \emph{et al.} (to be
  published).}\BibitemShut {Stop}%
\bibitem [{\citenamefont {Wang}\ \emph {et~al.}(2012)\citenamefont {Wang},
  \citenamefont {D'Incao}, \citenamefont {Esry},\ and\ \citenamefont
  {Greene}}]{PhysRevLett.108.263001}%
  \BibitemOpen
  \bibfield  {author} {\bibinfo {author} {\bibfnamefont {J.}~\bibnamefont
  {Wang}}, \bibinfo {author} {\bibfnamefont {J.~P.}\ \bibnamefont {D'Incao}},
  \bibinfo {author} {\bibfnamefont {B.~D.}\ \bibnamefont {Esry}}, \ and\
  \bibinfo {author} {\bibfnamefont {C.~H.}\ \bibnamefont {Greene}},\ }\href
  {\doibase 10.1103/PhysRevLett.108.263001} {\bibfield  {journal} {\bibinfo
  {journal} {Phys. Rev. Lett.}\ }\textbf {\bibinfo {volume} {108}},\ \bibinfo
  {pages} {263001} (\bibinfo {year} {2012})}\BibitemShut {NoStop}%
\bibitem [{\citenamefont {Naidon}\ \emph {et~al.}(2014)\citenamefont {Naidon},
  \citenamefont {Endo},\ and\ \citenamefont {Ueda}}]{PhysRevA.90.022106}%
  \BibitemOpen
  \bibfield  {author} {\bibinfo {author} {\bibfnamefont {P.}~\bibnamefont
  {Naidon}}, \bibinfo {author} {\bibfnamefont {S.}~\bibnamefont {Endo}}, \ and\
  \bibinfo {author} {\bibfnamefont {M.}~\bibnamefont {Ueda}},\ }\href {\doibase
  10.1103/PhysRevA.90.022106} {\bibfield  {journal} {\bibinfo  {journal} {Phys.
  Rev. A}\ }\textbf {\bibinfo {volume} {90}},\ \bibinfo {pages} {022106}
  (\bibinfo {year} {2014})}\BibitemShut {NoStop}%
\bibitem [{\citenamefont {Smith}\ \emph {et~al.}(2014)\citenamefont {Smith},
  \citenamefont {Braaten}, \citenamefont {Kang},\ and\ \citenamefont
  {Platter}}]{PhysRevLett.112.110402}%
  \BibitemOpen
  \bibfield  {author} {\bibinfo {author} {\bibfnamefont {D.~H.}\ \bibnamefont
  {Smith}}, \bibinfo {author} {\bibfnamefont {E.}~\bibnamefont {Braaten}},
  \bibinfo {author} {\bibfnamefont {D.}~\bibnamefont {Kang}}, \ and\ \bibinfo
  {author} {\bibfnamefont {L.}~\bibnamefont {Platter}},\ }\href {\doibase
  10.1103/PhysRevLett.112.110402} {\bibfield  {journal} {\bibinfo  {journal}
  {Phys. Rev. Lett.}\ }\textbf {\bibinfo {volume} {112}},\ \bibinfo {pages}
  {110402} (\bibinfo {year} {2014})}\BibitemShut {NoStop}%
\bibitem [{\citenamefont {Faddeev}\ and\ \citenamefont
  {Merkuriev}(2013)}]{faddeev2013quantum}%
  \BibitemOpen
  \bibfield  {author} {\bibinfo {author} {\bibfnamefont {L.~D.}\ \bibnamefont
  {Faddeev}}\ and\ \bibinfo {author} {\bibfnamefont {S.~P.}\ \bibnamefont
  {Merkuriev}},\ }\href@noop {} {\emph {\bibinfo {title} {Quantum Scattering
  Theory for Several Particle Systems}}},\ Vol.~\bibinfo {volume} {11}\
  (\bibinfo  {publisher} {Springer Science \& Business Media},\ \bibinfo {year}
  {2013})\BibitemShut {NoStop}%
\bibitem [{\citenamefont {Gl\"ockle}(1983)}]{glockle1983}%
  \BibitemOpen
  \bibfield  {author} {\bibinfo {author} {\bibfnamefont {W.}~\bibnamefont
  {Gl\"ockle}},\ }\href@noop {} {\emph {\bibinfo {title} {The Quantum
  Mechanical Few-Body Problem}}},\ edited by\ \bibinfo {editor} {\bibfnamefont
  {W.}~\bibnamefont {Beiglb\"ock}}\ (\bibinfo  {publisher} {Springer,
  Berline},\ \bibinfo {year} {1983})\BibitemShut {NoStop}%
\bibitem [{\citenamefont {D'Errico}\ \emph {et~al.}(2007)\citenamefont
  {D'Errico}, \citenamefont {Zaccanti}, \citenamefont {Fattori}, \citenamefont
  {Roati}, \citenamefont {Inguscio}, \citenamefont {Modugno},\ and\
  \citenamefont {Simoni}}]{D_Errico_2007}%
  \BibitemOpen
  \bibfield  {author} {\bibinfo {author} {\bibfnamefont {C.}~\bibnamefont
  {D'Errico}}, \bibinfo {author} {\bibfnamefont {M.}~\bibnamefont {Zaccanti}},
  \bibinfo {author} {\bibfnamefont {M.}~\bibnamefont {Fattori}}, \bibinfo
  {author} {\bibfnamefont {G.}~\bibnamefont {Roati}}, \bibinfo {author}
  {\bibfnamefont {M.}~\bibnamefont {Inguscio}}, \bibinfo {author}
  {\bibfnamefont {G.}~\bibnamefont {Modugno}}, \ and\ \bibinfo {author}
  {\bibfnamefont {A.}~\bibnamefont {Simoni}},\ }\href {\doibase
  10.1088/1367-2630/9/7/223} {\bibfield  {journal} {\bibinfo  {journal} {New
  Journal of Physics}\ }\textbf {\bibinfo {volume} {9}},\ \bibinfo {pages}
  {223} (\bibinfo {year} {2007})}\BibitemShut {NoStop}%
\bibitem [{\citenamefont {Schmidt}\ \emph {et~al.}(2012)\citenamefont
  {Schmidt}, \citenamefont {Rath},\ and\ \citenamefont
  {Zwerger}}]{Schmidt2012}%
  \BibitemOpen
  \bibfield  {author} {\bibinfo {author} {\bibfnamefont {R.}~\bibnamefont
  {Schmidt}}, \bibinfo {author} {\bibfnamefont {S.}~\bibnamefont {Rath}}, \
  and\ \bibinfo {author} {\bibfnamefont {W.}~\bibnamefont {Zwerger}},\ }\href
  {\doibase 10.1140/epjb/e2012-30841-3} {\bibfield  {journal} {\bibinfo
  {journal} {Eur. Phys. J. B}\ }\textbf {\bibinfo {volume} {85}},\ \bibinfo
  {pages} {386} (\bibinfo {year} {2012})}\BibitemShut {NoStop}%
\bibitem [{\citenamefont {Skorniakov}\ and\ \citenamefont
  {Ter-Martirosian}(1957)}]{skorniakov1957three}%
  \BibitemOpen
  \bibfield  {author} {\bibinfo {author} {\bibfnamefont {G.}~\bibnamefont
  {Skorniakov}}\ and\ \bibinfo {author} {\bibfnamefont {K.}~\bibnamefont
  {Ter-Martirosian}},\ }\href@noop {} {\bibfield  {journal} {\bibinfo
  {journal} {Soviet Phys. JETP}\ }\textbf {\bibinfo {volume} {4}} (\bibinfo
  {year} {1957})}\BibitemShut {NoStop}%
\bibitem [{\citenamefont {Press}\ \emph {et~al.}(1989)\citenamefont {Press},
  \citenamefont {Flannery}, \citenamefont {Teukolsky}, \citenamefont
  {Vetterling} \emph {et~al.}}]{press1989numerical}%
  \BibitemOpen
  \bibfield  {author} {\bibinfo {author} {\bibfnamefont {W.~H.}\ \bibnamefont
  {Press}}, \bibinfo {author} {\bibfnamefont {B.~P.}\ \bibnamefont {Flannery}},
  \bibinfo {author} {\bibfnamefont {S.~A.}\ \bibnamefont {Teukolsky}}, \bibinfo
  {author} {\bibfnamefont {W.~T.}\ \bibnamefont {Vetterling}},  \emph
  {et~al.},\ }\href@noop {} {\emph {\bibinfo {title} {Numerical Recipes}}},\
  Vol.~\bibinfo {volume} {3}\ (\bibinfo  {publisher} {Cambridge University
  Press, Cambridge},\ \bibinfo {year} {1989})\BibitemShut {NoStop}%
\bibitem [{\citenamefont {Kremp}\ \emph {et~al.}(1997)\citenamefont {Kremp},
  \citenamefont {Bonitz}, \citenamefont {Kraeft},\ and\ \citenamefont
  {Schlanges}}]{KREMP1997320}%
  \BibitemOpen
  \bibfield  {author} {\bibinfo {author} {\bibfnamefont {D.}~\bibnamefont
  {Kremp}}, \bibinfo {author} {\bibfnamefont {M.}~\bibnamefont {Bonitz}},
  \bibinfo {author} {\bibfnamefont {W.}~\bibnamefont {Kraeft}}, \ and\ \bibinfo
  {author} {\bibfnamefont {M.}~\bibnamefont {Schlanges}},\ }\href {\doibase
  https://doi.org/10.1006/aphy.1997.5703} {\bibfield  {journal} {\bibinfo
  {journal} {Annals of Physics}\ }\textbf {\bibinfo {volume} {258}},\ \bibinfo
  {pages} {320 } (\bibinfo {year} {1997})}\BibitemShut {NoStop}%
\bibitem [{\citenamefont {Blume}\ \emph {et~al.}(2018)\citenamefont {Blume},
  \citenamefont {Sze},\ and\ \citenamefont {Bohn}}]{PhysRevA.97.033621}%
  \BibitemOpen
  \bibfield  {author} {\bibinfo {author} {\bibfnamefont {D.}~\bibnamefont
  {Blume}}, \bibinfo {author} {\bibfnamefont {M.~W.~C.}\ \bibnamefont {Sze}}, \
  and\ \bibinfo {author} {\bibfnamefont {J.~L.}\ \bibnamefont {Bohn}},\ }\href
  {\doibase 10.1103/PhysRevA.97.033621} {\bibfield  {journal} {\bibinfo
  {journal} {Phys. Rev. A}\ }\textbf {\bibinfo {volume} {97}},\ \bibinfo
  {pages} {033621} (\bibinfo {year} {2018})}\BibitemShut {NoStop}%
\bibitem [{\citenamefont {Castin}\ and\ \citenamefont
  {Dum}(1998)}]{PhysRevA.57.3008}%
  \BibitemOpen
  \bibfield  {author} {\bibinfo {author} {\bibfnamefont {Y.}~\bibnamefont
  {Castin}}\ and\ \bibinfo {author} {\bibfnamefont {R.}~\bibnamefont {Dum}},\
  }\href {\doibase 10.1103/PhysRevA.57.3008} {\bibfield  {journal} {\bibinfo
  {journal} {Phys. Rev. A}\ }\textbf {\bibinfo {volume} {57}},\ \bibinfo
  {pages} {3008} (\bibinfo {year} {1998})}\BibitemShut {NoStop}%
\bibitem [{\citenamefont {Gardiner}(1997)}]{PhysRevA.56.1414}%
  \BibitemOpen
  \bibfield  {author} {\bibinfo {author} {\bibfnamefont {C.~W.}\ \bibnamefont
  {Gardiner}},\ }\href {\doibase 10.1103/PhysRevA.56.1414} {\bibfield
  {journal} {\bibinfo  {journal} {Phys. Rev. A}\ }\textbf {\bibinfo {volume}
  {56}},\ \bibinfo {pages} {1414} (\bibinfo {year} {1997})}\BibitemShut
  {NoStop}%
\bibitem [{\citenamefont {Yin}\ and\ \citenamefont
  {Radzihovsky}(2013)}]{PhysRevA.88.063611}%
  \BibitemOpen
  \bibfield  {author} {\bibinfo {author} {\bibfnamefont {X.}~\bibnamefont
  {Yin}}\ and\ \bibinfo {author} {\bibfnamefont {L.}~\bibnamefont
  {Radzihovsky}},\ }\href {\doibase 10.1103/PhysRevA.88.063611} {\bibfield
  {journal} {\bibinfo  {journal} {Phys. Rev. A}\ }\textbf {\bibinfo {volume}
  {88}},\ \bibinfo {pages} {063611} (\bibinfo {year} {2013})}\BibitemShut
  {NoStop}%
\bibitem [{\citenamefont {Ran\ifmmode~\mbox{\c{c}}\else \c{c}\fi{}on}\ and\
  \citenamefont {Levin}(2014)}]{PhysRevA.90.021602}%
  \BibitemOpen
  \bibfield  {author} {\bibinfo {author} {\bibfnamefont {A.}~\bibnamefont
  {Ran\ifmmode~\mbox{\c{c}}\else \c{c}\fi{}on}}\ and\ \bibinfo {author}
  {\bibfnamefont {K.}~\bibnamefont {Levin}},\ }\href {\doibase
  10.1103/PhysRevA.90.021602} {\bibfield  {journal} {\bibinfo  {journal} {Phys.
  Rev. A}\ }\textbf {\bibinfo {volume} {90}},\ \bibinfo {pages} {021602}
  (\bibinfo {year} {2014})}\BibitemShut {NoStop}%
\bibitem [{\citenamefont {Žitko}(2011)}]{Zitko2011}%
  \BibitemOpen
  \bibfield  {author} {\bibinfo {author} {\bibfnamefont {R.}~\bibnamefont
  {Žitko}},\ }\href {\doibase https://doi.org/10.1016/j.cpc.2011.05.013}
  {\bibfield  {journal} {\bibinfo  {journal} {Computer Physics Communications}\
  }\textbf {\bibinfo {volume} {182}},\ \bibinfo {pages} {2259 } (\bibinfo
  {year} {2011})}\BibitemShut {NoStop}%
\end{thebibliography}%

\end{document}